\documentclass[sigconf]{acmart}

\usepackage{booktabs}
\usepackage{multirow}
\usepackage{graphicx}
\usepackage{geometry}
\usepackage{hyperref}

\usepackage{array}

\newcolumntype{P}[1]{>{\raggedright\arraybackslash}p{#1}}

\usepackage{pdflscape}
\usepackage{afterpage}

\newcommand{\papercount}[1]{\textbf{\textcolor{blue}{(#1)}}}
\newcommand{\papercountnewline}[1]{\textbf{\newline\papercount{#1}}}

\AtBeginDocument{%
  \providecommand\BibTeX{{%
    \normalfont B\kern-0.5em{\scshape i\kern-0.25em b}\kern-0.8em\TeX}}}

\setcopyright{rightsretained}

\acmConference[CHI '21]{CHI '21: ACM CHI Conference on Human Factors in Computing Systems}{May 08--13, 2021}{Yokohama, Japan}
\acmISBN{978-1-4503-XXXX-X/21/05}

\begin{document}

\title[Beyond Expertise and Roles: Characterizing Interpretable ML Stakeholders and Needs]{Beyond Expertise and Roles: A Framework to Characterize the Stakeholders of Interpretable Machine Learning and their Needs}


\author{Harini Suresh}
\email{hsuresh@mit.edu}
\affiliation{%
  \institution{MIT}
}

\author{Steven R. Gomez}
\email{steven.gomez@mit.edu}
\affiliation{%
  \institution{MIT}
}

\author{Kevin K. Nam}
\email{kevin.nam@mit.edu}
\affiliation{%
  \institution{MIT}
}

\author{Arvind Satyanarayan}
\email{arvindsatya@mit.edu}
\affiliation{%
  \institution{MIT}
}

\renewcommand{\shortauthors}{Suresh et. al.}

\begin{abstract}
To ensure accountability and mitigate harm, it is critical that diverse stakeholders can interrogate black-box automated systems and find information that is understandable, relevant, and useful to them. In this paper, we eschew prior expertise- and role-based categorizations of interpretability stakeholders in favor of a more granular framework that decouples stakeholders' knowledge from their interpretability needs. We characterize stakeholders by their formal, instrumental, and personal knowledge and how it manifests in the contexts of machine learning, the data domain, and the general milieu. We additionally distill a hierarchical typology of stakeholder needs that distinguishes higher-level domain goals from lower-level interpretability tasks. 
In assessing the descriptive, evaluative, and generative powers of our framework, we find our more nuanced treatment of stakeholders reveals gaps and opportunities in the interpretability literature, adds precision to the design and comparison of user studies, and facilitates a more reflexive approach to conducting this research. 
\end{abstract}


\copyrightyear{2021}
\acmYear{2021}
\acmConference[CHI '21]{CHI Conference on Human Factors in Computing Systems}{May 8--13, 2021}{Yokohama, Japan}
\acmBooktitle{CHI Conference on Human Factors in Computing Systems (CHI '21), May 8--13, 2021, Yokohama, Japan}\acmDOI{10.1145/3411764.3445088}
\acmISBN{978-1-4503-8096-6/21/05}


\maketitle

\section{Introduction}Automated systems based on machine learning (ML) and artificial intelligence (AI) are often described as ``black boxes'' due to the difficulty of extracting the logic behind their outputs in an understandable way.   The inscrutability of such systems makes it difficult for them to facilitate effective trust-building and to be held accountable for the decisions they affect.  
A growing body of recent work has focused on tackling these issues through \textit{model interpretability}, which involves producing visual explanations about a model's behavior for its users.  But to design effective interpretability mechanisms, we need to first consider the question: who, exactly, are the stakeholders involved, and what are they trying to achieve? 

Take, for example, an ML-based medical decision-making tool.  The physicians using the system need to be able to align the output with their own clinical expectations and justify their recommendations to patients.  Patients, then, need to have some confidence in the validity of these recommendations, and may want to explain decisions to family members.  Other medical staff need to understand the decision-making processes insofar as it affects the treatment they administer to the patient. The developers who created the system should be able to monitor its performance and understand how to make improvements. Physicians and patients, as well, may want and be well-suited to provide feedback about on-the-ground errors the system makes.  And there are undoubtedly other people involved in or affected by this system, from external legal agencies to all the people whose data went into training the ML model. 

Existing methods for model interpretability, however, often do not explicitly identify or describe their intended user.  As a result, many of these methods inadvertently end up being most understandable to the people that build them (i.e., ML researchers or developers).  In other cases, the recipient of the interpretability system is described generally as a ``layperson'' or ``end user''; resulting methods may produce simpler visuals, but experimental studies have shown they too often are not useful for people in practice \cite{poursabzi-sangdeh_manipulating_2019, lai_human_2019, bussone_role_2015, sureshMisplaced}.  In our prior example, doctors, patients, and medical staff may all be considered ``end users,'' but have significantly different needs and goals when interpreting, understanding, and reacting to the output of the ML model.  
Indeed, when it comes down to it, many organizations say they want to give users insight into ML systems through interpretability mechanisms, but these methods are only actually used internally by developers~\cite{bhatt_explainable_2020}.

Part of this disconnect stems from the difficulty in identifying and characterizing different stakeholders and their interpretability needs.  
A growing body of work has engaged with this problem, proposing an ecosystem of stakeholders~\cite{tomsett_interpretable_2018, preece_stakeholders_2018}, and conducting literature surveys~\cite{yu_user-based_2018, hohman_visual_2019, ferreira_what_2020, mohseni_multidisciplinary_2020} and interview studies~\cite{cai_hello_2019, hong_human_2020, tonekaboni_what_2019} to understand their goals.
Resultant frameworks typically adopt one of two approaches: they either categorize stakeholders by their expertise (using labels such as ``experts'', ``novices'', or ``non-experts''~\cite{yu_user-based_2018, hohman_visual_2019, mohseni_multidisciplinary_2020}) or by their functional role in the ecosystem (e.g., ``executives'' and ``engineers''~\cite{bhatt_explainable_2020}, model ``breakers'' and ``consumers''~\cite{hong_human_2020}, or model ``operators'' and ``executors''~\cite{tomsett_interpretable_2018}).
Stakeholder needs and goals then follow from these categories.
While usefully advancing our understanding of the stakeholders involved, these initial frameworks are limited in their descriptive and generative powers~\cite{beaudouin2004designing}.
For instance, role-based frameworks implicitly conflate a person's expertise with what they need from the system, with roles often depicted as a relatively static constructs. 
And expertise-based categories typically portray stakeholders lying on roughly linear scales that only account for cognitive notions of expertise and, thus, do not acknowledge the rich tacit knowledge and lived experience they may possess.

 
In response, we introduce a framework with a more granular and composable vocabulary to characterize the stakeholders of interpretable machine learning, and their needs.
Our framework comprises two components. 
First, we decompose stakeholder expertise into two dimensions that describe the types of knowledge a stakeholder may possess (i.e., formal, instrumental, and personal knowledge), and the contexts in which this knowledge manifests (i.e., machine learning, the data domain, and the milieu).
Second, we define stakeholder needs using a three-level typology of long-term goals, shorter-term objectives that target these goals, and immediate tasks that stakeholders perform to meet their objectives.
 

To understand the implications of our framework, we assess its descriptive, evaluative, and generative powers\,---\,three properties of interaction models first described by Michel Beaudouin-Lafon~\cite{beaudouin2004designing}.
We code 58 papers describing interpretability systems or users, and find that our framework is consistently able to describe stakeholders' knowledge and interpretability needs while adding granularity and drawing new connections between them.  
We describe how our framework's abstractions can allow us to design more precise application-grounded evaluations~\cite{doshi-velez_towards_2017}, including bringing precision to participant recruiting and providing a structure to operationalize comparative studies.
And finally, we demonstrate that our framework generates a rich intersection of user expertise and needs for study, and can also be turned inwards to facilitate a more reflexive design process. 
\section{Background and Motivation}
Researchers have recognized that more precisely defining ``interpretability'' or ``explainability'' is a key challenge for the field~\cite{lipton_mythos_2018}. 
Although some work seeks to develop formal or technical definitions of interpretability \cite{doshi-velez_towards_2017,gilpin_explaining_2018,dhurandhar2017formal}, and though the burgeoning set of interpretability techniques often do not name specific target users or tasks \cite{ribeiro2016model,zilke2016deepred,fleet_visualizing_2014,bach2015pixel,du2019techniques}, there is a growing recognition for the need to approach this problem space in a human-oriented manner. 
In this section, we motivate our contribution by surveying prior work and describing the limitations we observe with current approaches for defining the \emph{why} and \emph{who} of machine learning interpretability. 

In early definitions, \citet{lipton_mythos_2018} and Doshi-Velez \& Kim~\cite{doshi-velez_towards_2017} identified that the need for interpretability primarily stems from a mismatch between the formal definition of the machine learning model, and its output and real-world impact. 
Lipton further expanded on this need by enumerating a set of desiderata for interpretability including building trust in the model, inferring causal relationships between the input and output, improving model tranferability and generalizability, providing introspection, and finally to facilitate fair and ethical decision-making. Others have since contributed to this list in a variety of ways including detailing how different interpretability methods can be chosen to mitigate particular cognitive biases~\cite{wang2019designing}, proposing taxonomies of questions used to arrive at an appropriate interpretability method~\cite{arya_one_2019}, discussing how applications with different contexts or levels of automation might necessitate different design decisions~\cite{shneiderman2020human,lim2009assessing}, and grounding the need for explanations in the social sciences~\cite{miller_explanation_2019}. 

Most relevant to our paper is a body of work that seeks to better define interpretability by studying the specific users involved. 
In surveying this work, we identified two distinct approaches to doing so. First are a group of papers that characterize users based on their expertise. 
For instance, both Yu \& Shi~\cite{yu_user-based_2018} and Hohman et al.~\cite{hohman_visual_2019} classify users on roughly linear scales of machine learning expertise (from beginner to expert for Yu \& Shi, while Hohman et al. adopt the terms ``model developers and builders,'' ``model users,'' and ``non-experts'').
Similarly, Mohseni et al.~\cite{mohseni_multidisciplinary_2020} identify ``AI Novices'' and ``AI Experts'' and add ``Data Experts'' to the mix. 
With all of these schemes, a user's needs then stem from their expertise. 
For instance, novices are typically described as needing educational or teaching tools, whereas experts require tools for debugging or deploying models, or assessing model performance. 

The second category of papers characterize users based on their functional role instead. 
For instance, Tomsett et al.~\cite{tomsett_interpretable_2018} posit an ecosystem of stakeholders including model creators, operators, executors and examiners, as well as the decision and data subjects that are affected by the model or whose data the model was trained with, respectively. 
Similarly, through semi-structured interviews, Bhatt et al.~\cite{bhatt_explainable_2020} identify four categories of stakeholders (executives, ML engineers, end users, and others) while Hong et al.~\cite{hong_human_2020} identify model builders, breakers, and consumers.
Across this work, the role a person inhabits within an organization (or the role they play during the human-AI interaction) determines their interpretability needs.  
For example, model creators/builders/engineers are said to want introspection of the level of individual instances and features, model operators/breakers may wish to monitor the performance of the model including authoring test cases, and finally model executors/executives/consumers want to be able to have confidence and trust in the model. 
Cai et al.~\cite{cai_hello_2019} and Tonekaboni et al.~\cite{tonekaboni_what_2019} follow this approach of role-based needfinding as well by interviewing clinicians. 

While both expertise-oriented and role-oriented frameworks have usefully brought further definition to the problem of machine learning interpretability, we can observe limitations to their descriptive and generative powers (i.e., the degree to which they describe \emph{existing} points, and help us identify \emph{new} points in the problem space, respectively~\cite{beaudouin2004designing}).
Role-based frameworks, for instance, do not break the problem space down into sufficiently granular and composable units. 
As a result, several roles appear to implicitly conflate expertise and interpretability needs\,---\,for example, model ``creators'' are likely most expert with machine learning, and thus need debugging tools at the level of individual instances or features; but, one could imagine ``auditors'' appreciating insight at this level of abstraction even if they lack an equivalent level of machine learning expertise.
Similarly, consider the domain of clinical diagnoses: model ``consumers'' could equally describe doctors and patients despite these users likely requiring different explanations of the model's output as a result of different levels of medical expertise.
Here, model ``executors'' does not provide much more precision as both doctors and patients are tasked with making decisions informed by the model\,---\,doctors about what treatment to prescribe, and patients about whether they do indeed wish to proceed with the treatment.
Finally, although most role-based frameworks explicitly note that roles are not mutually exclusive (i.e., a single role may map to more than one individual, and one person may play several roles), roles are nevertheless depicted as relatively static constructs.
Not only might an individual user's role change over time but, even if they remain in the same role(s), their interpretability needs may change through repeated exposure to and increased familiarity with the models they are working with, or the situations in which these models are deployed. 

Expertise-based frameworks exhibit similar limitations. 
In particular, a key concern is how these frameworks portray expertise as a linear scale from ``novice'' to ``expert.''
Several external literatures have articulated concerns with this framing of expertise. 
For instance, in critiquing the influential Dreyfus linear model of skill acquisition~\cite{dreyfus1986power}, Dall'Alba \& Sandberg note that \emph{``[s]tage models of development appear to assume we know what skillful performance entails for each area of skill''} and that the \emph{``focus on stages veils more fundamental aspects of development; it directs attention away from the skill that is being developed''}~\cite{dall2006unveiling}. 
Moreover, Dall'Alba \& Sandberg point to the fact that such models are primarily concerned with cognitive development and fail to acknowledge expertise gained through embodied practice of a skill~\cite{dall2006unveiling}.
We see a form of this latter critique in the literature on participatory design as well, which advocates that all stakeholders in a design process possess valuable expertise through their lived experience and tacit knowledge~\cite{spinuzzi2005methodology, paml}.
Finally, although recent frameworks usefully consider domain expertise in addition to machine learning expertise, such a clean decoupling does not account for the ways expertise may transfer.
For example, Cai et al. find that while medical practitioners express a desire for an ``AI primer'', they are nevertheless able to bring some of their training and experience working with other clinical technologies to bear\,---\,for instance, in understanding that the output of a model will not be perfect, or in enumerating ``test cases'' for an AI assistant~\cite{cai_hello_2019}.
Similarly, as AI/ML-enabled technologies increasingly permeate every day life, this ubiquity and familiarity will shape users' interpretability needs in ways that current expertise-based frameworks leave unaddressed.

And, across the two types of frameworks, interpretability needs or goals are determined primarily by the category a user falls within. 
While many frameworks allow for categories to overlap, this approach nevertheless obscures the fact that many goals can cut across several roles or expertise.
For instance, almost every stakeholder involved will likely want to have trust in the model, and want to be able to assess the degree to which it may be biased\,---\,we see explicit evidence for this for machine learning experts~\cite{hohman_visual_2019} and data experts~\cite{mohseni_multidisciplinary_2020}, model creators and breakers~\cite{hong_human_2020}, model operators~\cite{cai_hello_2019, tonekaboni_what_2019}, as well as for decision- and data-subjects who may wish to contest a decision or otherwise seek recourse~\cite{alkhatib2019street, paml}.
Similarly, while current frameworks primarily pose debugging and improving the model as goals model creators, builders, or any other traditionally-``expert'' stakeholders may have, one could imagine that activists and other groups with non-traditional expertise may also wish to assess the outcome of domain-specific test cases. 

In summary, recent work has recognized that better defining the problem space is a key challenge for machine learning interpretability, and has advanced our understanding by contributing frameworks for describing the stakeholders involved and their goals or needs. 
However, in analyzing the descriptive and generative powers of these frameworks, we see several limitations.
In particular, by not providing a sufficiently granular or composable vocabulary, existing frameworks poorly distinguish the rich intersection that exists between attributes of the stakeholder (e.g., their expertise), the role that they may play (e.g., model creator or consumer), and their ultimate goals or needs with regards to interpretability (e.g., debugging the model, or building trust).

\section{A Framework to Characterize the Stakeholders of Interpretable ML}
\label{sec:framework}

To develop a more granular and composable vocabulary for describing the stakeholders of interpretable machine learning, we engaged in an iterative process with alternating phases to diverge and converge our thinking.
In particular, we began by surveying the literature on interpretability summarized in the previous section, and extracting passages that described users and stakeholders, as well as their needs, actions, and goals. 
To diverge our thinking, we looked to domains outside of interpretability and computer science, including the literatures on expertise and pedagogy~\cite{fleck1998expertise, eraut2010knowledge, yielder_integrated_2004, yielder_professional_2001, williams_exploring_1998, billett_learning_2010, kinchin_reconsidering_2010, hartelius_2008, hartelius_review_2012, hartelius_rhetorics_2011, dall2006unveiling}, critical theory~\cite{ogbonnaya-ogburu_critical_2020, thatcher_data_2016, mohamed_decolonial_2020, alkhatib2019street, le_dantec_strangers_2015}, law \cite{wachter_counterfactual_2018,zarsky2013transparent,hildebrandt2012,citron2014scored,doshi2017accountability}, and participatory action research \cite{greenwood_introduction_2007,jull_community-based_2017}.
To converge our thinking, we reflected on how concepts from these external domains could be adapted within interpretability.
This reflection process involved alternating phases of open coding to map external concepts to the passages we had initially extracted, affinity diagramming to identify recurring groupings and patterns between codes, analytic memo writing, and weekly hour-long conversations between all authors. 

Our framework comprises two halves. First, it describes the knowledge stakeholders may possess and the contexts this knowledge may manifest in. And, second, it enumerates the long-term goals stakeholders may have, and breaks these goals down into shorter-term objectives and specific tasks they can perform.

\subsection{Decomposing Stakeholder Expertise into Knowledge and Context}

\renewcommand{\arraystretch}{1.7}
\begin{table*}[tb]
    \centering
    \caption{Examples of how the three types of knowledge manifest in the three contexts described by our framework.}
    \label{tab:my-table}
    \resizebox{\textwidth}{!}{%
        \begin{tabular}{l p{.35\linewidth} p{.35\linewidth} p{.35\linewidth}} \toprule
            & \multicolumn{3}{c}{\textbf{Knowledge}} \\ \cmidrule(lr){2-4} 
            \textbf{Context} & \multicolumn{1}{c}{\textbf{Formal}} & \multicolumn{1}{c}{\textbf{Instrumental}} & \multicolumn{1}{c}{\textbf{Personal}} \\ \midrule
            \textbf{ML}
            & The math behind model architectures, optimization and training processes, etc.
            & Familiarity with ML toolkits,  off-the-shelf models, etc.
            & Tricks of the trade (e.g., hyperparameter values, feature engineering, etc.) \\ \hline
            
            \textbf{Data Domain}
            & Theories relevant to the data domain (e.g., symptoms and treatments, case law and legal precedent, etc.)
            & Experience working with other related technology (e.g., medical devices, document mining tools, etc.)
            & Lived experience (e.g., prior memories of similar events) \\ \hline
            
            \textbf{Milieu}
            & Sociocultural theories (e.g., redlining, gerrymandering, mass incarceration, etc.)
            & Familiarity with broader ML-enabled systems (e.g., virtual assistants, recommendation algorithms, etc.)
            & Lived experience and cultural knowledge (e.g., values, attitudes) \\
            \bottomrule
        \end{tabular}
    }
\end{table*}

Prior work has identified, either explicitly~\cite{hohman_visual_2019} or implicitly via roles~\cite{tomsett_interpretable_2018,hong_human_2020,mohseni_multidisciplinary_2020}, that expertise is a defining attribute of interpretable ML stakeholders.
To provide a more granular treatment of expertise, we adapt models of expertise from Fleck~\cite{fleck1998expertise} and Eraut~\cite{eraut2010knowledge} to decompose a singular notion of expertise into three constituent types of \emph{knowledge}.
\textbf{Formal} knowledge comprises an understanding of codified theories, embodied in text or diagrams such as those found in textbooks, and is acquired through a prolonged educational process.
\textbf{Instrumental} knowledge is an understanding of how to ``apply'' formal knowledge.
It is embodied in the use of tools or other instruments, and is learnt through demonstration and practice. 
Finally, \textbf{personal} knowledge describes information that is entirely embodied in individual people, and is gained through their participation in specific domains. 
It is difficult to codify~\cite{eraut2010knowledge} as it consists of a person's lived experience (e.g., memories of specific events, self-knowledge about the way they may react in certain scenarios, etc.) as well as values that may be distributed in the cultures and societies they are a member of.

These types of knowledge manifest in \emph{contexts}, or the domains or situations that determine what knowledge is relevant.
We identify three contexts: \textbf{machine learning}, or the knowledge required to research, develop, operate, or deploy machine learning models; the \textbf{data domain}, or the knowledge necessary to collect, organize, analyze, and communicate the data the model was trained with or makes decisions about; and \textbf{milieus}, which refer to the environments that the human-AI interaction may be occurring within. 
These environments include both the physical surroundings (e.g., a home, bank, courthouse, doctor's office, etc.) as well as the broader sociocultural context (e.g., mass incarceration, redlining, gerrymandering, etc.).

Our framework provides a more expansive yet precise treatment of expertise in interpretable ML. 
While prior expertise- or role-based approaches latently encode notions of formal and instrumental knowledge, by explicitly articulating these concepts, our framework facilitates teasing apart differences and understanding the implications on interpretability design.
For example, ``model users''~\cite{hohman_visual_2019} and ``model breakers''~\cite{hong_human_2020} cover an extremely broad range of possible stakeholders including model architects, trainers, engineers, data scientists, and machine learning artists~\cite{hohman_visual_2019}, as well as domain experts, product managers, and auditors~\cite{hong_human_2020}, respectively.
These categories appear primarily focused on stakeholders' instrumental knowledge and, by analyzing contexts, we can separate machine learning instrumentalists (model architects, trainers) from data domain instrumentalists (artists, domain experts, product managers), and those that may span the two (data scientists, auditors).
Doing so suggests that these stakeholder groups may have different interpretability needs that the broader categories of ``model users'' or ``model breakers'' obscured. 
For instance, perhaps interfaces for machine learning instrumentalists should be articulated in terms of the components exposed by popular toolkits. 
Similarly, for data domain instrumentalists, how might we analogize interpretability to tools and systems that they already work with in order to enable expertise transfer (akin to how Cai et al. found medical practitioners reasoning about uncertainty~\cite{cai_hello_2019})?

Moreover, our framework explicitly recognizes the personal knowledge stakeholders may have\,---\,including ``tricks of the trade'' a person may have acquired, their experiences and memories, or the more distributed values of the cultures and societies they are a member of\,---\,as an important consideration when designing for interpretability.
Critically, by placing it alongside formal and instrumental knowledge, our framework identifies it as an \emph{equally important} form of knowledge.
As a result, one might consider designing \emph{for} stakeholders' personal knowledge\,---\,for instance, using example-based explanations such that a stakeholder might better ``see themselves'' in the data~\cite{aamodt1994case, renkl2014toward, peck2019data}.
But our framework also suggests designing \emph{with} stakeholders, to better account for personal knowledge that designers do not have\,---\,a position advocated for by various communities including participatory action research~\cite{greenwood_introduction_2007} \& design~\cite{spinuzzi2005methodology, paml}.
For instance, members of the general public might have different notions of what constitutes an ``error'' based on their personal knowledge \cite{haraway1988situated}.

Our framework also highlights that interpretability design must attend to more than the immediate contexts of machine learning and the data domain\,---\,explanations must be situated in stakeholders' milieu.
Here, we draw an analogy to data visualization.
Researchers and data journalists consider annotations to be a crucial component of effective visualization design because it helps readers understand the broader context associated with the visualized data.
As Amanda Cox, Data Editor for The New York Times, says, \textit{``the annotation layer is the most important thing we do ... otherwise it's a case of here it is, you go figure it out}~\cite{amandacox-eyeo}.
We believe this property holds true for interpretability as well\,---\,it is insufficient for an explanation to be articulated purely in terms of the model or data if it misses critical aspects of the milieu.
For instance, consider an ML-backed loan evaluation system:
explanations in the ML context would articulate the output decision in terms of model components, while explanations in the data domain might also discuss distributions in the training or test set and how this may lead to biased output.
However, under our framework, we would consider these explanations to be incomplete if they were not situated the broader sociocultural milieu\,---\,for instance, how disparities in data distributions have occurred through policies such as redlining, or in the difficulty ex-offenders have in finding employment.

Finally, by decoupling knowledge and context as two orthogonal dimensions of the problem space, our framework enables a more systematic analysis of the stakeholders of interpretability.
It eschews prior easily-quantifiable linear scales in favor of more descriptive treatments of expertise. 
Designers can work with each dimension individually\,---\,for example, how might interpretability help stakeholders formalize their personal knowledge in the data domain by scaffolding example-based explanation with featured-based saliency methods akin to faded worked examples~\cite{atkinson2000learning}; or, as described previously, how might instrumental knowledge in the data domain transfer to the machine learning context~\cite{cai_hello_2019}. 
And, by considering the intersection of the two dimensions as well, our framework can help us identify the ways in which expertise recurs in the interpretability ecosystem.





\subsection{Distilling Stakeholder Needs into Goals, Objectives, and Tasks} \label{goal_objective_task}

\begin{figure*}
    \centering
    \includegraphics[width=.75\textwidth]{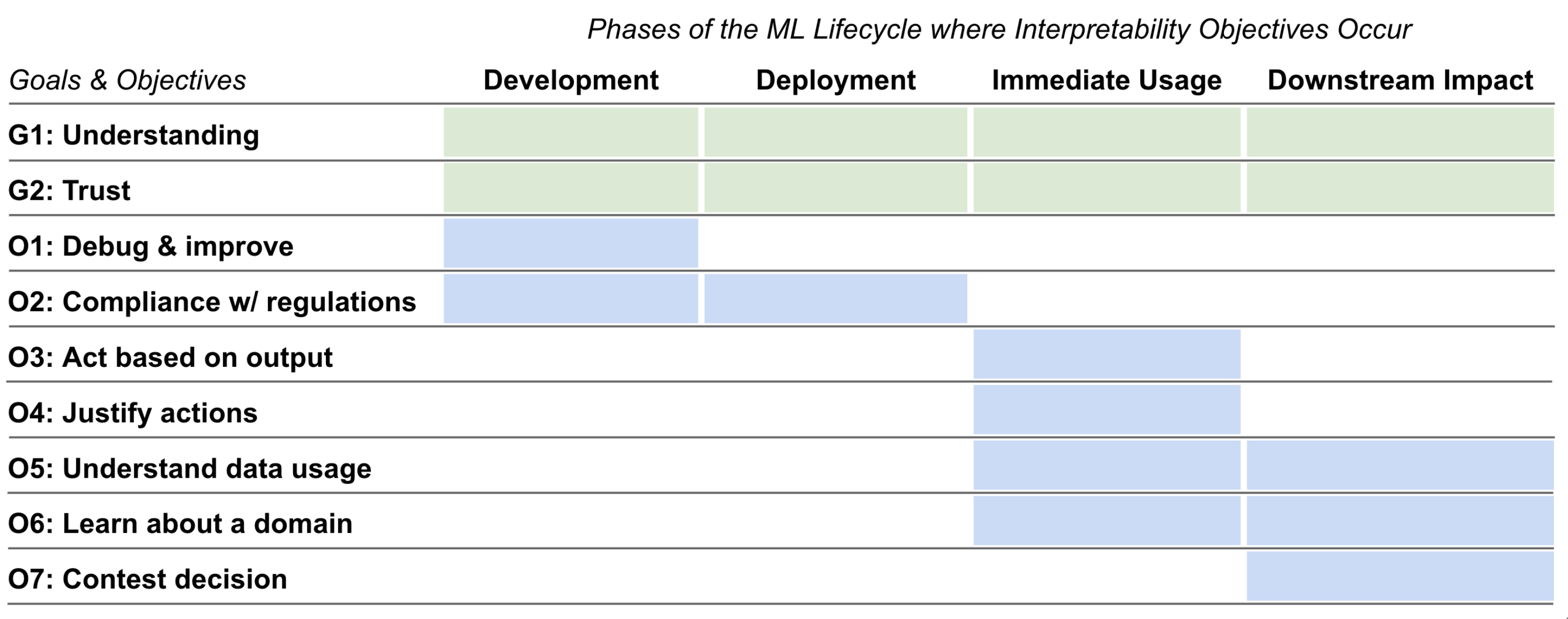}
    \Description[A visualization of the latent chronology in goals and objectives.]{A table structure where rows list goals and objectives, and columns list four phases of the ML process (Development, Deployment, Immediate Usage and Downstream Impact). The following phases for each goal or objective are colored: G1 (all phases), G2 (all phases), O1 (Development), O2 (Development & Deployment), O3 (Immediate Usage), O4 (Immediate Usage), O5 (Immediate Usage & Downstream Impact), O6 (Immediate Usage & Downstream Impact), O7 (Downstream Impact).}
    \caption{A visualization of the latent chronology in goals and objectives. Categories along the horizontal axis are relevant phases of the ML process. Colored cells indicate the phase in which a particular goal or objective typically occurs. Phases need not occur linearly, may be iteratively revisited, and many different stakeholders may be involved at any given phase.}
    \label{fig:chronology}
\end{figure*}

Through our open coding and reflection process, we distilled a three-level typology of interpretability needs. 
The first level identifies two long-term \emph{goals}: \textbf{understanding the model (G1)} and \textbf{building trust in the model (G2)}. 
These goals are high-level and difficult to define precisely, but we include them in our framework to acknowledge that they underlie almost every single piece of work we read.
We do not expect stakeholders to be able to directly accomplish these goals, nor do we imagine that future methods or systems will address them squarely.
Rather, these goals function like substrates which inform and influence the two lower levels of needs we describe below. 
For instance, we expect stakeholders to develop an understanding of models over time\,---\,through repeated exposure and interactions.
And much work has framed trust as something society as a whole needs in order to accept new technologies \cite{wachter_counterfactual_2018,brennen_what_2020,zarsky2013transparent}\,---\,indeed, trust may grow as stakeholders better understand the model, but may also develop in a proxied or deferred fashion through increased regulations and standards.

The second level of our typology describes the shorter-term \emph{objectives} stakeholders might target to achieve their longer-term goals. We give real examples for each objective, and demonstrate how they can be relevant to a diverse range of stakeholders.
These objectives are grounded in stakeholders' current real-world needs, but as ML tools continue to be deployed in new domains, we expect this typology will continue to evolve. 

\begin{enumerate}
    \item [\textbf{(O1)}] \textbf{Debug or improve a model.} The objective of improving a system or correcting its mistakes appears frequently in the literature, and is often posed in terms of the needs of developers~\cite{brennen_what_2020, liao_questioning_2020,hepenstal2020}.  For example, \citet{bhatt_explainable_2020} describe how internal members of organizations try to use interpretability techniques to uncover inconsistencies between the model’s logic and their intuition or expectations, in order to guide further improvements. 
    However, it is critical to acknowledge that developers are not the only stakeholder group to which this need applies.
    For instance, \citet{tonekaboni_what_2019} and \citet{zarsky2013transparent} both highlight the value of allowing a larger group of stakeholders, including the general public, to provide feedback for improving systems. Indeed, theories from Participatory Action Research also hold that people on the ground in a specific context are often much better suited to realize errors and devise appropriate fixes, as opposed to developers for whom the errors typically have no direct consequence~\cite{greenwood_introduction_2007}. 
    
    \item [\textbf{(O2)}] \textbf{Ensure compliance with standards or regulations.}  Auditing, or ensuring that the development, deployment, and results of a certain system are compliant with a particular set of standards (whether they are legal, ethical, safety, or other) is already necessary in other areas such as finance or aerospace, and is emerging as an important objective for ML systems as well.  The introduction of the GDPR, for example, has established a set of legal standards that automated systems must comply with.  And it is not only external watchdog agencies or governments that are interested in ensuring such compliance. Individuals or groups within an organization may also have their own internal standards they want to ensure are met\,---\,for example, \citet{raji2020closing} describe the design of an internal auditing pipeline.

    \item [\textbf{(O3)}] \textbf{Understand how to incorporate the model's output into downstream actions.}  Several prior papers mention the need for guidance on whether and how to incorporate model predictions into further actions\,---\,whether that involves relating the model’s output to relevant and actionable decisions, or understanding how much weight to place on the model’s prediction ~\cite{hong_human_2020, tonekaboni_what_2019, brennen_what_2020, bhatt_explainable_2020, liao_questioning_2020}.  This objective emerged as important for a number of different types of stakeholders, such as doctors using a diagnostic aid~\cite{hong_human_2020, tonekaboni_what_2019}, people applying for health insurance that involves automated screening~\cite{bhatt_explainable_2020}, or for those subjected to automated decisions more generally \cite{wachter_counterfactual_2018}.

    \item [\textbf{(O4)}] \textbf{Justify or explain actions influenced by a model’s output.}  Through interviews with Intensive Care Unit and Emergency Department clinicians, \citet{tonekaboni_what_2019} describe clinicians' desire to justify decisions influenced by a model’s output to patients or colleagues.  
    Similarly, by interviewing the head of AI at a bank using automated credit evaluations, \citet{hong_human_2020} identify the need to justify to customers decisions that were influenced by the model. 
    In addition to the immediate stakeholders acting on the model output (\emph{executors} according to \citet{tomsett_interpretable_2018}), this goal can also stem from people about whom a decision was made (\emph{decision subjects}). 
    For example, \citet{zarsky2013transparent} frames the need to provide someone with an explanation of a decision or action that affects them as one that is necessary in order to respect their autonomy.
    
    
    \item [\textbf{(O5)}]  \textbf{Understand how one’s data is being used.} \citet{zarsky2013transparent} grounds this objective in the theoretical premise of information privacy rights, framing it as an extension of individual autonomy.  The need to have control over one’s personal data has also been broadly accepted in European data protection law. \citet{hildebrandt2012} makes a further argument that people should understand not only what data about them is being utilized, but the potential consequences of this usage as well. And \citet{buneman2001and} distinguish between different types of data provenance users may be interested in. Clearly, given the prevalence of data mining, this objective is relevant to a wide range of stakeholders.
    
    \item [\textbf{(O6)}] \textbf{Learn about a domain.} Through interview studies, both \citet{hong_human_2020} and \citet{liao_questioning_2020} describe how stakeholders across different domains use interpretability to generate new hypotheses or insights about a domain. For example, one participant aimed to use a model predicting surgeons’ future performances as a tool to better understand what factors drive good performance, rather than using it as a predictive system. \citet{hohman2019gamut} focus specifically on data scientists, describing how interpretability helped them find ``valuable nuggets of information'' in the data. Similarly, \citet{doshi-velez_towards_2017} identify the use of interpretability to advance scientific understanding. Indeed, there is a growing subfield of machine learning investigating how interpretability mechanisms can aid in scientific discovery \cite{workshop}. 
    
    \item [\textbf{(O7)}] \textbf{Contest a decision made based on the model's output.} \citet{citron2014scored} posit that the right to challenge a decision affecting oneself should be ensured under due process, and \citet{doshi2017accountability} draw a comparison to the legal system, where mechanisms for redress serve as a powerful form of accountability. \citet{wachter_counterfactual_2018} also notes that the right to contest an automated decision is provided in Art. 22(3) of the European General Data Protect Regulation (GDPR).  An individual affected by an algorithmic system may not be the only one who wishes to contest it, either.  We can imagine that external stakeholders like lawyers, judges, or activists may also be interested in pushing back against model outputs that seem incorrect, arbitrary, or unfair. 
    
\end{enumerate}

The third and final level of our typology identifies the specific \emph{tasks} a stakeholder can perform to achieve the goals described above.
We break tasks out as a separate level of the typology to make clear that tasks do not map to objectives in a one-to-one fashion; rather, the same task may be used to accomplish several different objectives. 
For instance, detecting discrimination or other undesirable behavior in a model's prediction is likely to be a necessary task for both contesting a decision (O7) but also for understanding whether or how to incorporate model output in downstream actions (O3). 
Although the task is shared across these objectives, the specific type of discriminatory behavior a stakeholder may wish to detect, and the manner in which is it exposed and communicated, may differ based on the domain, the higher-level objective, and the stakeholder's knowledge.  
Here, we describe several such underlying tasks that we found to recur in the literature and give examples of how they can be relevant for multiple objectives.  
As with objectives, we expect this level of the typology to grow as ML continues to be deployed in new situations.

\begin{enumerate}
    \item [\textbf{(T1)}] \textbf{Assess reliability of a given prediction.}  Understanding the reliability of a given prediction is important for deciding how (or whether) to incorporate the model’s output into further actions (O3), to prevent harmful outcomes or over-reliance \cite{zhang2020effect,bussone_role_2015}. 
    Similarly, the ability to assess a given prediction and show, for example, that it may not have been reliable, is likely to provide important evidence for contesting a decision (O7).
    
    \item [\textbf{(T2)}] \textbf{Detect mistaken, discriminatory, or arbitrary behavior.}  
    The ability to detect discrimination or other unwanted logic codified in a model is considered a crucial tool for being able to contest an automated decision (O7) \cite{doshi2017accountability}.  Similarly, ensuring that predictions are not being made arbitrarily is likely necessary to ensure compliance with ethical or legal standards (O2). 
    In other cases, detecting incorrect reasoning was a way to guide model debugging and elucidate areas for improvement (O1) \cite{cai_hello_2019}.  Some papers also frame this task as its converse, i.e., verifying that predictions are sensible and/or fair (by some definition) \cite{ribera2019can}.  
    
    \item [\textbf{(T3)}] \textbf{Understand the extent of the information the model is using.} 
    Understanding details and extent of features used emerged as important for explaining actions influenced by the model (O4). \citet{tonekaboni_what_2019}, for example, describe how doctors felt that understanding the clinically relevant model features that were used was critical to first rationalizing the predictions to themselves, and then explaining them to patients.  
    Depending on the context, recognizing higher-level groups of features (e.g., ``demographic information,'' ``patient medical history'') may be more understandable and feasible than individual features.
    We can also imagine that developing this understanding will also be an important way for stakeholders to identify what aspects of their personal data are being incorporated into a specific system (O5) \cite{hildebrandt2012}.
    
    \item [\textbf{(T4)}] \textbf{Understand the influence of different factors on the model's output.}   
    For stakeholders who are interested in generating new insights about a domain (O6), understanding how different factors influence the output is key. 
    \citet{roscher_explainable_2020} provide several examples of deriving scientific or medical insights by investigating the impact of scientifically-meaningful factors on predictive outcomes.
    This task is also important for ensuring compliance with particular standards or regulations (O2), which may detail when/how it is acceptable to use certain features. 
    Unlike T3, this task may not provide a comprehensive understanding of the features used (e.g., perhaps just listing the most important) and is more focused on the ways those features influence the output.
    
    \item [\textbf{(T5)}] \textbf{Understand model strengths and limitations.} 
    Understanding the model’s overall potential weaknesses is critical for understanding how to incorporate its output into further actions (O3).  
    For example, \citet{cai_hello_2019} describe how doctors consistently wanted to know the proposed AI tool's specific limitations so that they could anticipate and account for them during decision-making. 
    Understanding areas of weakness is likely to also be useful for debugging and improving the model (O1), e.g., by guiding additional data collection or training.
    
\end{enumerate}

Note that specific implementations (e.g., counterfactual explanations \cite{wachter_counterfactual_2018}) are not included at the task-level; rather, they are used to \textit{implement} a particular task. For example, counterfactual explanations might be one way to implement the task ``detect discriminatory behavior'' (T2), but might be more or less appropriate depending on the stakeholder's knowledge, their overarching objective, and the surrounding context. Section 4.3 (Generative Power) further discusses the implications our framework might have on choosing particular methods.

While several prior literature surveys have sought to collate and organize a list of interpretability needs, our framework makes some key advances to provide a more nuanced understanding these needs.
First, where prior surveys focus primarily on computer science subdisciplines~\cite{hohman_visual_2019, mohseni_multidisciplinary_2020, ferreira_what_2020}, our framework incorporates these insights and extends them by looking to the legal literature \cite{wachter_counterfactual_2018,hildebrandt2012,zarsky2013transparent,citron2014scored} and research on participatory action and design~\cite{greenwood_introduction_2007, spinuzzi2005methodology}.
As a result, our framework is able to surface objectives such as ``contesting a decision'' (O7) or ``understanding how one's data is being used'' (O5) that prior surveys did not identify.


Second, and more importantly, where prior approaches define interpretability needs as a function of stakeholder expertise or role, our framework defines these needs as an independent component of the problem space. 
As a result, and as the examples above illustrate, our framework helps reveal that interpretability needs can cut across several different stakeholders.
For instance, model debugging (O1) is one of the most frequently identified interpretability goals; but, prior work has primarily categorized it as a need machine learning experts (or model builders and developers) have.
In contrast, our framework identifies that although certain stakeholders may not have much formal or instrumental machine learning knowledge, their personal knowledge may be crucial for identifying or fixing model errors.
Similarly, while it may have previously been tempting to think that contesting a decision (O7) is a need primarily expressed by decision subjects, our framework highlights that other stakeholders (including lawyers, judges, and activists) may wish to do so as well to affect systematic change.

\begin{table*}[ht!]
    \centering
    \caption{Knowledge types and contexts for interpretability stakeholders, with examples identified in our literature survey. We found a range of expertise and backgrounds under our framework, highlighting information that might be lost if XAI designers only consider a small set of roles like ``ML expert" and ``non-expert", as we have widely observed in past work. \label{tab:knowledge-survey}}
    \resizebox{\textwidth}{!}{%
        \begin{tabular}{l p{.4\textwidth} p{.4\textwidth} p{.4\textwidth}} \toprule
         & \multicolumn{3}{c}{\textbf{Knowledge}} \\
        \cmidrule(lr){2-4}
        \textbf{Context} & \multicolumn{1}{c}{\textbf{Formal}} & \multicolumn{1}{c}{\textbf{Instrumental}} & \multicolumn{1}{c}{\textbf{Personal}} \\ \midrule
        \textbf{ML}
        &  model developers~\cite{spinner_explainer_2020}, computer science students~\cite{wexler2020what},
        machine learning scientists and researchers~\cite{mohseni_multidisciplinary_2020, miller_explanation_2019, doshi-velez_towards_2017, ribera2019can}, model builders/engineers~\cite{boukhelifa_exploratory_2019, roscher_explainable_2020, bhatt_explainable_2020, ras_explanation_2018, tomsett_interpretable_2018, yang_re-examining_2020, ferreira_what_2020}, model analyst~\cite{kim2018interpretability}, general~\cite{yu_user-based_2018}
        \papercountnewline{15/58}

        & model users~\cite{spinner_explainer_2020}, data scientists/ML practitioners~\cite{kaur_interpreting_2020, arya_one_2019, wexler2020what}, autonomous robot developers~\cite{theodorou2017designing},
         data scientists~\cite{bhatt_explainable_2020, hong_human_2020, abdul_trends_2018, brennen_what_2020}, ``medical experts' increasing familiarity with [computer-aided diagnosis] systems"~\cite{cai_hello_2019}, ``domain experts who use machine learning for analysis"~\cite{mohseni_multidisciplinary_2020, ferreira_what_2020, tonekaboni_what_2019}, practitioners~\cite{yu_user-based_2018, liao_questioning_2020}, optimization expertise~\cite{boukhelifa_exploratory_2019}, students with some ML familiarity~\cite{bucinca_proxy_2020}, greater machine learning community~\cite{gilpin_explaining_2018}, developers/implementers~\cite{ras_explanation_2018, tomsett_interpretable_2018, schneider_personalized_2019, glass_toward_2008, weller_transparency_2019}
         \papercountnewline{23/58}

        & 
         ``intuition of how the network looks"~\cite{yu_user-based_2018}, ``[ML] researchers' intuition of what constitutes a 'good' explanation"~\cite{miller_explanation_2019}
        \papercountnewline{2/58}
        \\
        \textbf{Data Domain}
        & biologists~\cite{spinner_explainer_2020}, robotics~\cite{theodorou2017designing}, 
         scientists~\cite{roscher_explainable_2020}, doctors~\cite{cai_hello_2019, vayena_machine_2018, doshi-velez_towards_2017, sundararajan2019exploring, xie2019outlining, kim2018interpretability}, ``enrolled in law school"~\cite{lakkaraju_how_2020}, judges~\cite{bansal_does_2020}, agronomic engineers~\cite{boukhelifa_exploratory_2019}, regulators~\cite{bhatt_explainable_2020}, HR managers who produce expert estimates~\cite{tullio_how_2007}, energy data operators~\cite{bhatia2019explainable}, ``business logic"~\cite{tomsett_interpretable_2018}, game theorists~\cite{weller_transparency_2019}, general~\cite{bansal_beyond_2019, ribera2019can}
         \papercountnewline{19/58}
        
        & ``model novices" interested in applying ML to specific domains~\cite{spinner_explainer_2020}, ``deep knowledge of the circumstances for employee retention"~\cite{arya_one_2019}, sign-language learners~\cite{paudyal2019learn2sign}, domain knowledge to verify ML results qualitatively~\cite{schlegel_towards_2019}, 
         ``only specialists in part of the underlying process"~\cite{boukhelifa_exploratory_2019}, internal financial auditors~\cite{raji2020closing}, clinicians~\cite{cai_hello_2019, schneider_personalized_2019, doshi-velez_towards_2017, tonekaboni_what_2019}, ``increasingly adopt ML for optimizing and producing scientific outcomes"~\cite{roscher_explainable_2020}, operators~\cite{tomsett_interpretable_2018, yin_understanding_2019}, peer grading in online education~\cite{kizilcec2016much}, general~\cite{tullio_how_2007, hong_human_2020, cheng2019explaining}
        \papercountnewline{17/58}
        &   ``without accurate mental models, social factors can rationalize suspicious observations [about explanations]"~\cite{kaur_interpreting_2020}, 
         ``how well the system's conceptual model fits their mental model"~\cite{chromik_dark_2019}, mental models of the system to generalize the AI behavior~\cite{miller_explanation_2019, yin_understanding_2019}, patient/client/decision subject~\cite{tomsett_interpretable_2018, vayena_machine_2018, schneider_personalized_2019, weller_transparency_2019, brennen_what_2020, ribera2019can}, ``hold preconception of what constitutes useful explanations for decisions"~\cite{liao_questioning_2020}, ``prognosticating their patient's condition in their personal experience"~\cite{tonekaboni_what_2019}
        \papercountnewline{12/58}
        \\
        \textbf{Milieu}
        & students studying information policy~\cite{wexler2020what},
         ethicists~\cite{ras_explanation_2018}, bodies like institutional review boards or ethics committees~\cite{vayena_machine_2018}, ``understanding requirements arising from social contexts other than just from usability or human cognitive psychology"~\cite{abdul_trends_2018}
        \papercountnewline{4/58}
        &
         ``community hospital small groups, to academic medical centers"~\cite{cai_hello_2019}, ``use AI products in daily life"~\cite{mohseni_multidisciplinary_2020}, UX/design practitioners~\cite{liao_questioning_2020, yang_re-examining_2020}, data subject~\cite{ras_explanation_2018, tomsett_interpretable_2018}, product managers~\cite{hong_human_2020}, examiners/auditors~\cite{tomsett_interpretable_2018}, departments adopting decision-support technologies~\cite{tonekaboni_what_2019}, use of AI in government and industry~\cite{brennen_what_2020}
         \papercountnewline{10/58}
        
        &  loan applicant~\cite{arya_one_2019, ustun2019recourse}, 
        ``different cultural, demographic or phenotypic groups"~\cite{raji2020closing}, recommender system users ~\cite{zhao2019transparency, krebs2019tell}, ``actors bring their own points of view and own priorities"~\cite{wolf_explainability_2019}, ``people employ certain biases and social expectations"~\cite{miller_explanation_2019}, ``anticipating the situated, user-encountered capability of AI is difficult"~\cite{yang_re-examining_2020}, familiarity with privacy and personal data issues~\cite{ferreira_what_2020}, individual fatigue and workflow issues in heathcare~\cite{tonekaboni_what_2019}, general~\cite{tullio_how_2007, schneider_personalized_2019,bansal_does_2020}
        \papercountnewline{13/58}\\
        \bottomrule
        \end{tabular} %
        }
\end{table*}



Finally, in contrast to the uniform treatment of prior interpretability surveys, our framework provides new levels of abstraction for discussing interpretability needs.
In doing so, we can distinguish that these needs form a hierarchy: immediate tasks help stakeholders accomplish short-term objectives which, over time, achieve long-term goals.
As with other multi-level typologies~\cite{brehmer2013multi}, this structure surfaces the compositionality latent in this space.
For instance, as described above, there is a many-to-many relationship between goals, objectives, and tasks: one task may apply to several objectives; many tasks may be required to accomplish a single objective; and, together, they are all necessary to achieve goals.
Similarly, our three-level sequence allows for describing interpretability needs as sequences of action.
For example, to improve a model (O1), a stakeholder may wish to understand its strengths and limitations (T5) by repeatedly assessing the reliability of individual predictions (T1); or, a stakeholder's trust in the model (G2) may increase or decrease as a result of better understanding how it works (G1).
 
While we do not ascribe objectives to specific roles or expertise levels as in prior work, we note that they nevertheless exhibit a latent temporal structure\,---\,for example, the need to understand how a model's output should be incorporated into a decision (O3) occurs before someone wishing to contest that decision (O7). However, formalizing this latent chronology is not straightforward, as a given objective may (re)occur at several different stages during the ML process. 
And, there is a risk of unintentionally recapitulating prior stakeholder categorizations as particular roles or expertise may be implicitly associated with different stages of the ML process.

As a result, the chronology we settle on, shown in Figure~\ref{fig:chronology}, is more flexible and refers to broad phases of the ML process. Rather than provide a precise ordering, it is meant to lend some helpful structure to the many stakeholder objectives. We indicate the phase(s) in which a particular objective typically occurs, 
and note that these phases are likely to unfold iteratively. For example, the development and deployment stages may be revisited after observing a system's downstream impact. Furthermore, many different stakeholders may be involved in each phase. For example, beyond engineers with formal ML or data knowledge, downstream users with significant personal knowledge may provide input to the development phase of a particular system if they report bugs or provide feedback that is used to retrain the model.  We omit tasks from this chronology to preserve the many-to-many mapping between objectives and tasks.
 



\section{Evaluation \& Example Applications of the Framework}
\label{sec:evaluation}

\renewcommand{\arraystretch}{1.7}

\begin{table*}[tb]
    \centering
    \caption{Goals, objectives, and tasks for interpretability stakeholders. Our literature survey identified instances of theoretical and systems work that discuss or address these needs. \label{tab:goals-survey}}
    \resizebox{\textwidth}{!}{%
        \begin{tabular}{P{4cm} p{1.0\linewidth}} \toprule
        \textbf{Stakeholder Need} & \textbf{References} \\ \midrule
        \textbf{G1: Understanding the model}
        & 
        ``machine models"~\cite{mohseni_human-grounded_2020, wexler2020what},
         ``understand the agent's behavior and responses enough to participate in the mixed-initiative execution process"~\cite{glass_toward_2008}, ``to attain scientific outcomes with ML one wants an understanding"~\cite{roscher_explainable_2020}, ``understand the `algorithmic decision model'"~\cite{cheng2019explaining}, general~\cite{cai_hello_2019, bansal_beyond_2019, gilpin_explaining_2018, das_opportunities_2020, weller_transparency_2019,doshi-velez_towards_2017, abdul_trends_2018, yin_understanding_2019, brennen_what_2020, nunes2017systematic, cai2019effects} 
         \papercount{16/58} 
        \\ \hline
        \textbf{G2: building trust in the model}
        & 
        mechanisms for steering trust building~\cite{spinner_explainer_2020},
         build appropriate trust~\cite{bansal_does_2020, bucinca_proxy_2020}, mechanistic interpretation needed for trust building~\cite{vayena_machine_2018}, trust for tool adoption and continued use~\cite{kizilcec2016much}, ensure that ML models reflect appropriate values~\cite{kim2018interpretability}, general~\cite{lakkaraju_how_2020, mohseni_multidisciplinary_2020, byrne_counterfactuals_2019, gilpin_explaining_2018, miller_explanation_2019, hong_human_2020, tomsett_interpretable_2018, das_opportunities_2020, liao_questioning_2020, lipton_mythos_2018, glass_toward_2008, abdul_trends_2018, yin_understanding_2019, balog_measuring_2020, ferreira_what_2020, tonekaboni_what_2019, brennen_what_2020, cheng2019explaining, nunes2017systematic, cai2019effects}
         \papercount{26/58}
        \\\hline \hline
                \textbf{O1: Debug or improve a model}
        &
        model refinement~\cite{spinner_explainer_2020, kim2018interpretability},
         help data experts to tune ML parameters for the data~\cite{mohseni_multidisciplinary_2020, yu_user-based_2018, sundararajan2019exploring}, ``identify issues with a model and how to fix it" or debug and optimize~\cite{hong_human_2020, hohman_visual_2019, weller_transparency_2019, nunes2017systematic, samek2017explainable, schlegel_towards_2019, theodorou2017designing, wexler2020what}, improve an aspect or part of a system~\cite{ferreira_what_2020}, general~\cite{chromik_dark_2019, roscher_explainable_2020, bhatt_explainable_2020, ras_explanation_2018, adadi_peeking_2018, liao_questioning_2020, ribera2019can}
         \papercount{21/58}
        \\\hline
                \textbf{O2: Ensure compliance with standards or regulations}
        &
         adherence to standards and laws like GDPR and ``right to explanation"~\cite{raji2020closing, gilpin_explaining_2018, adadi_peeking_2018, doshi-velez_towards_2017, ferreira_what_2020, ribera2019can, krebs2019tell, samek2017explainable}, forensics~\cite{tomsett_interpretable_2018}, justify clinical validation of ML in medical studies~\cite{vayena_machine_2018}, facilitate monitoring for safety standards~\cite{weller_transparency_2019}, general~\cite{spinner_explainer_2020, brennen_what_2020, roscher_explainable_2020, bhatt_explainable_2020, lipton_mythos_2018, ras_explanation_2018}
         \papercount{17/58}
        \\\hline
        \textbf{O3: Understand how to incorporate the model’s output into downstream actions}
        & 
        learn ``factors that could be changed to improve their profile for possible approval in the future"~\cite{arya_one_2019}, learn how to correct actions based on model feedback~\cite{paudyal2019learn2sign},
         apply own domain-related decision-making using the XAI or not~\cite{bucinca_proxy_2020}, make better or faster decisions~\cite{nunes2017systematic}, understand impact of prediction on other system components~\cite{liao_questioning_2020}, understand how to get a desired outcome~\cite{ustun2019recourse}, understand consequences~\cite{ribera2019can}, understand errors for safety-oriented task~\cite{doshi-velez_towards_2017}, directing use in patient or medical work practice~\cite{tonekaboni_what_2019, sundararajan2019exploring, xie2019outlining}, general~\cite{tullio_how_2007, balog_measuring_2020}
         \papercount{13/58} 
         \\\hline
        \textbf{O4: Justify or explain actions influenced by a model’s output.}
        & 
        justify the user's decision-making~\cite{spinner_explainer_2020, arya_one_2019, tonekaboni_what_2019}, reason about data outputs~\cite{kaur_interpreting_2020},
         explaining findings to collaborators~\cite{boukhelifa_exploratory_2019}, ``enables the user to consider contrastive explanations... why one decision was made instead of another"~\cite{byrne_counterfactuals_2019}, explain causes of an event~\cite{miller_explanation_2019, weller_transparency_2019, abdul_trends_2018}, recommend treatment options to patient~\cite{tomsett_interpretable_2018}, ``justify the result"~\cite{adadi_peeking_2018, bucinca_proxy_2020, zhao2019transparency, cai2019effects}, general~\cite{cai_hello_2019, bansal_does_2020, hohman_visual_2019, yu_user-based_2018, hong_human_2020, roscher_explainable_2020, das_opportunities_2020}
         \papercount{21/58}
        \\\hline
        \textbf{O5: Understand how one’s data is being used}
        & 
        ``disclose what user data is being used in algorithmic decision-making"~\cite{mohseni_multidisciplinary_2020}, know how one's data is being used to make decisions about others~\cite{tomsett_interpretable_2018}, understand why certain user data is collected~\cite{zhao2019transparency}, general~\cite{arya_one_2019, chromik_dark_2019, lipton_mythos_2018}
        \papercount{6/58}
        \\\hline
        \textbf{O6: Learn about a domain}
        &
        learn about sign language and how to use it correctly~\cite{paudyal2019learn2sign},
         learning about ML~\cite{yu_user-based_2018}, explanation ``as a vehicle to generate insights about the phenomena described by the data"~\cite{hong_human_2020, lipton_mythos_2018}, learn how to solve a task~\cite{schneider_personalized_2019}, learn game strategy (Go)~\cite{samek2017explainable}, learn new facts/gain knowledge~\cite{adadi_peeking_2018, doshi-velez_towards_2017, ribera2019can}, learn design strategies~\cite{bhatia2019explainable}
         \papercount{10/58}
        \\\hline
        \textbf{O7: Contest a decision made based on the model’s output}
        & 
         ``when I see things I don't completely agree with"~\cite{cai_hello_2019}, ``present an incontestable subset of reasons to the bank employee"~\cite{chromik_dark_2019}, contest a discriminatory decision~\cite{ustun2019recourse}, general~\cite{wolf_explainability_2019, tomsett_interpretable_2018, lipton_mythos_2018, weller_transparency_2019}
         \papercount{7/58}
        \\\hline
         \hline
        \textbf{T1: Assess reliability of a given prediction}
        & 
         identify and explain an outlier~\cite{boukhelifa_exploratory_2019}, increase or decrease trust in the model based on observed accuracy, relative or not to one's own performance~\cite{bansal_does_2020, yin_understanding_2019}, ``to ensure the scientific value of the outcome"~\cite{roscher_explainable_2020}, assess the AI's judgment~\cite{liao_questioning_2020}
         \papercount{5/58}
        \\\hline
        \textbf{T2: Detect mistaken, discriminatory, or arbitrary behavior.}
        & 
         ``anticipate ethics-related failures before launch"~\cite{raji2020closing}, bias or mistake detection~\cite{wolf_explainability_2019, ribera2019can, nunes2017systematic, samek2017explainable}, understand skewness and biases in input data~\cite{das_opportunities_2020}, find unknown vulnerabilities and flaws~\cite{adadi_peeking_2018, doshi-velez_towards_2017}
         \papercount{8/58}
        \\\hline
        \textbf{T3: Understand the extent of the information the model is using}
        & 
         data entanglement~\cite{raji2020closing}, be informed when the ML is not suitable for particular systems~\cite{roscher_explainable_2020}, understand ``what the system was sensing to make its inferences"~\cite{tullio_how_2007, glass_toward_2008, cheng2019explaining, zhao2019transparency}
         \papercount{6/58}
        \\\hline
        \textbf{T4: Understand the influence of different factors on the model’s output.}
        & 
         explore counterfactuals and how changes to data points affect predictions~\cite{wexler2020what},
         understand model prediction mechanisms~\cite{hong_human_2020, cheng2019explaining}, ``factors influencing their individual decision"~\cite{chromik_dark_2019, cai2019effects, ustun2019recourse}, ``inspect how output changes with instance changes"~\cite{liao_questioning_2020, lipton_mythos_2018}, ``how drift in feature distributions would impact model outcomes"~\cite{bhatt_explainable_2020}, ``did the factor 'race' influence the outcome of the system"~\cite{ras_explanation_2018}, ``feedforward can help people understand and predict what is going to happen"~\cite{abdul_trends_2018}
         \papercount{11/58}
        \\\hline
        \textbf{T5: Understand model strengths and limitations}
        & 
         understand model error from predictions~\cite{bansal_beyond_2019}, know when to trust the prediction or be skeptical~\cite{bansal_does_2020, hong_human_2020}, understand limitations~\cite{liao_questioning_2020}, ``clarity around why the model under-performs'' \cite{tonekaboni_what_2019}
         \papercount{5/58}
        
        \\\bottomrule
        \end{tabular}
    }
\end{table*}

To assess the implications of our framework, we look to the three powers of interaction models described by \citet{beaudouin2004designing}: the \emph{descriptive} power, or how much coverage the framework achieves over existing points in the problem space; the \emph{evaluative} power, or how well the framework helps us compare two points in the problem space; and, the \emph{generative} power, or how the framework helps us envision new or previously unexplored points in the problem space. In addition to an evaluation of the framework, the evaluative and generative powers also serve as a demonstration of ways in which the framework can be used.

We find that our framework is able to describe over 50 existing papers on interpretability, and further provides a more granular treatment of relevant stakeholders.
We then illustrate how our framework gives us a language with which to more carefully evaluate interpretability systems.
Finally, we demonstrate how our framework can be used to generate new combinations of personas and needs, how it may suggest ways of designing future interpretability interfaces, and how it may be turned inwards to facilitate a more reflexive design process.


\subsection{Descriptive Power}


We assess our framework's descriptive power by using it to characterize the users and goals described by existing work on interpretable ML.
We collected papers using a mix of explicit keyword searches in academic search engines and libraries (e.g., Google Scholar and arXiv), following the citation graph of collected entries, and by compiling the bibliographies of previous literature surveys~\cite{hohman_visual_2019, ferreira_what_2020}.
Our final list of papers span several research contribution types~\cite{lee2019broadening, wobbrock2016research} including frameworks that define interpretability desiderata or key considerations (e.g., \citet{tomsett_interpretable_2018,arya_one_2019,lipton_mythos_2018}), evaluations of specific interpretability techniques (e.g., \citet{balog_measuring_2020,cheng2019explaining,cai2019effects}) and user/case studies that provide insights into pertinent human factors in interpretability (e.g., \citet{tonekaboni_what_2019,hong_human_2020,liao_questioning_2020}).
We excluded any papers that introduced novel interpretability techniques without discussing target users or user-centric considerations (e.g., \cite{Zhang_2018_CVPR,fleet_visualizing_2014}). Similarly, we excluded papers that included only a passing reference to user characteristics (e.g., ``interpretability is important for doctors'') without explicitly discussing them. We aimed collect a representative sample of current interpretability research directions, going beyond the papers we used to initially develop the framework, and sought out references across different applications, data domains, and computer science disciplines including machine learning, data visualization, human-computer interaction, and scientific computing.


In total, we selected 58 papers, and each paper was coded by at least two authors of this paper. 
Each coder used the framework to identify instances of stakeholder knowledge types and contexts, as well as goals, objectives, and tasks. 
When the coders disagreed on a designation for the entry, they discussed the conflicts until there was agreement on the code.
Where possible, we collected snippets of the papers corresponding to a description or discussion of stakeholder knowledge or needs.
The outcome of this coding process, including snippets\footnote{Quotations in the snippets may be paraphrased, and should be interpreted as describing a common theme in cases where multiple references are grouped together.}, is shown in Tables~\ref{tab:knowledge-survey} and~\ref{tab:goals-survey}. 

\begin{figure*}
    
    \includegraphics[width=\textwidth]{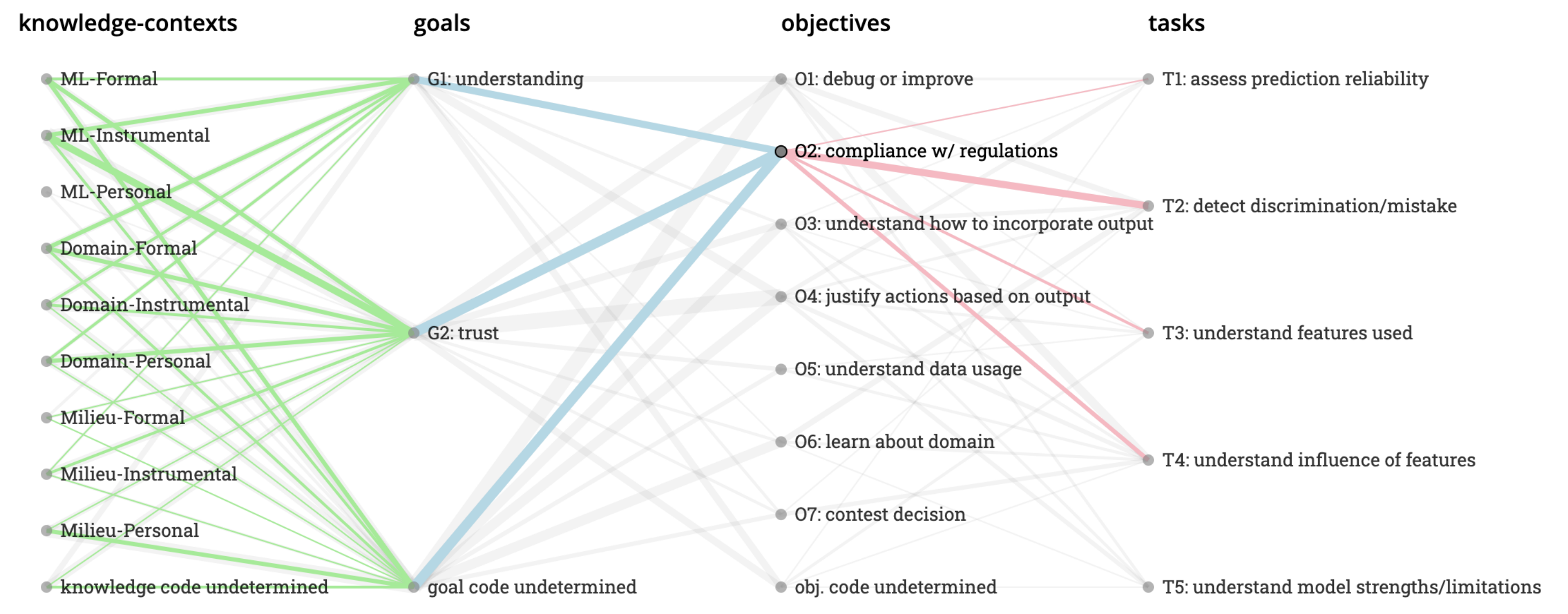}
    \Description[A state of an interactive figure that visualizes the results of the descriptive power analysis.]{The figure has four columns of nodes corresponding to parts of the framework (knowledge-contexts, goals, objectives, and tasks). Each column contains nodes corresponding to the specific categorizations within that part (for example, the tasks column contains T1 through T5 as nodes. Light grey lines connect nodes in consecutive columns. One of the nodes (O2: compliance with regulations) is being hovered over in the screenshot. Smaller, colored lines emerging from this node are overlaid on the background grey lines, indicating the codes that are present in papers with the O2 code.)}
    \caption{A state of an interactive figure,  included in supplementary material\textsuperscript{2}, that visualizes the results of the analysis of our framework's descriptive power. We see how the two halves of the framework (knowledge-contexts and goals-objectives-tasks) provide a more granular and composable vocabulary with which to describe 58 papers from the literature on ML interpretability. Light grey links represent the set of all papers, and connect codes that appear together. The width of the link corresponds to the number of papers it represents. We use "code undetermined" to indicate cases where we were not able to code a particular category (e.g., if a paper did not explicitly specify a knowledge-context). In the interactive figure, hovering over a code selects all papers that contain the code, and highlights links to visualize the co-occurrence of other codes (e.g., "O2" shown here).}
    \label{fig:framework_connections}
\end{figure*}

We found that the vocabulary provided by our framework was able to describe stakeholders and needs that appeared in prior work.  All knowledge type-context intersections and goal/objective/task categories appeared in more than one paper. The most observed knowledge categories were ML-Instrumental (23 of 58 papers), Data Domain-Formal (19/58), and Data Domain-Instrumental (17/58), which capture substantial technical expertise. The most observed objectives were justifying or explain decisions influenced by model output (O4, 21/58) and debugging and model improvement (O1, 21/58). The most common tasks were understanding factors that influence the model output (T4, 11/58) and detecting mistaken/arbitrary behavior (T2, 8/58). 
We note, however, that there were some goals or objectives that arose which we were not able to categorize given our current framework, such as persuading the user \cite{nunes2017systematic,schneider_personalized_2019}. In the current version, we have chosen not to explicitly integrate this into the framework because it was presented as a need imposed upon a stakeholder by, e.g., the company deploying a particular system, while the current needs we describe are driven by the stakeholder themselves.

Beyond its comprehensiveness, the framework is often able to add an additional layer of granularity. For example, whereas many papers describe people with various types of milieu knowledge as ``lay users,'' we are able to recognize and tease apart the different types of expertise they possess.  In other cases, we are able to provide consistency and draw similarities between concepts that were previously obscured\,---\,for example, from looking at Table~\ref{tab:goals-survey}, we can see that there are many papers that use different terminology to ultimately refer to the same goal.

In using our framework as a descriptive instrument, we found that users' personal knowledge is often the most challenging to define due to the subtle ways it may interact with its context. For example, personal knowledge in a particular domain may cross over into the milieu in cases where a decision domain intersects with everyday functions in society, like finance (``loan applicants", who learn to interact with banking ecosystems). 
At the same time, another decision subject such as a medical patient might acquire personal domain knowledge by learning about a medical condition that affects them and relating that knowledge to their own symptoms and experiences. 
Personal knowledge may also be less observable by experimenters who are evaluating and developing systems using traditional processes; however, we believe developing methods for eliciting this knowledge is an under-explored opportunity for human-centered design.

Our framework also helps reveal patterns of under- and over-representation of certain stakeholders and needs in current interpretability research. For example, we noticed a relative lack of interpretability methods specifically focused on objectives like understanding how one's data is being used (O5) and contesting a decision based on the model's output (O7). The papers that did cover these areas were primarily \emph{justifying} these needs from a legal perspective, rather than systems or methods development to help meet them. Similarly, while users with formal ML and data domain knowledge are mentioned frequently, there is significantly less attention paid to those with formal knowledge in the milieu. These stakeholders, however, are likely to have a deep understanding of the broader sociotechnical context in which systems are deployed.  Our framework gives us the vocabulary to consciously recognize these gaps, and subsequently, work towards filling them. 

Figure~\ref{fig:framework_connections} uses a parallel coordinates display to summarize the framework's descriptive power, and connect its two halves.
Axes correspond to the four components of the framework (knowledge-contexts, goals, objectives, and tasks) and nodes correspond to individual codes used during our qualitative process. 
Lines connect nodes to represent papers which contain these codes, with line width encoding the number of papers. 
In a supplemental interactive version of the figure\footnote{\label{note1}The interactive figure is also available online at \texttt{\url{vis.csail.mit.edu/pubs/beyond-expertise-roles/framework-connections}}.}, hovering over specific nodes selects all papers that contain the code, and highlights links to visualize the co-occurrence of other codes.
In doing so, the figure demonstrates the composable, many-to-many nature of our framework.

\subsection{Evaluative Power}

Our framework's evaluative power comes from it's ability to help design more ecologically valid and appropriately-scoped evaluations of interpretability systems/methods.

Similar to prior work \cite{doshi-velez_towards_2017}, we noticed three methods that are primarily used to evaluate interpretability techniques when coding the interpretability literature. The most popular approach does not involve human evaluation.
Instead, example outputs generated by the technique are used to illustrate its performance~\cite{goyal_counterfactual_2019,fleet_visualizing_2014,du2019techniques,chen2019looks,Zhang_2018_CVPR,Fong_2017_ICCV,lim2010toolkit} and capabilities including how expressive the technique is~\cite{olah2017feature, olah2018building}.
The next-most frequent style of evaluation are user studies conducted with \emph{proxy stakeholders} (e.g., sourced from Amazon Mechanical Turk or a similar online platform) and/or \emph{proxy tasks} (e.g., guessing a model's outcome)~\cite{poursabzi-sangdeh_manipulating_2019,cheng2019explaining,cai2019effects,lage_human_2019,yin_understanding_2019, lim2009why, dodge2019explaining}.
Recent work, however, has demonstrated the fallibility of using proxies\,---\,as \citet{bucinca_proxy_2020} describe, proxies induce a different style of cognition, forcing participants to explicitly attend to explanations and the AI rather than implicitly incorporating them as part of the overall process.

The current gold standard evaluative methodology are application-grounded studies~\cite{doshi-velez_towards_2017} in which domain experts engage with real-world tasks which include interpretable ML.
For example, \citet{bussone_role_2015},  \citet{wang2019designing}, and \citet{lundberg2018explainable} evaluate interpretability methods by asking healthcare professionals to engage in hypothetical diagnostic scenarios.
Although these studies often elicit richer and more relevant feedback about a system's real-world implications, they remain relatively rarely used.
We posit that this lack of adoption is due, in part, to the difficulty of designing such studies\,---\,there is little principled guidance on how to recruit participants, particularly when the real-world situation requires specific types of expertise.
Moreover, when comparing interpretability techniques, it can be difficult to design a task that is equitable for the various conditions.
And, even when studies are successfully conducted, it can be difficult to understand how the results generalize or inform future work.

We believe our framework's vocabulary begins to make addressing these issues more tractable.
In particular, stakeholder knowledge and context offers a more precise way of defining the participant pool, including identifying vectors along which it may be acceptable to introduce proxying.
For instance, consider an ML-enabled clinical diagnosis; if it were difficult to recruit a sufficient number of doctors to participate in the study, our framework suggests that residents or medical students might also be viable participants because of their shared formal data domain knowledge (medicine).
Similarly, consider evaluating explanations for loan applications; our framework helps identify that a study may not be ecologically valid if participants do not draw from similar pools of personal knowledge of the milieu.

Finally, our three-level typology of goals, objectives, and tasks provides a structure to operationalize comparative studies.
For example, to evaluate the relative effectiveness of similar interpretability techniques (e.g., the plethora of saliency and attribution methods) in a human-centric manner, our framework's \emph{tasks} may be the appropriate level of abstraction to target\,---\,they describe operations that can be performed directly and measured through quantitative and qualitative means, and are thus conducive for A/B testing style experiments.
On the other hand, for larger-scale interpretability systems, it may be more appropriate to target \emph{objectives} as the specific sequence of operations a participant performs will likely vary significantly between conditions; thus, results will likely be generated qualitatively, through observation and conversation.

\subsection{Generative Power}



Finally, we operationalize our framework to imagine either novel futures or futures underexplored by the existing work.

\textbf{Generating new personas.} Where prior literature has treated ``non-experts'' (i.e., stakeholders without expertise in either machine learning or the data domain) as a single, homogeneous group, and has designed for them as such, our framework makes explicit how heterogenous this group may be.
By introducing the context of the general milieu, and considering how instrumental and personal knowledge may manifest in it, we can generate new stakeholder personas and imagine the implications on interpretability design.
For instance, consider everyday people who have some familiarity or exposure to coding, or have tinkered with the Maker movement; we might describe them as having developed mental models for ``computational thinking,'' or instrumental knowledge in the milieu under our framework.
As a result of this knowledge, perhaps they would be more amenable to interpretability interfaces that promoted interactive question-answering by manipulating inputs.
Similarly, consider someone who has been closely following mainstream media reporting on the propensity of social media recommendation algorithms to radicalize individuals\,---\,our framework would describe them as having rich personal knowledge in the milieu.
Perhaps as a result of this knowledge, this person would be initially suspicious of an ML model.
In this case, instead of starting with a tabula rasa, perhaps an interpretability interface would be initialized with summaries of the model's strengths and weaknesses (akin to a model card~\cite{mitchell2019model}).

\textbf{Generating new persona-need combinations.}
Deriving needs from definitions of stakeholders has led to a relatively rigid set of interpretability needs that are recognized and developed for. 
But, by decoupling needs from stakeholder attributes, our framework allows for a much richer intersection of concerns than prior approaches.
For example, consider the objective of debugging or improving a model (O1)\,---\,prior work has typically viewed this as a need faced by ML experts, and debugging tools are thus built to primarily serve them.
Under our framework, we might describe these prior target users as having formal or instrumental ML knowledge; but our framework also exposes other stakeholders who might wish to address this objective as well: people with personal knowledge in the data domain or milieu.
Indeed, this aligns with theories of personal and formal knowledge in Participatory Action Research.
As Greenwood and Levin say, \textit{``[p]recisely because local stakeholders take action in their own environments, the consequences of errors are both significant to them and often rapidly apparent''}~\cite{greenwood_introduction_2007}.
In contrast, researchers with more formal knowledge may \textit{``rarely know whether they are right or not, as their findings seldom are acted upon and the practical results from their research rarely have direct consequences for them.''} 
Feminist standpoint theory \cite{haraway1988situated} would further posit that what is even considered an ``error'' or harmful might differ depending on the stakeholder's personal knowledge. 


\textbf{Generating new designs.} One might initially consider our framework to be silent on \emph{how} to design for particular stakeholders\,---\,for instance, it does not explicitly prescribe when to use local explanations~\cite{ribeiro2016model}, saliency maps, or feature visualization~\cite{olah2017feature, olah2018building}.
However, our more granular definitions of expertise allows us to adapt theories of knowledge development from cognitive science and pedagogy.
In particular, the cognitive science literature describes a process of ``chunking'', where people organize and think about information in terms of high-level concepts (or ``chunks'') which develop through experience, familiarity, and with increased knowledge  \cite{miller1956magical, narayanan2018humans, abdul2020cogam}. 
Similarly, the literature on expertise offers several models of decision making that posit two modes: analytic or deliberative thinking that is based in formal or instrumental knowledge, and intuitive thinking that is based on informal or personal knowledge. 
Thus, when designing interfaces for ML interpretability, designers could begin by first eliciting and characterizing the knowledge their stakeholders have, and the associated cognitive chunks and/or modes of thinking.
Standardized instruments\,---\,such as the Preference for Intuition and Deliberation (PID) scale~\cite{betsch2004praferenz}\,---\,could also be used.
These results could then inform what features are used in an explanation (e.g., raw features or higher level combinations of features that align more with the stakeholder's cognitive chunks) or what types of explanations are given (e.g., more intuitive example-based explanations versus more analytical or mechanistic explanations). Prior work by \citet{wang2019designing} provides further guidance on how particular types of explanations can be more/less suited to different modes of reasoning.

Situating stakeholders' knowledge and goals within broader societal power dynamics can also help inform what sorts of interpretability methods do or do not work towards subverting existing hierarchies.  
Indeed, the literature on expertise from which we derive our framework inextricably links types of knowledge with issues of power.
In particular, as Fleck notes, \textit{``the view of knowledge as being disinterested or value neutral is idealistic''} and \textit{``the possession of formal knowledge confers status and consequently a measure of power or influence within organizations''}~\cite{fleck1998expertise}.
Interpretability can play a key role here, addressing the \textit{``pathology of beneficence''} that Yielder describes~\cite{yielder_professional_2001}\,---\,where experts have a tendency to make decisions \textit{for} people rather than allowing them to decide for themselves\,---\,and reducing the ability of experts to merely ``rent'' out their knowledge~\cite{labaree2000nature}.
However, this work must be conducted carefully for, as \citet{thatcher_data_2016} observes, \textit{``[t]he very obscurity of transformation from individual data point to commodified, aggregate big data also masks the asymmetrical power relations between users of technology and the almost exclusively corporate entities which algorithmically collect, link, and analyze the data points of many users.''} 
Take the example of a mortgage applicant living in a redlined neighborhood, who wishes to contest the ML-based decision to reject their application. 
Their relative lack of power in this situation may be further compounded if they have little formal ML or data domain knowledge.  
We can begin to see, then, that interpretability methods that put the onus on the individual to change things about themselves in order to receive a better outcome (e.g., ``Had your income been \$3000 higher, you would have received the loan'') help uphold, rather than subvert, existing hierarchies.  
Interpretability methods that instead shift their gaze upwards and focus on alerting affected stakeholders to potentially discriminatory or arbitrary behaviors by the algorithm might provide much stronger evidence to fight against the reigning power differentials~\cite{barabas2020studying}.

\textbf{Generating a more reflexive design process.} 
The scope of interpretable ML should not be imagined by researchers or engineers alone\,---\,building interpretability systems that challenge power hinge on the involvement of stakeholders with different goals and knowledge.  
Indeed, people with formal knowledge (e.g., interpretability researchers, developers and designers, and the institutions within which they work) are often precisely the ones in positions of power over those with more personal knowledge (often those most directly affected by algorithmic systems).  
The concept of \textit{interest convergence}, which stems from critical race theory, holds that those in power tend to support goals that serve their own interests~\cite{ogbonnaya-ogburu_critical_2020}.  
In other words, without actively involving stakeholders whose interests are in opposition to existing power structures, and considering their input crucial, resultant interpretability systems will fit the standards and needs of those in power\,---\,for example, executives with a vested interest in maintaining the status quo, or engineers and researchers who might communicate about model decisions in a way that is not understandable to people without formal ML knowledge.  
Involving stakeholders with different interests first requires \textit{reflexivity}, or explicitly acknowledging what \textit{our own} backgrounds and interests are.  
However, doing so in the abstract can be difficult.
While our framework was primarily designed to describe the external stakeholders of interpretable ML, we believe it can also be turned to focus internally on the participants of the interpretability design process. 
By using it to describe our knowledge and goals, we can more clearly recognize gaps in our own knowledge and, thus, the additional people we must deliberately include.





\section{Limitations and Future Work}

In this paper, we present a framework to characterize the stakeholders of interpretable ML, and their needs. 
Our framework depicts stakeholder expertise as a two-dimensional space that describes the knowledge they possess (formal, instrumental, and personal knowledge), and the contexts in which this knowledge manifests (ML, the data domain, and the milieu).
Our framework also details stakeholder needs as a three-level typology of long-term goals (understanding the model, and building trust in it), shorter-term objectives that build towards these goals (e.g., debugging a model, or contesting a decision), and finally immediate tasks that stakeholders can perform to meet their objectives (e.g., assess prediction reliability, and detect mistakes). 
In evaluating our framework, we find that it suitably covers a sample of 58 papers on ML interpretability, and its granular structure reveals gaps in the literature.
Moreover, while speculative, we believe the framework offers the necessary vocabulary to assist in more precisely comparing and conducting user-focused evaluations of interpretability systems. Finally, we find that the framework offers a richer intersection of stakeholder expertise and needs than prior approaches, and that it can be turned inward to facilitate a more reflexive design process.



Our framework takes the next step in better defining who the users of interpretable ML are, and its limitations point to promising opportunities for future work.
In particular, we do not consider our framework to be an exhaustive description of the problem space, but rather a ``living'' artifact that will grow and adapt as interpretability matures as a research field.
For example, we expect new goals, objectives, and tasks to be added to the framework as ML is deployed more deeply in existing domains, and as it reaches new domains. 
Indeed, when coding the interpretability literature, we found occasional instances of needs that do not precisely fit into our current framework (e.g., persuasion, adoption). But, more evidence is needed to determine at what level of the typology these needs fit into, and whether they are specific instances of a more general or fundamental need.


Similarly, while our framework begins to decompose expertise into knowledge and contexts, the models we base it on provide even more granularity.
For example, Fleck~\cite{fleck1998expertise} names several additional types of knowledge including informal knowledge, contingent knowledge, tacit knowledge, and meta-knowledge; and Eraut~\cite{eraut2010knowledge} identifies cultural and tacit knowledge, and the degree to which either have or have not been codified.
Under our framework, these different types all lie within personal knowledge as we did not find sufficient evidence in the interpretability literature to warrant the additional granularity.
The milieu context is similarly broad\,---\,covering physical, social, and cultural contexts in the literature.
As additional work on interpretable ML is conducted, these two broad categories may come under the same pressure we initially identified with prior expertise- and role-based approaches: they may begin to conflate otherwise independent concerns.
By identifying recurring instances of these tensions, we can begin to disentangle them and enumerate other knowledge types and contexts that are meaningful for interpretability.


Finally, our framework's model of expertise is grounded only in epistemology; but the literature on expertise has also argued that expertise is constructed rhetorically.
As Johanna Hartelius describes, \textit{``[a] speaker is only able to exercise expertise and enjoy expert status to the extent that she can persuade an audience to grant such things''}~\cite{hartelius_2008}.
Rhetoric undoubtedly plays an important role in interpretability, and we can see evidence for this in the adjacent domain of data visualization~\cite{correll2019ethical}.
Researchers have argued that the clean, minimalist aesthetic of modern visualizations lends them an air of authority and certainty~\cite{kennedy2016work} that contributes to their \textit{``persuasive and seductive rhetorical force''}~\cite{drucker2012humanistic}.
Through close readings of visualizations, researchers have shown how citing sources and representing uncertainty can signal transparency and impartiality~\cite{hullman2011visualization}, and with empirical studies, researchers have demonstrated that even seemingly-innocuous elements like titles can frame or slant reading visualizations~\cite{kong2018frames} and can impact trust and recall~\cite{kong2019trust}.
How to adapt and replicate these findings for interpretability is a fertile ground for future work, and interpretability poses its own unique considerations.
In particular, unlike visualizations, the rhetorical performance of an interpretability interface may sometimes be shared with or mediated by a human (e.g., an ``operator''~\cite{tomsett_interpretable_2018} or through reports~\cite{hong_human_2020}, respectively).


\begin{acks}
This research was sponsored by NSF Award \#1900991, and by the United States Air Force Research Laboratory under Cooperative Agreement Number FA8750-19-2-1000. The views and conclusions contained in this document are those of the authors and should not be interpreted as representing the official policies, either expressed or implied, of the United States Air Force or the U.S. Government. The U.S. Government is authorized to reproduce and distribute reprints for Government purposes notwithstanding any copyright notation herein.
\end{acks}




\bibliographystyle{ACM-Reference-Format}
\bibliography{bibliography}


\begin{thebibliography}{137}


\ifx \showCODEN    \undefined \def \showCODEN     #1{\unskip}     \fi
\ifx \showDOI      \undefined \def \showDOI       #1{#1}\fi
\ifx \showISBNx    \undefined \def \showISBNx     #1{\unskip}     \fi
\ifx \showISBNxiii \undefined \def \showISBNxiii  #1{\unskip}     \fi
\ifx \showISSN     \undefined \def \showISSN      #1{\unskip}     \fi
\ifx \showLCCN     \undefined \def \showLCCN      #1{\unskip}     \fi
\ifx \shownote     \undefined \def \shownote      #1{#1}          \fi
\ifx \showarticletitle \undefined \def \showarticletitle #1{#1}   \fi
\ifx \showURL      \undefined \def \showURL       {\relax}        \fi
\providecommand\bibfield[2]{#2}
\providecommand\bibinfo[2]{#2}
\providecommand\natexlab[1]{#1}
\providecommand\showeprint[2][]{arXiv:#2}

\bibitem[\protect\citeauthoryear{??}{wor}{2020}]%
        {workshop}
 \bibinfo{year}{2020}\natexlab{}.
\newblock \bibinfo{title}{{ML Interpretability for Scientific Discovery
  (MLI4SD) Workshop}}.
\newblock
  \bibinfo{howpublished}{\url{https://sites.google.com/view/mli4sd-icml2020/home}}.
\newblock
\newblock
\shownote{Accessed: 2020-09-16.}


\bibitem[\protect\citeauthoryear{Aamodt and Plaza}{Aamodt and Plaza}{1994}]%
        {aamodt1994case}
\bibfield{author}{\bibinfo{person}{Agnar Aamodt} {and} \bibinfo{person}{Enric
  Plaza}.} \bibinfo{year}{1994}\natexlab{}.
\newblock \showarticletitle{Case-based reasoning: Foundational issues,
  methodological variations, and system approaches}.
\newblock \bibinfo{journal}{\emph{AI communications}} \bibinfo{volume}{7},
  \bibinfo{number}{1} (\bibinfo{year}{1994}), \bibinfo{pages}{39--59}.
\newblock


\bibitem[\protect\citeauthoryear{Abdul, Vermeulen, Wang, Lim, and
  Kankanhalli}{Abdul et~al\mbox{.}}{2018}]%
        {abdul_trends_2018}
\bibfield{author}{\bibinfo{person}{Ashraf Abdul}, \bibinfo{person}{Jo
  Vermeulen}, \bibinfo{person}{Danding Wang}, \bibinfo{person}{Brian~Y. Lim},
  {and} \bibinfo{person}{Mohan Kankanhalli}.} \bibinfo{year}{2018}\natexlab{}.
\newblock \showarticletitle{Trends and {Trajectories} for {Explainable},
  {Accountable} and {Intelligible} {Systems}: {An} {HCI} {Research} {Agenda}}.
  In \bibinfo{booktitle}{\emph{Proceedings of the 2018 {CHI} {Conference} on
  {Human} {Factors} in {Computing} {Systems} - {CHI} '18}}.
  \bibinfo{publisher}{ACM Press}, \bibinfo{address}{Montreal QC, Canada},
  \bibinfo{pages}{1--18}.
\newblock
\showISBNx{978-1-4503-5620-6}
\urldef\tempurl%
\url{https://doi.org/10.1145/3173574.3174156}
\showDOI{\tempurl}


\bibitem[\protect\citeauthoryear{Abdul, von~der Weth, Kankanhalli, and
  Lim}{Abdul et~al\mbox{.}}{2020}]%
        {abdul2020cogam}
\bibfield{author}{\bibinfo{person}{Ashraf Abdul}, \bibinfo{person}{Christian
  von~der Weth}, \bibinfo{person}{Mohan Kankanhalli}, {and}
  \bibinfo{person}{Brian~Y Lim}.} \bibinfo{year}{2020}\natexlab{}.
\newblock \showarticletitle{COGAM: Measuring and Moderating Cognitive Load in
  Machine Learning Model Explanations}. In
  \bibinfo{booktitle}{\emph{Proceedings of the 2020 CHI Conference on Human
  Factors in Computing Systems}}. \bibinfo{pages}{1--14}.
\newblock


\bibitem[\protect\citeauthoryear{Adadi and Berrada}{Adadi and Berrada}{2018}]%
        {adadi_peeking_2018}
\bibfield{author}{\bibinfo{person}{Amina Adadi} {and} \bibinfo{person}{Mohammed
  Berrada}.} \bibinfo{year}{2018}\natexlab{}.
\newblock \showarticletitle{Peeking {Inside} the {Black}-{Box}: {A} {Survey} on
  {Explainable} {Artificial} {Intelligence} ({XAI})}.
\newblock \bibinfo{journal}{\emph{IEEE Access}}  \bibinfo{volume}{6}
  (\bibinfo{year}{2018}), \bibinfo{pages}{52138--52160}.
\newblock
\showISSN{2169-3536}
\urldef\tempurl%
\url{https://doi.org/10.1109/ACCESS.2018.2870052}
\showDOI{\tempurl}
\newblock
\shownote{Conference Name: IEEE Access.}


\bibitem[\protect\citeauthoryear{Alkhatib and Bernstein}{Alkhatib and
  Bernstein}{2019}]%
        {alkhatib2019street}
\bibfield{author}{\bibinfo{person}{Ali Alkhatib} {and} \bibinfo{person}{Michael
  Bernstein}.} \bibinfo{year}{2019}\natexlab{}.
\newblock \showarticletitle{Street-level algorithms: A theory at the gaps
  between policy and decisions}. In \bibinfo{booktitle}{\emph{Proceedings of
  the 2019 CHI Conference on Human Factors in Computing Systems}}.
  \bibinfo{pages}{1--13}.
\newblock


\bibitem[\protect\citeauthoryear{Arya, Bellamy, Chen, Dhurandhar, Hind,
  Hoffman, Houde, Liao, Luss, Mojsilović, Mourad, Pedemonte, Raghavendra,
  Richards, Sattigeri, Shanmugam, Singh, Varshney, Wei, and Zhang}{Arya
  et~al\mbox{.}}{2019}]%
        {arya_one_2019}
\bibfield{author}{\bibinfo{person}{Vijay Arya}, \bibinfo{person}{Rachel K.~E.
  Bellamy}, \bibinfo{person}{Pin-Yu Chen}, \bibinfo{person}{Amit Dhurandhar},
  \bibinfo{person}{Michael Hind}, \bibinfo{person}{Samuel~C. Hoffman},
  \bibinfo{person}{Stephanie Houde}, \bibinfo{person}{Q.~Vera Liao},
  \bibinfo{person}{Ronny Luss}, \bibinfo{person}{Aleksandra Mojsilović},
  \bibinfo{person}{Sami Mourad}, \bibinfo{person}{Pablo Pedemonte},
  \bibinfo{person}{Ramya Raghavendra}, \bibinfo{person}{John Richards},
  \bibinfo{person}{Prasanna Sattigeri}, \bibinfo{person}{Karthikeyan
  Shanmugam}, \bibinfo{person}{Moninder Singh}, \bibinfo{person}{Kush~R.
  Varshney}, \bibinfo{person}{Dennis Wei}, {and} \bibinfo{person}{Yunfeng
  Zhang}.} \bibinfo{year}{2019}\natexlab{}.
\newblock \showarticletitle{One {Explanation} {Does} {Not} {Fit} {All}: {A}
  {Toolkit} and {Taxonomy} of {AI} {Explainability} {Techniques}}.
\newblock \bibinfo{journal}{\emph{arXiv:1909.03012 [cs, stat]}}
  (\bibinfo{date}{Sept.} \bibinfo{year}{2019}).
\newblock
\urldef\tempurl%
\url{http://arxiv.org/abs/1909.03012}
\showURL{%
\tempurl}
\newblock
\shownote{arXiv: 1909.03012.}


\bibitem[\protect\citeauthoryear{Atkinson, Derry, Renkl, and Wortham}{Atkinson
  et~al\mbox{.}}{2000}]%
        {atkinson2000learning}
\bibfield{author}{\bibinfo{person}{Robert~K Atkinson},
  \bibinfo{person}{Sharon~J Derry}, \bibinfo{person}{Alexander Renkl}, {and}
  \bibinfo{person}{Donald Wortham}.} \bibinfo{year}{2000}\natexlab{}.
\newblock \showarticletitle{Learning from examples: Instructional principles
  from the worked examples research}.
\newblock \bibinfo{journal}{\emph{Review of educational research}}
  \bibinfo{volume}{70}, \bibinfo{number}{2} (\bibinfo{year}{2000}),
  \bibinfo{pages}{181--214}.
\newblock


\bibitem[\protect\citeauthoryear{Bach, Binder, Montavon, Klauschen, M{\"u}ller,
  and Samek}{Bach et~al\mbox{.}}{2015}]%
        {bach2015pixel}
\bibfield{author}{\bibinfo{person}{Sebastian Bach}, \bibinfo{person}{Alexander
  Binder}, \bibinfo{person}{Gr{\'e}goire Montavon}, \bibinfo{person}{Frederick
  Klauschen}, \bibinfo{person}{Klaus-Robert M{\"u}ller}, {and}
  \bibinfo{person}{Wojciech Samek}.} \bibinfo{year}{2015}\natexlab{}.
\newblock \showarticletitle{On pixel-wise explanations for non-linear
  classifier decisions by layer-wise relevance propagation}.
\newblock \bibinfo{journal}{\emph{PloS one}} \bibinfo{volume}{10},
  \bibinfo{number}{7} (\bibinfo{year}{2015}), \bibinfo{pages}{e0130140}.
\newblock


\bibitem[\protect\citeauthoryear{Balog and Radlinski}{Balog and
  Radlinski}{2020}]%
        {balog_measuring_2020}
\bibfield{author}{\bibinfo{person}{Krisztian Balog} {and}
  \bibinfo{person}{Filip Radlinski}.} \bibinfo{year}{2020}\natexlab{}.
\newblock \showarticletitle{Measuring {Recommendation} {Explanation}
  {Quality}:{The} {Conflicting} {Goals} of {Explanations}}. In
  \bibinfo{booktitle}{\emph{Proceedings of the 43rd {International} {ACM}
  {SIGIR} {Conference} on {Research} and {Development} in {Information}
  {Retrieval} ({SIGIR} ’20)}}. \bibinfo{publisher}{ACM, New York, NY, USA},
  \bibinfo{address}{Virtual Event}, \bibinfo{pages}{10}.
\newblock


\bibitem[\protect\citeauthoryear{Bansal, Nushi, Kamar, Lasecki, Weld, and
  Horvitz}{Bansal et~al\mbox{.}}{[n.d.]}]%
        {bansal_beyond_2019}
\bibfield{author}{\bibinfo{person}{Gagan Bansal}, \bibinfo{person}{Besmira
  Nushi}, \bibinfo{person}{Ece Kamar}, \bibinfo{person}{Walter Lasecki},
  \bibinfo{person}{Daniel~S Weld}, {and} \bibinfo{person}{Eric Horvitz}.}
  \bibinfo{year}{[n.d.]}\natexlab{}.
\newblock \showarticletitle{Beyond {Accuracy}: {The} {Role} of {Mental}
  {Models} in {Human}-{AI} {Team} {Performance}}.
\newblock  (\bibinfo{year}{[n.\,d.]}), \bibinfo{pages}{10}.
\newblock


\bibitem[\protect\citeauthoryear{Bansal, Wu, Zhou, Fok, Nushi, Kamar, Ribeiro,
  and Weld}{Bansal et~al\mbox{.}}{2020}]%
        {bansal_does_2020}
\bibfield{author}{\bibinfo{person}{Gagan Bansal}, \bibinfo{person}{Tongshuang
  Wu}, \bibinfo{person}{Joyce Zhou}, \bibinfo{person}{Raymond Fok},
  \bibinfo{person}{Besmira Nushi}, \bibinfo{person}{Ece Kamar},
  \bibinfo{person}{Marco~Tulio Ribeiro}, {and} \bibinfo{person}{Daniel~S.
  Weld}.} \bibinfo{year}{2020}\natexlab{}.
\newblock \showarticletitle{Does the {Whole} {Exceed} its {Parts}? {The}
  {Effect} of {AI} {Explanations} on {Complementary} {Team} {Performance}}.
\newblock \bibinfo{journal}{\emph{arXiv:2006.14779 [cs]}} (\bibinfo{date}{June}
  \bibinfo{year}{2020}).
\newblock
\urldef\tempurl%
\url{http://arxiv.org/abs/2006.14779}
\showURL{%
\tempurl}
\newblock
\shownote{arXiv: 2006.14779.}


\bibitem[\protect\citeauthoryear{Barabas, Doyle, Rubinovitz, and
  Dinakar}{Barabas et~al\mbox{.}}{2020}]%
        {barabas2020studying}
\bibfield{author}{\bibinfo{person}{Chelsea Barabas}, \bibinfo{person}{Colin
  Doyle}, \bibinfo{person}{JB Rubinovitz}, {and} \bibinfo{person}{Karthik
  Dinakar}.} \bibinfo{year}{2020}\natexlab{}.
\newblock \showarticletitle{Studying up: reorienting the study of algorithmic
  fairness around issues of power}. In \bibinfo{booktitle}{\emph{Proceedings of
  the 2020 Conference on Fairness, Accountability, and Transparency}}.
  \bibinfo{pages}{167--176}.
\newblock


\bibitem[\protect\citeauthoryear{Beaudouin-Lafon}{Beaudouin-Lafon}{2004}]%
        {beaudouin2004designing}
\bibfield{author}{\bibinfo{person}{Michel Beaudouin-Lafon}.}
  \bibinfo{year}{2004}\natexlab{}.
\newblock \showarticletitle{Designing interaction, not interfaces}. In
  \bibinfo{booktitle}{\emph{Proceedings of the working conference on Advanced
  visual interfaces}}. \bibinfo{pages}{15--22}.
\newblock


\bibitem[\protect\citeauthoryear{Betsch}{Betsch}{2004}]%
        {betsch2004praferenz}
\bibfield{author}{\bibinfo{person}{Cornelia Betsch}.}
  \bibinfo{year}{2004}\natexlab{}.
\newblock \showarticletitle{Pr{\"a}ferenz f{\"u}r intuition und deliberation
  (PID)}.
\newblock \bibinfo{journal}{\emph{Zeitschrift f{\"u}r Differentielle und
  Diagnostische Psychologie}} \bibinfo{volume}{25}, \bibinfo{number}{4}
  (\bibinfo{year}{2004}), \bibinfo{pages}{179--197}.
\newblock


\bibitem[\protect\citeauthoryear{Bhatia, Garg, Haves, and Pudi}{Bhatia
  et~al\mbox{.}}{2019}]%
        {bhatia2019explainable}
\bibfield{author}{\bibinfo{person}{Aviruch Bhatia}, \bibinfo{person}{Vishal
  Garg}, \bibinfo{person}{Philip Haves}, {and} \bibinfo{person}{Vikram Pudi}.}
  \bibinfo{year}{2019}\natexlab{}.
\newblock \showarticletitle{Explainable Clustering Using Hyper-Rectangles for
  Building Energy Simulation Data}.
\newblock \bibinfo{journal}{\emph{E\&ES}} \bibinfo{volume}{238},
  \bibinfo{number}{1} (\bibinfo{year}{2019}), \bibinfo{pages}{012068}.
\newblock


\bibitem[\protect\citeauthoryear{Bhatt, Xiang, Sharma, Weller, Taly, Jia,
  Ghosh, Puri, Moura, and Eckersley}{Bhatt et~al\mbox{.}}{2020}]%
        {bhatt_explainable_2020}
\bibfield{author}{\bibinfo{person}{Umang Bhatt}, \bibinfo{person}{Alice Xiang},
  \bibinfo{person}{Shubham Sharma}, \bibinfo{person}{Adrian Weller},
  \bibinfo{person}{Ankur Taly}, \bibinfo{person}{Yunhan Jia},
  \bibinfo{person}{Joydeep Ghosh}, \bibinfo{person}{Ruchir Puri},
  \bibinfo{person}{José M.~F. Moura}, {and} \bibinfo{person}{Peter
  Eckersley}.} \bibinfo{year}{2020}\natexlab{}.
\newblock \showarticletitle{Explainable machine learning in deployment}. In
  \bibinfo{booktitle}{\emph{Proceedings of the 2020 {Conference} on {Fairness},
  {Accountability}, and {Transparency}}} \emph{(\bibinfo{series}{{FAT}* '20})}.
  \bibinfo{publisher}{Association for Computing Machinery},
  \bibinfo{address}{Barcelona, Spain}, \bibinfo{pages}{648--657}.
\newblock
\showISBNx{978-1-4503-6936-7}
\urldef\tempurl%
\url{https://doi.org/10.1145/3351095.3375624}
\showDOI{\tempurl}


\bibitem[\protect\citeauthoryear{Billett}{Billett}{2010}]%
        {billett_learning_2010}
\bibfield{editor}{\bibinfo{person}{Stephen Billett}} (Ed.).
  \bibinfo{year}{2010}\natexlab{}.
\newblock \bibinfo{booktitle}{\emph{Learning {Through} {Practice}}}.
\newblock \bibinfo{publisher}{Springer Netherlands},
  \bibinfo{address}{Dordrecht}.
\newblock
\showISBNx{978-90-481-3938-5 978-90-481-3939-2}
\urldef\tempurl%
\url{https://doi.org/10.1007/978-90-481-3939-2}
\showDOI{\tempurl}


\bibitem[\protect\citeauthoryear{Boukhelifa, Bezerianos, Trelea, Perrot, and
  Lutton}{Boukhelifa et~al\mbox{.}}{2019}]%
        {boukhelifa_exploratory_2019}
\bibfield{author}{\bibinfo{person}{Nadia Boukhelifa},
  \bibinfo{person}{Anastasia Bezerianos}, \bibinfo{person}{Ioan~Cristian
  Trelea}, \bibinfo{person}{Nathalie~Méjean Perrot}, {and}
  \bibinfo{person}{Evelyne Lutton}.} \bibinfo{year}{2019}\natexlab{}.
\newblock \showarticletitle{An {Exploratory} {Study} on {Visual} {Exploration}
  of {Model} {Simulations} by {Multiple} {Types} of {Experts}}. In
  \bibinfo{booktitle}{\emph{Proceedings of the 2019 {CHI} {Conference} on
  {Human} {Factors} in {Computing} {Systems} - {CHI} '19}}.
  \bibinfo{publisher}{ACM Press}, \bibinfo{address}{Glasgow, Scotland Uk},
  \bibinfo{pages}{1--14}.
\newblock
\showISBNx{978-1-4503-5970-2}
\urldef\tempurl%
\url{https://doi.org/10.1145/3290605.3300874}
\showDOI{\tempurl}


\bibitem[\protect\citeauthoryear{Brehmer and Munzner}{Brehmer and
  Munzner}{2013}]%
        {brehmer2013multi}
\bibfield{author}{\bibinfo{person}{Matthew Brehmer} {and}
  \bibinfo{person}{Tamara Munzner}.} \bibinfo{year}{2013}\natexlab{}.
\newblock \showarticletitle{A multi-level typology of abstract visualization
  tasks}.
\newblock \bibinfo{journal}{\emph{IEEE transactions on visualization and
  computer graphics}} \bibinfo{volume}{19}, \bibinfo{number}{12}
  (\bibinfo{year}{2013}), \bibinfo{pages}{2376--2385}.
\newblock


\bibitem[\protect\citeauthoryear{Brennen}{Brennen}{2020}]%
        {brennen_what_2020}
\bibfield{author}{\bibinfo{person}{Andrea Brennen}.}
  \bibinfo{year}{2020}\natexlab{}.
\newblock \showarticletitle{What {Do} {People} {Really} {Want} {When} {They}
  {Say} {They} {Want} "{Explainable} {AI}?" {We} {Asked} 60 {Stakeholders}.}.
  In \bibinfo{booktitle}{\emph{Extended {Abstracts} of the 2020 {CHI}
  {Conference} on {Human} {Factors} in {Computing} {Systems}}}
  \emph{(\bibinfo{series}{{CHI} {EA} '20})}. \bibinfo{publisher}{Association
  for Computing Machinery}, \bibinfo{address}{Honolulu, HI, USA},
  \bibinfo{pages}{1--7}.
\newblock
\showISBNx{9781450368193}
\urldef\tempurl%
\url{https://doi.org/10.1145/3334480.3383047}
\showDOI{\tempurl}


\bibitem[\protect\citeauthoryear{Buneman, Khanna, and Wang-Chiew}{Buneman
  et~al\mbox{.}}{2001}]%
        {buneman2001and}
\bibfield{author}{\bibinfo{person}{Peter Buneman}, \bibinfo{person}{Sanjeev
  Khanna}, {and} \bibinfo{person}{Tan Wang-Chiew}.}
  \bibinfo{year}{2001}\natexlab{}.
\newblock \showarticletitle{Why and where: A characterization of data
  provenance}. In \bibinfo{booktitle}{\emph{International conference on
  database theory}}. Springer, \bibinfo{pages}{316--330}.
\newblock


\bibitem[\protect\citeauthoryear{Bussone, Stumpf, and O'Sullivan}{Bussone
  et~al\mbox{.}}{2015}]%
        {bussone_role_2015}
\bibfield{author}{\bibinfo{person}{Adrian Bussone}, \bibinfo{person}{Simone
  Stumpf}, {and} \bibinfo{person}{Dympna O'Sullivan}.}
  \bibinfo{year}{2015}\natexlab{}.
\newblock \showarticletitle{The {Role} of {Explanations} on {Trust} and
  {Reliance} in {Clinical} {Decision} {Support} {Systems}}. In
  \bibinfo{booktitle}{\emph{2015 {International} {Conference} on {Healthcare}
  {Informatics}}}. \bibinfo{publisher}{IEEE}, \bibinfo{address}{Dallas, TX,
  USA}, \bibinfo{pages}{160--169}.
\newblock
\showISBNx{978-1-4673-9548-9}
\urldef\tempurl%
\url{https://doi.org/10.1109/ICHI.2015.26}
\showDOI{\tempurl}


\bibitem[\protect\citeauthoryear{Buçinca, Lin, Gajos, and Glassman}{Buçinca
  et~al\mbox{.}}{2020}]%
        {bucinca_proxy_2020}
\bibfield{author}{\bibinfo{person}{Zana Buçinca}, \bibinfo{person}{Phoebe
  Lin}, \bibinfo{person}{Krzysztof~Z. Gajos}, {and} \bibinfo{person}{Elena~L.
  Glassman}.} \bibinfo{year}{2020}\natexlab{}.
\newblock \showarticletitle{Proxy {Tasks} and {Subjective} {Measures} {Can}
  {Be} {Misleading} in {Evaluating} {Explainable} {AI} {Systems}}.
\newblock \bibinfo{journal}{\emph{Proceedings of the 25th International
  Conference on Intelligent User Interfaces}} (\bibinfo{date}{March}
  \bibinfo{year}{2020}), \bibinfo{pages}{454--464}.
\newblock
\urldef\tempurl%
\url{https://doi.org/10.1145/3377325.3377498}
\showDOI{\tempurl}
\newblock
\shownote{arXiv: 2001.08298.}


\bibitem[\protect\citeauthoryear{Byrne}{Byrne}{2019}]%
        {byrne_counterfactuals_2019}
\bibfield{author}{\bibinfo{person}{Ruth M.~J. Byrne}.}
  \bibinfo{year}{2019}\natexlab{}.
\newblock \showarticletitle{Counterfactuals in {Explainable} {Artificial}
  {Intelligence} ({XAI}): {Evidence} from {Human} {Reasoning}}. In
  \bibinfo{booktitle}{\emph{Proceedings of the {Twenty}-{Eighth}
  {International} {Joint} {Conference} on {Artificial} {Intelligence}}}.
  \bibinfo{publisher}{International Joint Conferences on Artificial
  Intelligence Organization}, \bibinfo{address}{Macao, China},
  \bibinfo{pages}{6276--6282}.
\newblock
\showISBNx{978-0-9992411-4-1}
\urldef\tempurl%
\url{https://doi.org/10.24963/ijcai.2019/876}
\showDOI{\tempurl}


\bibitem[\protect\citeauthoryear{Cai, Jongejan, and Holbrook}{Cai
  et~al\mbox{.}}{2019a}]%
        {cai2019effects}
\bibfield{author}{\bibinfo{person}{Carrie~J Cai}, \bibinfo{person}{Jonas
  Jongejan}, {and} \bibinfo{person}{Jess Holbrook}.}
  \bibinfo{year}{2019}\natexlab{a}.
\newblock \showarticletitle{The effects of example-based explanations in a
  machine learning interface}. In \bibinfo{booktitle}{\emph{Proceedings of the
  24th International Conference on Intelligent User Interfaces}}.
  \bibinfo{pages}{258--262}.
\newblock


\bibitem[\protect\citeauthoryear{Cai, Winter, Steiner, Wilcox, and Terry}{Cai
  et~al\mbox{.}}{2019b}]%
        {cai_hello_2019}
\bibfield{author}{\bibinfo{person}{Carrie~J. Cai}, \bibinfo{person}{Samantha
  Winter}, \bibinfo{person}{David Steiner}, \bibinfo{person}{Lauren Wilcox},
  {and} \bibinfo{person}{Michael Terry}.} \bibinfo{year}{2019}\natexlab{b}.
\newblock \showarticletitle{"{Hello} {AI}": {Uncovering} the {Onboarding}
  {Needs} of {Medical} {Practitioners} for {Human}-{AI} {Collaborative}
  {Decision}-{Making}}.
\newblock \bibinfo{journal}{\emph{Proceedings of the ACM on Human-Computer
  Interaction}} \bibinfo{volume}{3}, \bibinfo{number}{CSCW}
  (\bibinfo{date}{Nov.} \bibinfo{year}{2019}), \bibinfo{pages}{1--24}.
\newblock
\showISSN{2573-0142, 2573-0142}
\urldef\tempurl%
\url{https://doi.org/10.1145/3359206}
\showDOI{\tempurl}


\bibitem[\protect\citeauthoryear{Chen, Li, Tao, Barnett, Rudin, and Su}{Chen
  et~al\mbox{.}}{2019}]%
        {chen2019looks}
\bibfield{author}{\bibinfo{person}{Chaofan Chen}, \bibinfo{person}{Oscar Li},
  \bibinfo{person}{Daniel Tao}, \bibinfo{person}{Alina Barnett},
  \bibinfo{person}{Cynthia Rudin}, {and} \bibinfo{person}{Jonathan~K Su}.}
  \bibinfo{year}{2019}\natexlab{}.
\newblock \showarticletitle{This looks like that: deep learning for
  interpretable image recognition}. In \bibinfo{booktitle}{\emph{Advances in
  neural information processing systems}}. \bibinfo{pages}{8930--8941}.
\newblock


\bibitem[\protect\citeauthoryear{Cheng, Wang, Zhang, O'Connell, Gray, Harper,
  and Zhu}{Cheng et~al\mbox{.}}{2019}]%
        {cheng2019explaining}
\bibfield{author}{\bibinfo{person}{Hao-Fei Cheng}, \bibinfo{person}{Ruotong
  Wang}, \bibinfo{person}{Zheng Zhang}, \bibinfo{person}{Fiona O'Connell},
  \bibinfo{person}{Terrance Gray}, \bibinfo{person}{F~Maxwell Harper}, {and}
  \bibinfo{person}{Haiyi Zhu}.} \bibinfo{year}{2019}\natexlab{}.
\newblock \showarticletitle{Explaining decision-making algorithms through UI:
  Strategies to help non-expert stakeholders}. In
  \bibinfo{booktitle}{\emph{Proceedings of the 2019 chi conference on human
  factors in computing systems}}. \bibinfo{pages}{1--12}.
\newblock


\bibitem[\protect\citeauthoryear{Chromik, Eiband, Völkel, and Buschek}{Chromik
  et~al\mbox{.}}{2019}]%
        {chromik_dark_2019}
\bibfield{author}{\bibinfo{person}{Michael Chromik}, \bibinfo{person}{Malin
  Eiband}, \bibinfo{person}{Sarah~Theres Völkel}, {and}
  \bibinfo{person}{Daniel Buschek}.} \bibinfo{year}{2019}\natexlab{}.
\newblock \showarticletitle{Dark {Patterns} of {Explainability},
  {Transparency}, and {User} {Control} for {Intelligent} {Systems}}.
\newblock \bibinfo{journal}{\emph{Los Angeles}} (\bibinfo{year}{2019}),
  \bibinfo{pages}{6}.
\newblock


\bibitem[\protect\citeauthoryear{Citron and Pasquale}{Citron and
  Pasquale}{2014}]%
        {citron2014scored}
\bibfield{author}{\bibinfo{person}{Danielle~Keats Citron} {and}
  \bibinfo{person}{Frank Pasquale}.} \bibinfo{year}{2014}\natexlab{}.
\newblock \showarticletitle{The scored society: Due process for automated
  predictions}.
\newblock \bibinfo{journal}{\emph{Wash. L. Rev.}}  \bibinfo{volume}{89}
  (\bibinfo{year}{2014}), \bibinfo{pages}{1}.
\newblock


\bibitem[\protect\citeauthoryear{Correll}{Correll}{2019}]%
        {correll2019ethical}
\bibfield{author}{\bibinfo{person}{Michael Correll}.}
  \bibinfo{year}{2019}\natexlab{}.
\newblock \showarticletitle{Ethical dimensions of visualization research}. In
  \bibinfo{booktitle}{\emph{Proceedings of the 2019 CHI Conference on Human
  Factors in Computing Systems}}. \bibinfo{pages}{1--13}.
\newblock


\bibitem[\protect\citeauthoryear{Cox}{Cox}{2011}]%
        {amandacox-eyeo}
\bibfield{author}{\bibinfo{person}{Amanda Cox}.}
  \bibinfo{year}{2011}\natexlab{}.
\newblock \bibinfo{title}{Shaping Data for News}.
\newblock
\newblock
\urldef\tempurl%
\url{https://vimeo.com/29391942}
\showURL{%
\tempurl}


\bibitem[\protect\citeauthoryear{Dall’Alba and Sandberg}{Dall’Alba and
  Sandberg}{2006}]%
        {dall2006unveiling}
\bibfield{author}{\bibinfo{person}{Gloria Dall’Alba} {and}
  \bibinfo{person}{J{\"o}rgen Sandberg}.} \bibinfo{year}{2006}\natexlab{}.
\newblock \showarticletitle{Unveiling professional development: A critical
  review of stage models}.
\newblock \bibinfo{journal}{\emph{Review of educational research}}
  \bibinfo{volume}{76}, \bibinfo{number}{3} (\bibinfo{year}{2006}),
  \bibinfo{pages}{383--412}.
\newblock


\bibitem[\protect\citeauthoryear{Das and Rad}{Das and Rad}{2020}]%
        {das_opportunities_2020}
\bibfield{author}{\bibinfo{person}{Arun Das} {and} \bibinfo{person}{Paul Rad}.}
  \bibinfo{year}{2020}\natexlab{}.
\newblock \showarticletitle{Opportunities and {Challenges} in {Explainable}
  {Artificial} {Intelligence} ({XAI}): {A} {Survey}}.
\newblock \bibinfo{journal}{\emph{arXiv:2006.11371 [cs]}} (\bibinfo{date}{June}
  \bibinfo{year}{2020}).
\newblock
\urldef\tempurl%
\url{http://arxiv.org/abs/2006.11371}
\showURL{%
\tempurl}
\newblock
\shownote{arXiv: 2006.11371.}


\bibitem[\protect\citeauthoryear{Dhurandhar, Iyengar, Luss, and
  Shanmugam}{Dhurandhar et~al\mbox{.}}{2017}]%
        {dhurandhar2017formal}
\bibfield{author}{\bibinfo{person}{Amit Dhurandhar}, \bibinfo{person}{Vijay
  Iyengar}, \bibinfo{person}{Ronny Luss}, {and} \bibinfo{person}{Karthikeyan
  Shanmugam}.} \bibinfo{year}{2017}\natexlab{}.
\newblock \showarticletitle{A formal framework to characterize interpretability
  of procedures}.
\newblock \bibinfo{journal}{\emph{arXiv preprint arXiv:1707.03886}}
  (\bibinfo{year}{2017}).
\newblock


\bibitem[\protect\citeauthoryear{Dodge, Liao, Zhang, Bellamy, and Dugan}{Dodge
  et~al\mbox{.}}{2019}]%
        {dodge2019explaining}
\bibfield{author}{\bibinfo{person}{Jonathan Dodge}, \bibinfo{person}{Q~Vera
  Liao}, \bibinfo{person}{Yunfeng Zhang}, \bibinfo{person}{Rachel~KE Bellamy},
  {and} \bibinfo{person}{Casey Dugan}.} \bibinfo{year}{2019}\natexlab{}.
\newblock \showarticletitle{Explaining models: an empirical study of how
  explanations impact fairness judgment}. In
  \bibinfo{booktitle}{\emph{Proceedings of the 24th International Conference on
  Intelligent User Interfaces}}. \bibinfo{pages}{275--285}.
\newblock


\bibitem[\protect\citeauthoryear{Doshi-Velez and Kim}{Doshi-Velez and
  Kim}{2017}]%
        {doshi-velez_towards_2017}
\bibfield{author}{\bibinfo{person}{Finale Doshi-Velez} {and}
  \bibinfo{person}{Been Kim}.} \bibinfo{year}{2017}\natexlab{}.
\newblock \showarticletitle{Towards {A} {Rigorous} {Science} of {Interpretable}
  {Machine} {Learning}}.
\newblock \bibinfo{journal}{\emph{arXiv:1702.08608 [cs, stat]}}
  (\bibinfo{date}{March} \bibinfo{year}{2017}).
\newblock
\urldef\tempurl%
\url{http://arxiv.org/abs/1702.08608}
\showURL{%
\tempurl}
\newblock
\shownote{arXiv: 1702.08608.}


\bibitem[\protect\citeauthoryear{Doshi-Velez, Kortz, Budish, Bavitz, Gershman,
  O'Brien, Schieber, Waldo, Weinberger, and Wood}{Doshi-Velez
  et~al\mbox{.}}{2017}]%
        {doshi2017accountability}
\bibfield{author}{\bibinfo{person}{Finale Doshi-Velez}, \bibinfo{person}{Mason
  Kortz}, \bibinfo{person}{Ryan Budish}, \bibinfo{person}{Chris Bavitz},
  \bibinfo{person}{Sam Gershman}, \bibinfo{person}{David O'Brien},
  \bibinfo{person}{Stuart Schieber}, \bibinfo{person}{James Waldo},
  \bibinfo{person}{David Weinberger}, {and} \bibinfo{person}{Alexandra Wood}.}
  \bibinfo{year}{2017}\natexlab{}.
\newblock \showarticletitle{Accountability of AI under the law: The role of
  explanation}.
\newblock \bibinfo{journal}{\emph{arXiv preprint arXiv:1711.01134}}
  (\bibinfo{year}{2017}).
\newblock


\bibitem[\protect\citeauthoryear{Dreyfus and Dreyfus}{Dreyfus and
  Dreyfus}{1986}]%
        {dreyfus1986power}
\bibfield{author}{\bibinfo{person}{Hubert~L Dreyfus} {and}
  \bibinfo{person}{Stuart~E Dreyfus}.} \bibinfo{year}{1986}\natexlab{}.
\newblock \showarticletitle{The power of human intuition and expertise in the
  era of the computer}.
\newblock \bibinfo{journal}{\emph{Mind over machine. Nueva York: The Free
  Press}} (\bibinfo{year}{1986}).
\newblock


\bibitem[\protect\citeauthoryear{Drucker}{Drucker}{2012}]%
        {drucker2012humanistic}
\bibfield{author}{\bibinfo{person}{Johanna Drucker}.}
  \bibinfo{year}{2012}\natexlab{}.
\newblock \showarticletitle{Humanistic theory and digital scholarship}.
\newblock \bibinfo{journal}{\emph{Debates in the digital humanities}}
  \bibinfo{volume}{150} (\bibinfo{year}{2012}), \bibinfo{pages}{85--95}.
\newblock


\bibitem[\protect\citeauthoryear{Du, Liu, and Hu}{Du et~al\mbox{.}}{2019}]%
        {du2019techniques}
\bibfield{author}{\bibinfo{person}{Mengnan Du}, \bibinfo{person}{Ninghao Liu},
  {and} \bibinfo{person}{Xia Hu}.} \bibinfo{year}{2019}\natexlab{}.
\newblock \showarticletitle{Techniques for interpretable machine learning}.
\newblock \bibinfo{journal}{\emph{Commun. ACM}} \bibinfo{volume}{63},
  \bibinfo{number}{1} (\bibinfo{year}{2019}), \bibinfo{pages}{68--77}.
\newblock


\bibitem[\protect\citeauthoryear{Eraut}{Eraut}{2010}]%
        {eraut2010knowledge}
\bibfield{author}{\bibinfo{person}{Michael Eraut}.}
  \bibinfo{year}{2010}\natexlab{}.
\newblock \showarticletitle{Knowledge, working practices, and learning}.
\newblock In \bibinfo{booktitle}{\emph{Learning through practice}}.
  \bibinfo{publisher}{Springer}, \bibinfo{pages}{37--58}.
\newblock


\bibitem[\protect\citeauthoryear{Ferreira and Monteiro}{Ferreira and
  Monteiro}{2020}]%
        {ferreira_what_2020}
\bibfield{author}{\bibinfo{person}{Juliana~J. Ferreira} {and}
  \bibinfo{person}{Mateus~S. Monteiro}.} \bibinfo{year}{2020}\natexlab{}.
\newblock \showarticletitle{What {Are} {People} {Doing} {About} {XAI} {User}
  {Experience}? {A} {Survey} on {AI} {Explainability} {Research} and
  {Practice}}. In \bibinfo{booktitle}{\emph{Design, {User} {Experience}, and
  {Usability}. {Design} for {Contemporary} {Interactive} {Environments}}}
  \emph{(\bibinfo{series}{Lecture {Notes} in {Computer} {Science}})},
  \bibfield{editor}{\bibinfo{person}{Aaron Marcus} {and}
  \bibinfo{person}{Elizabeth Rosenzweig}} (Eds.). \bibinfo{publisher}{Springer
  International Publishing}, \bibinfo{address}{Cham}, \bibinfo{pages}{56--73}.
\newblock
\showISBNx{978-3-030-49760-6}
\urldef\tempurl%
\url{https://doi.org/10.1007/978-3-030-49760-6_4}
\showDOI{\tempurl}


\bibitem[\protect\citeauthoryear{Fleck}{Fleck}{1998}]%
        {fleck1998expertise}
\bibfield{author}{\bibinfo{person}{James Fleck}.}
  \bibinfo{year}{1998}\natexlab{}.
\newblock \showarticletitle{Expertise: knowledge, power and tradeability}.
\newblock In \bibinfo{booktitle}{\emph{Exploring expertise}}.
  \bibinfo{publisher}{Springer}, \bibinfo{pages}{143--171}.
\newblock


\bibitem[\protect\citeauthoryear{Fong and Vedaldi}{Fong and Vedaldi}{2017}]%
        {Fong_2017_ICCV}
\bibfield{author}{\bibinfo{person}{Ruth~C. Fong} {and} \bibinfo{person}{Andrea
  Vedaldi}.} \bibinfo{year}{2017}\natexlab{}.
\newblock \showarticletitle{Interpretable Explanations of Black Boxes by
  Meaningful Perturbation}. In \bibinfo{booktitle}{\emph{Proceedings of the
  IEEE International Conference on Computer Vision (ICCV)}}.
\newblock


\bibitem[\protect\citeauthoryear{Gilpin, Bau, Yuan, Bajwa, Specter, and
  Kagal}{Gilpin et~al\mbox{.}}{2018}]%
        {gilpin_explaining_2018}
\bibfield{author}{\bibinfo{person}{Leilani~H. Gilpin}, \bibinfo{person}{David
  Bau}, \bibinfo{person}{Ben~Z. Yuan}, \bibinfo{person}{Ayesha Bajwa},
  \bibinfo{person}{Michael Specter}, {and} \bibinfo{person}{Lalana Kagal}.}
  \bibinfo{year}{2018}\natexlab{}.
\newblock \showarticletitle{Explaining {Explanations}: {An} {Overview} of
  {Interpretability} of {Machine} {Learning}}. In
  \bibinfo{booktitle}{\emph{2018 {IEEE} 5th {International} {Conference} on
  {Data} {Science} and {Advanced} {Analytics} ({DSAA})}}.
  \bibinfo{pages}{80--89}.
\newblock
\urldef\tempurl%
\url{https://doi.org/10.1109/DSAA.2018.00018}
\showDOI{\tempurl}


\bibitem[\protect\citeauthoryear{Glass, McGuinness, and Wolverton}{Glass
  et~al\mbox{.}}{2008}]%
        {glass_toward_2008}
\bibfield{author}{\bibinfo{person}{Alyssa Glass}, \bibinfo{person}{Deborah~L.
  McGuinness}, {and} \bibinfo{person}{Michael Wolverton}.}
  \bibinfo{year}{2008}\natexlab{}.
\newblock \showarticletitle{Toward establishing trust in adaptive agents}. In
  \bibinfo{booktitle}{\emph{Proceedings of the 13th international conference on
  {Intelligent} user interfaces - {IUI} '08}}. \bibinfo{publisher}{ACM Press},
  \bibinfo{address}{Gran Canaria, Spain}, \bibinfo{pages}{227}.
\newblock
\showISBNx{978-1-59593-987-6}
\urldef\tempurl%
\url{https://doi.org/10.1145/1378773.1378804}
\showDOI{\tempurl}


\bibitem[\protect\citeauthoryear{Goyal, Wu, Ernst, Batra, Parikh, and
  Lee}{Goyal et~al\mbox{.}}{2019}]%
        {goyal_counterfactual_2019}
\bibfield{author}{\bibinfo{person}{Yash Goyal}, \bibinfo{person}{Ziyan Wu},
  \bibinfo{person}{Jan Ernst}, \bibinfo{person}{Dhruv Batra},
  \bibinfo{person}{Devi Parikh}, {and} \bibinfo{person}{Stefan Lee}.}
  \bibinfo{year}{2019}\natexlab{}.
\newblock \showarticletitle{Counterfactual {Visual} {Explanations}}. In
  \bibinfo{booktitle}{\emph{Proceedings of the 36th {International}
  {Conference} on {Machine} {Learning}}}, Vol.~\bibinfo{volume}{97}.
  \bibinfo{address}{Long Beach, California, USA}.
\newblock
\urldef\tempurl%
\url{http://proceedings.mlr.press/v97/goyal19a.html}
\showURL{%
\tempurl}
\newblock
\shownote{arXiv: 1904.07451.}


\bibitem[\protect\citeauthoryear{Greenwood and Levin}{Greenwood and
  Levin}{2007}]%
        {greenwood_introduction_2007}
\bibfield{author}{\bibinfo{person}{Davydd Greenwood} {and}
  \bibinfo{person}{Morten Levin}.} \bibinfo{year}{2007}\natexlab{}.
\newblock \bibinfo{booktitle}{\emph{Introduction to {Action} {Research}}}.
\newblock \bibinfo{publisher}{SAGE Publications, Inc.}, \bibinfo{address}{2455
  Teller Road, Thousand Oaks California 91320 United States of America}.
\newblock
\showISBNx{978-1-4129-2597-6 978-1-4129-8461-4}
\urldef\tempurl%
\url{https://doi.org/10.4135/9781412984614}
\showDOI{\tempurl}


\bibitem[\protect\citeauthoryear{Haraway}{Haraway}{1988}]%
        {haraway1988situated}
\bibfield{author}{\bibinfo{person}{Donna Haraway}.}
  \bibinfo{year}{1988}\natexlab{}.
\newblock \showarticletitle{Situated knowledges: The science question in
  feminism and the privilege of partial perspective}.
\newblock \bibinfo{journal}{\emph{Feminist studies}} \bibinfo{volume}{14},
  \bibinfo{number}{3} (\bibinfo{year}{1988}), \bibinfo{pages}{575--599}.
\newblock


\bibitem[\protect\citeauthoryear{Hartelius}{Hartelius}{2008}]%
        {hartelius_2008}
\bibfield{author}{\bibinfo{person}{Johanna Hartelius}.}
  \bibinfo{year}{2008}\natexlab{}.
\newblock \emph{\bibinfo{title}{The Rhetoric of Expertise}}.
\newblock \bibinfo{thesistype}{Ph.D. Dissertation}. \bibinfo{school}{The
  University of Texas at Austin}.
\newblock


\bibitem[\protect\citeauthoryear{Hartelius}{Hartelius}{2011}]%
        {hartelius_rhetorics_2011}
\bibfield{author}{\bibinfo{person}{Johanna Hartelius}.}
  \bibinfo{year}{2011}\natexlab{}.
\newblock \showarticletitle{Rhetorics of {Expertise}}.
\newblock \bibinfo{journal}{\emph{Social Epistemology}} \bibinfo{volume}{25},
  \bibinfo{number}{3} (\bibinfo{date}{July} \bibinfo{year}{2011}),
  \bibinfo{pages}{211--215}.
\newblock
\showISSN{0269-1728}
\urldef\tempurl%
\url{https://doi.org/10.1080/02691728.2011.578301}
\showDOI{\tempurl}
\newblock
\shownote{Publisher: Routledge \_eprint:
  https://doi.org/10.1080/02691728.2011.578301.}


\bibitem[\protect\citeauthoryear{Hepenstal and McNeish}{Hepenstal and
  McNeish}{2020}]%
        {hepenstal2020}
\bibfield{author}{\bibinfo{person}{Sam Hepenstal} {and} \bibinfo{person}{David
  McNeish}.} \bibinfo{year}{2020}\natexlab{}.
\newblock \showarticletitle{Explainable Artificial Intelligence: What Do You
  Need to Know?}. In \bibinfo{booktitle}{\emph{Augmented Cognition. Theoretical
  and Technological Approaches}}, \bibfield{editor}{\bibinfo{person}{Dylan~D.
  Schmorrow} {and} \bibinfo{person}{Cali~M. Fidopiastis}} (Eds.).
  \bibinfo{publisher}{Springer International Publishing},
  \bibinfo{address}{Cham}, \bibinfo{pages}{266--275}.
\newblock
\showISBNx{978-3-030-50353-6}


\bibitem[\protect\citeauthoryear{Hildebrandt}{Hildebrandt}{2012}]%
        {hildebrandt2012}
\bibfield{author}{\bibinfo{person}{Mireille Hildebrandt}.}
  \bibinfo{year}{2012}\natexlab{}.
\newblock \showarticletitle{The Dawn of a Critical Transparency Right for the
  Profiling Era}.
\newblock \bibinfo{journal}{\emph{Astronomy \& Astrophysics - ASTRON
  ASTROPHYS}} (\bibinfo{date}{06} \bibinfo{year}{2012}).
\newblock
\urldef\tempurl%
\url{https://doi.org/10.3233/978-1-61499-057-4-41}
\showDOI{\tempurl}


\bibitem[\protect\citeauthoryear{Hohman, Head, Caruana, DeLine, and
  Drucker}{Hohman et~al\mbox{.}}{2019a}]%
        {hohman2019gamut}
\bibfield{author}{\bibinfo{person}{Fred Hohman}, \bibinfo{person}{Andrew Head},
  \bibinfo{person}{Rich Caruana}, \bibinfo{person}{Robert DeLine}, {and}
  \bibinfo{person}{Steven~M. Drucker}.} \bibinfo{year}{2019}\natexlab{a}.
\newblock \showarticletitle{Gamut: A Design Probe to Understand How Data
  Scientists Understand Machine Learning Models}. In
  \bibinfo{booktitle}{\emph{Proceedings of the 2019 CHI Conference on Human
  Factors in Computing Systems}} (Glasgow, Scotland Uk)
  \emph{(\bibinfo{series}{CHI '19})}. \bibinfo{publisher}{Association for
  Computing Machinery}, \bibinfo{address}{New York, NY, USA},
  \bibinfo{pages}{1–13}.
\newblock
\showISBNx{9781450359702}
\urldef\tempurl%
\url{https://doi.org/10.1145/3290605.3300809}
\showDOI{\tempurl}


\bibitem[\protect\citeauthoryear{Hohman, Kahng, Pienta, and Chau}{Hohman
  et~al\mbox{.}}{2019b}]%
        {hohman_visual_2019}
\bibfield{author}{\bibinfo{person}{Fred Hohman}, \bibinfo{person}{Minsuk
  Kahng}, \bibinfo{person}{Robert Pienta}, {and} \bibinfo{person}{Duen~Horng
  Chau}.} \bibinfo{year}{2019}\natexlab{b}.
\newblock \showarticletitle{Visual {Analytics} in {Deep} {Learning}: {An}
  {Interrogative} {Survey} for the {Next} {Frontiers}}.
\newblock \bibinfo{journal}{\emph{IEEE Transactions on Visualization and
  Computer Graphics}} \bibinfo{volume}{25}, \bibinfo{number}{8}
  (\bibinfo{date}{Aug.} \bibinfo{year}{2019}), \bibinfo{pages}{2674--2693}.
\newblock
\showISSN{1941-0506}
\urldef\tempurl%
\url{https://doi.org/10.1109/TVCG.2018.2843369}
\showDOI{\tempurl}


\bibitem[\protect\citeauthoryear{Hong, Hullman, and Bertini}{Hong
  et~al\mbox{.}}{2020}]%
        {hong_human_2020}
\bibfield{author}{\bibinfo{person}{Sungsoo~Ray Hong}, \bibinfo{person}{Jessica
  Hullman}, {and} \bibinfo{person}{Enrico Bertini}.}
  \bibinfo{year}{2020}\natexlab{}.
\newblock \showarticletitle{Human {Factors} in {Model} {Interpretability}:
  {Industry} {Practices}, {Challenges}, and {Needs}}.
\newblock \bibinfo{journal}{\emph{Proceedings of the ACM on Human-Computer
  Interaction}} \bibinfo{volume}{4}, \bibinfo{number}{CSCW1}
  (\bibinfo{date}{May} \bibinfo{year}{2020}), \bibinfo{pages}{1--26}.
\newblock
\showISSN{2573-0142, 2573-0142}
\urldef\tempurl%
\url{https://doi.org/10.1145/3392878}
\showDOI{\tempurl}


\bibitem[\protect\citeauthoryear{Hullman and Diakopoulos}{Hullman and
  Diakopoulos}{2011}]%
        {hullman2011visualization}
\bibfield{author}{\bibinfo{person}{Jessica Hullman} {and} \bibinfo{person}{Nick
  Diakopoulos}.} \bibinfo{year}{2011}\natexlab{}.
\newblock \showarticletitle{Visualization rhetoric: Framing effects in
  narrative visualization}.
\newblock \bibinfo{journal}{\emph{IEEE transactions on visualization and
  computer graphics}} \bibinfo{volume}{17}, \bibinfo{number}{12}
  (\bibinfo{year}{2011}), \bibinfo{pages}{2231--2240}.
\newblock


\bibitem[\protect\citeauthoryear{Jull, Giles, and Graham}{Jull
  et~al\mbox{.}}{2017}]%
        {jull_community-based_2017}
\bibfield{author}{\bibinfo{person}{Janet Jull}, \bibinfo{person}{Audrey Giles},
  {and} \bibinfo{person}{Ian~D. Graham}.} \bibinfo{year}{2017}\natexlab{}.
\newblock \showarticletitle{Community-based participatory research and
  integrated knowledge translation: advancing the co-creation of knowledge}.
\newblock \bibinfo{journal}{\emph{Implementation Science}}
  \bibinfo{volume}{12}, \bibinfo{number}{1} (\bibinfo{date}{Dec.}
  \bibinfo{year}{2017}), \bibinfo{pages}{150}.
\newblock
\showISSN{1748-5908}
\urldef\tempurl%
\url{https://doi.org/10.1186/s13012-017-0696-3}
\showDOI{\tempurl}


\bibitem[\protect\citeauthoryear{Kaur, Nori, Jenkins, Caruana, Wallach, and
  Wortman~Vaughan}{Kaur et~al\mbox{.}}{2020}]%
        {kaur_interpreting_2020}
\bibfield{author}{\bibinfo{person}{Harmanpreet Kaur}, \bibinfo{person}{Harsha
  Nori}, \bibinfo{person}{Samuel Jenkins}, \bibinfo{person}{Rich Caruana},
  \bibinfo{person}{Hanna Wallach}, {and} \bibinfo{person}{Jennifer
  Wortman~Vaughan}.} \bibinfo{year}{2020}\natexlab{}.
\newblock \showarticletitle{Interpreting {Interpretability}: {Understanding}
  {Data} {Scientists}' {Use} of {Interpretability} {Tools} for {Machine}
  {Learning}}. In \bibinfo{booktitle}{\emph{Proceedings of the 2020 {CHI}
  {Conference} on {Human} {Factors} in {Computing} {Systems}}}.
  \bibinfo{publisher}{ACM}, \bibinfo{address}{Honolulu HI USA},
  \bibinfo{pages}{1--14}.
\newblock
\showISBNx{978-1-4503-6708-0}
\urldef\tempurl%
\url{https://doi.org/10.1145/3313831.3376219}
\showDOI{\tempurl}


\bibitem[\protect\citeauthoryear{Kennedy, Hill, Aiello, and Allen}{Kennedy
  et~al\mbox{.}}{2016}]%
        {kennedy2016work}
\bibfield{author}{\bibinfo{person}{Helen Kennedy},
  \bibinfo{person}{Rosemary~Lucy Hill}, \bibinfo{person}{Giorgia Aiello}, {and}
  \bibinfo{person}{William Allen}.} \bibinfo{year}{2016}\natexlab{}.
\newblock \showarticletitle{The work that visualisation conventions do}.
\newblock \bibinfo{journal}{\emph{Information, Communication \& Society}}
  \bibinfo{volume}{19}, \bibinfo{number}{6} (\bibinfo{year}{2016}),
  \bibinfo{pages}{715--735}.
\newblock


\bibitem[\protect\citeauthoryear{Kim, Wattenberg, Gilmer, Cai, Wexler, Viegas,
  et~al\mbox{.}}{Kim et~al\mbox{.}}{2018}]%
        {kim2018interpretability}
\bibfield{author}{\bibinfo{person}{Been Kim}, \bibinfo{person}{Martin
  Wattenberg}, \bibinfo{person}{Justin Gilmer}, \bibinfo{person}{Carrie Cai},
  \bibinfo{person}{James Wexler}, \bibinfo{person}{Fernanda Viegas},
  {et~al\mbox{.}}} \bibinfo{year}{2018}\natexlab{}.
\newblock \showarticletitle{Interpretability beyond feature attribution:
  Quantitative testing with concept activation vectors (tcav)}. In
  \bibinfo{booktitle}{\emph{International conference on machine learning}}.
  PMLR, \bibinfo{pages}{2668--2677}.
\newblock


\bibitem[\protect\citeauthoryear{Kinchin and Cabot}{Kinchin and Cabot}{2010}]%
        {kinchin_reconsidering_2010}
\bibfield{author}{\bibinfo{person}{Ian~M. Kinchin} {and} \bibinfo{person}{B.
  Cabot}.} \bibinfo{year}{2010}\natexlab{}.
\newblock \showarticletitle{Reconsidering the dimensions of expertise: from
  linear stages towards dual processing}.
\newblock \bibinfo{journal}{\emph{London Review of Education}}
  (\bibinfo{date}{July} \bibinfo{year}{2010}).
\newblock
\showISSN{1474-8460}
\urldef\tempurl%
\url{https://doi.org/10.1080/14748460.2010.487334}
\showDOI{\tempurl}


\bibitem[\protect\citeauthoryear{Kizilcec}{Kizilcec}{2016}]%
        {kizilcec2016much}
\bibfield{author}{\bibinfo{person}{Ren{\'e}~F Kizilcec}.}
  \bibinfo{year}{2016}\natexlab{}.
\newblock \showarticletitle{How much information? Effects of transparency on
  trust in an algorithmic interface}. In \bibinfo{booktitle}{\emph{Proceedings
  of the 2016 CHI Conference on Human Factors in Computing Systems}}.
  \bibinfo{pages}{2390--2395}.
\newblock


\bibitem[\protect\citeauthoryear{Kong, Liu, and Karahalios}{Kong
  et~al\mbox{.}}{2018}]%
        {kong2018frames}
\bibfield{author}{\bibinfo{person}{Ha-Kyung Kong}, \bibinfo{person}{Zhicheng
  Liu}, {and} \bibinfo{person}{Karrie Karahalios}.}
  \bibinfo{year}{2018}\natexlab{}.
\newblock \showarticletitle{Frames and slants in titles of visualizations on
  controversial topics}. In \bibinfo{booktitle}{\emph{Proceedings of the 2018
  CHI Conference on Human Factors in Computing Systems}}.
  \bibinfo{pages}{1--12}.
\newblock


\bibitem[\protect\citeauthoryear{Kong, Liu, and Karahalios}{Kong
  et~al\mbox{.}}{2019}]%
        {kong2019trust}
\bibfield{author}{\bibinfo{person}{Ha-Kyung Kong}, \bibinfo{person}{Zhicheng
  Liu}, {and} \bibinfo{person}{Karrie Karahalios}.}
  \bibinfo{year}{2019}\natexlab{}.
\newblock \showarticletitle{Trust and recall of information across varying
  degrees of title-visualization misalignment}. In
  \bibinfo{booktitle}{\emph{Proceedings of the 2019 CHI Conference on Human
  Factors in Computing Systems}}. \bibinfo{pages}{1--13}.
\newblock


\bibitem[\protect\citeauthoryear{Krebs, Alvarado~Rodriguez, Dewitte, Ausloos,
  Geerts, Naudts, and Verbert}{Krebs et~al\mbox{.}}{2019}]%
        {krebs2019tell}
\bibfield{author}{\bibinfo{person}{Luciana~Monteiro Krebs},
  \bibinfo{person}{Oscar~Luis Alvarado~Rodriguez}, \bibinfo{person}{Pierre
  Dewitte}, \bibinfo{person}{Jef Ausloos}, \bibinfo{person}{David Geerts},
  \bibinfo{person}{Laurens Naudts}, {and} \bibinfo{person}{Katrien Verbert}.}
  \bibinfo{year}{2019}\natexlab{}.
\newblock \showarticletitle{Tell me what you know: GDPR implications on
  designing transparency and accountability for news recommender systems}. In
  \bibinfo{booktitle}{\emph{Extended Abstracts of the 2019 CHI Conference on
  Human Factors in Computing Systems}}. \bibinfo{pages}{1--6}.
\newblock


\bibitem[\protect\citeauthoryear{Kulynych, Madras, Milli, Raji, Zhou, and
  Zemel}{Kulynych et~al\mbox{.}}{2020}]%
        {paml}
\bibfield{author}{\bibinfo{person}{Bogdan Kulynych}, \bibinfo{person}{David
  Madras}, \bibinfo{person}{Smitha Milli}, \bibinfo{person}{Inioluwa~Deborah
  Raji}, \bibinfo{person}{Angela Zhou}, {and} \bibinfo{person}{Richard Zemel}.}
  \bibinfo{year}{2020}\natexlab{}.
\newblock \bibinfo{title}{Participatory Approaches to Machine Learning}.
\newblock \bibinfo{howpublished}{International Conference on Machine Learning
  Workshop}.
\newblock


\bibitem[\protect\citeauthoryear{Labaree}{Labaree}{2000}]%
        {labaree2000nature}
\bibfield{author}{\bibinfo{person}{David~F Labaree}.}
  \bibinfo{year}{2000}\natexlab{}.
\newblock \showarticletitle{On the nature of teaching and teacher education:
  Difficult practices that look easy}.
\newblock \bibinfo{journal}{\emph{Journal of teacher education}}
  \bibinfo{volume}{51}, \bibinfo{number}{3} (\bibinfo{year}{2000}),
  \bibinfo{pages}{228--233}.
\newblock


\bibitem[\protect\citeauthoryear{Lage, Chen, He, Narayanan, Kim, Gershman, and
  Doshi-Velez}{Lage et~al\mbox{.}}{2019}]%
        {lage_human_2019}
\bibfield{author}{\bibinfo{person}{Isaac Lage}, \bibinfo{person}{Emily Chen},
  \bibinfo{person}{Jeffrey He}, \bibinfo{person}{Menaka Narayanan},
  \bibinfo{person}{Been Kim}, \bibinfo{person}{Samuel~J. Gershman}, {and}
  \bibinfo{person}{Finale Doshi-Velez}.} \bibinfo{year}{2019}\natexlab{}.
\newblock \showarticletitle{Human {Evaluation} of {Models} {Built} for
  {Interpretability}}.
\newblock \bibinfo{journal}{\emph{Proceedings of the AAAI Conference on Human
  Computation and Crowdsourcing}} \bibinfo{volume}{7}, \bibinfo{number}{1}
  (\bibinfo{date}{Oct.} \bibinfo{year}{2019}), \bibinfo{pages}{59--67}.
\newblock
\urldef\tempurl%
\url{https://www.aaai.org/ojs/index.php/HCOMP/article/view/5280}
\showURL{%
\tempurl}
\newblock
\shownote{Number: 1.}


\bibitem[\protect\citeauthoryear{Lai and Tan}{Lai and Tan}{2019}]%
        {lai_human_2019}
\bibfield{author}{\bibinfo{person}{Vivian Lai} {and} \bibinfo{person}{Chenhao
  Tan}.} \bibinfo{year}{2019}\natexlab{}.
\newblock \showarticletitle{On {Human} {Predictions} with {Explanations} and
  {Predictions} of {Machine} {Learning} {Models}: {A} {Case} {Study} on
  {Deception} {Detection}}. In \bibinfo{booktitle}{\emph{Proceedings of the
  {Conference} on {Fairness}, {Accountability}, and {Transparency} - {FAT}*
  '19}}. \bibinfo{publisher}{ACM Press}, \bibinfo{address}{Atlanta, GA, USA},
  \bibinfo{pages}{29--38}.
\newblock
\showISBNx{978-1-4503-6125-5}
\urldef\tempurl%
\url{https://doi.org/10.1145/3287560.3287590}
\showDOI{\tempurl}


\bibitem[\protect\citeauthoryear{Lakkaraju and Bastani}{Lakkaraju and
  Bastani}{2020}]%
        {lakkaraju_how_2020}
\bibfield{author}{\bibinfo{person}{Himabindu Lakkaraju} {and}
  \bibinfo{person}{Osbert Bastani}.} \bibinfo{year}{2020}\natexlab{}.
\newblock \showarticletitle{"{How} do {I} fool you?": {Manipulating} {User}
  {Trust} via {Misleading} {Black} {Box} {Explanations}}. In
  \bibinfo{booktitle}{\emph{Proceedings of the {AAAI}/{ACM} {Conference} on
  {AI}, {Ethics}, and {Society}}} \emph{(\bibinfo{series}{{AIES} '20})}.
  \bibinfo{publisher}{Association for Computing Machinery},
  \bibinfo{address}{New York, NY, USA}, \bibinfo{pages}{79--85}.
\newblock
\showISBNx{978-1-4503-7110-0}
\urldef\tempurl%
\url{https://doi.org/10.1145/3375627.3375833}
\showDOI{\tempurl}


\bibitem[\protect\citeauthoryear{Le~Dantec and Fox}{Le~Dantec and Fox}{2015}]%
        {le_dantec_strangers_2015}
\bibfield{author}{\bibinfo{person}{Christopher~A. Le~Dantec} {and}
  \bibinfo{person}{Sarah Fox}.} \bibinfo{year}{2015}\natexlab{}.
\newblock \showarticletitle{Strangers at the {Gate}: {Gaining} {Access},
  {Building} {Rapport}, and {Co}-{Constructing} {Community}-{Based}
  {Research}}. In \bibinfo{booktitle}{\emph{Proceedings of the 18th {ACM}
  {Conference} on {Computer} {Supported} {Cooperative} {Work} \& {Social}
  {Computing} - {CSCW} '15}}. \bibinfo{publisher}{ACM Press},
  \bibinfo{address}{Vancouver, BC, Canada}, \bibinfo{pages}{1348--1358}.
\newblock
\showISBNx{978-1-4503-2922-4}
\urldef\tempurl%
\url{https://doi.org/10.1145/2675133.2675147}
\showDOI{\tempurl}


\bibitem[\protect\citeauthoryear{Lee, Isaacs, Szafir, Marai, Turkay, Tory,
  Carpendale, and Endert}{Lee et~al\mbox{.}}{2019}]%
        {lee2019broadening}
\bibfield{author}{\bibinfo{person}{Bongshin Lee}, \bibinfo{person}{Kate
  Isaacs}, \bibinfo{person}{Danielle~Albers Szafir},
  \bibinfo{person}{G~Elisabeta Marai}, \bibinfo{person}{Cagatay Turkay},
  \bibinfo{person}{Melanie Tory}, \bibinfo{person}{Sheelagh Carpendale}, {and}
  \bibinfo{person}{Alex Endert}.} \bibinfo{year}{2019}\natexlab{}.
\newblock \showarticletitle{Broadening intellectual diversity in visualization
  research papers}.
\newblock \bibinfo{journal}{\emph{IEEE computer graphics and applications}}
  \bibinfo{volume}{39}, \bibinfo{number}{4} (\bibinfo{year}{2019}),
  \bibinfo{pages}{78--85}.
\newblock


\bibitem[\protect\citeauthoryear{Liao, Gruen, and Miller}{Liao
  et~al\mbox{.}}{2020}]%
        {liao_questioning_2020}
\bibfield{author}{\bibinfo{person}{Q.~Vera Liao}, \bibinfo{person}{Daniel
  Gruen}, {and} \bibinfo{person}{Sarah Miller}.}
  \bibinfo{year}{2020}\natexlab{}.
\newblock \showarticletitle{Questioning the {AI}: {Informing} {Design}
  {Practices} for {Explainable} {AI} {User} {Experiences}}. In
  \bibinfo{booktitle}{\emph{Proceedings of the 2020 {CHI} {Conference} on
  {Human} {Factors} in {Computing} {Systems}}} \emph{(\bibinfo{series}{{CHI}
  '20})}. \bibinfo{publisher}{Association for Computing Machinery},
  \bibinfo{address}{Honolulu, HI, USA}, \bibinfo{pages}{1--15}.
\newblock
\showISBNx{978-1-4503-6708-0}
\urldef\tempurl%
\url{https://doi.org/10.1145/3313831.3376590}
\showDOI{\tempurl}


\bibitem[\protect\citeauthoryear{Lim and Dey}{Lim and Dey}{2009}]%
        {lim2009assessing}
\bibfield{author}{\bibinfo{person}{Brian~Y. Lim} {and}
  \bibinfo{person}{Anind~K. Dey}.} \bibinfo{year}{2009}\natexlab{}.
\newblock \showarticletitle{Assessing Demand for Intelligibility in
  Context-Aware Applications}. In \bibinfo{booktitle}{\emph{Proceedings of the
  11th International Conference on Ubiquitous Computing}} (Orlando, Florida,
  USA) \emph{(\bibinfo{series}{UbiComp '09})}. \bibinfo{publisher}{Association
  for Computing Machinery}, \bibinfo{address}{New York, NY, USA},
  \bibinfo{pages}{195–204}.
\newblock
\showISBNx{9781605584317}
\urldef\tempurl%
\url{https://doi.org/10.1145/1620545.1620576}
\showDOI{\tempurl}


\bibitem[\protect\citeauthoryear{Lim and Dey}{Lim and Dey}{2010}]%
        {lim2010toolkit}
\bibfield{author}{\bibinfo{person}{Brian~Y. Lim} {and}
  \bibinfo{person}{Anind~K. Dey}.} \bibinfo{year}{2010}\natexlab{}.
\newblock \showarticletitle{Toolkit to Support Intelligibility in Context-Aware
  Applications}. In \bibinfo{booktitle}{\emph{Proceedings of the 12th ACM
  International Conference on Ubiquitous Computing}} (Copenhagen, Denmark)
  \emph{(\bibinfo{series}{UbiComp '10})}. \bibinfo{publisher}{Association for
  Computing Machinery}, \bibinfo{address}{New York, NY, USA},
  \bibinfo{pages}{13–22}.
\newblock
\showISBNx{9781605588438}
\urldef\tempurl%
\url{https://doi.org/10.1145/1864349.1864353}
\showDOI{\tempurl}


\bibitem[\protect\citeauthoryear{Lim, Dey, and Avrahami}{Lim
  et~al\mbox{.}}{2009}]%
        {lim2009why}
\bibfield{author}{\bibinfo{person}{Brian~Y. Lim}, \bibinfo{person}{Anind~K.
  Dey}, {and} \bibinfo{person}{Daniel Avrahami}.}
  \bibinfo{year}{2009}\natexlab{}.
\newblock \showarticletitle{Why and Why Not Explanations Improve the
  Intelligibility of Context-Aware Intelligent Systems}. In
  \bibinfo{booktitle}{\emph{Proceedings of the SIGCHI Conference on Human
  Factors in Computing Systems}} (Boston, MA, USA) \emph{(\bibinfo{series}{CHI
  '09})}. \bibinfo{publisher}{Association for Computing Machinery},
  \bibinfo{address}{New York, NY, USA}, \bibinfo{pages}{2119–2128}.
\newblock
\showISBNx{9781605582467}
\urldef\tempurl%
\url{https://doi.org/10.1145/1518701.1519023}
\showDOI{\tempurl}


\bibitem[\protect\citeauthoryear{Lipton}{Lipton}{2018}]%
        {lipton_mythos_2018}
\bibfield{author}{\bibinfo{person}{Zachary~C. Lipton}.}
  \bibinfo{year}{2018}\natexlab{}.
\newblock \showarticletitle{The mythos of model interpretability}.
\newblock \bibinfo{journal}{\emph{Commun. ACM}} \bibinfo{volume}{61},
  \bibinfo{number}{10} (\bibinfo{date}{Sept.} \bibinfo{year}{2018}),
  \bibinfo{pages}{36--43}.
\newblock
\showISSN{0001-0782, 1557-7317}
\urldef\tempurl%
\url{https://doi.org/10.1145/3233231}
\showDOI{\tempurl}


\bibitem[\protect\citeauthoryear{Lundberg, Nair, Vavilala, Horibe, Eisses,
  Adams, Liston, Low, Newman, Kim, et~al\mbox{.}}{Lundberg
  et~al\mbox{.}}{2018}]%
        {lundberg2018explainable}
\bibfield{author}{\bibinfo{person}{Scott~M Lundberg}, \bibinfo{person}{Bala
  Nair}, \bibinfo{person}{Monica~S Vavilala}, \bibinfo{person}{Mayumi Horibe},
  \bibinfo{person}{Michael~J Eisses}, \bibinfo{person}{Trevor Adams},
  \bibinfo{person}{David~E Liston}, \bibinfo{person}{Daniel King-Wai Low},
  \bibinfo{person}{Shu-Fang Newman}, \bibinfo{person}{Jerry Kim},
  {et~al\mbox{.}}} \bibinfo{year}{2018}\natexlab{}.
\newblock \showarticletitle{Explainable machine-learning predictions for the
  prevention of hypoxaemia during surgery}.
\newblock \bibinfo{journal}{\emph{Nature biomedical engineering}}
  \bibinfo{volume}{2}, \bibinfo{number}{10} (\bibinfo{year}{2018}),
  \bibinfo{pages}{749--760}.
\newblock


\bibitem[\protect\citeauthoryear{Miller}{Miller}{1956}]%
        {miller1956magical}
\bibfield{author}{\bibinfo{person}{George~A Miller}.}
  \bibinfo{year}{1956}\natexlab{}.
\newblock \showarticletitle{The magical number seven, plus or minus two: Some
  limits on our capacity for processing information.}
\newblock \bibinfo{journal}{\emph{Psychological review}} \bibinfo{volume}{63},
  \bibinfo{number}{2} (\bibinfo{year}{1956}), \bibinfo{pages}{81}.
\newblock


\bibitem[\protect\citeauthoryear{Miller}{Miller}{2019}]%
        {miller_explanation_2019}
\bibfield{author}{\bibinfo{person}{Tim Miller}.}
  \bibinfo{year}{2019}\natexlab{}.
\newblock \showarticletitle{Explanation in artificial intelligence: {Insights}
  from the social sciences}.
\newblock \bibinfo{journal}{\emph{Artificial Intelligence}}
  \bibinfo{volume}{267} (\bibinfo{date}{Feb.} \bibinfo{year}{2019}),
  \bibinfo{pages}{1--38}.
\newblock
\showISSN{00043702}
\urldef\tempurl%
\url{https://doi.org/10.1016/j.artint.2018.07.007}
\showDOI{\tempurl}


\bibitem[\protect\citeauthoryear{Mitchell, Wu, Zaldivar, Barnes, Vasserman,
  Hutchinson, Spitzer, Raji, and Gebru}{Mitchell et~al\mbox{.}}{2019}]%
        {mitchell2019model}
\bibfield{author}{\bibinfo{person}{Margaret Mitchell}, \bibinfo{person}{Simone
  Wu}, \bibinfo{person}{Andrew Zaldivar}, \bibinfo{person}{Parker Barnes},
  \bibinfo{person}{Lucy Vasserman}, \bibinfo{person}{Ben Hutchinson},
  \bibinfo{person}{Elena Spitzer}, \bibinfo{person}{Inioluwa~Deborah Raji},
  {and} \bibinfo{person}{Timnit Gebru}.} \bibinfo{year}{2019}\natexlab{}.
\newblock \showarticletitle{Model cards for model reporting}. In
  \bibinfo{booktitle}{\emph{Proceedings of the conference on fairness,
  accountability, and transparency}}. \bibinfo{pages}{220--229}.
\newblock


\bibitem[\protect\citeauthoryear{Mohamed, Png, and Isaac}{Mohamed
  et~al\mbox{.}}{2020}]%
        {mohamed_decolonial_2020}
\bibfield{author}{\bibinfo{person}{Shakir Mohamed},
  \bibinfo{person}{Marie-Therese Png}, {and} \bibinfo{person}{William Isaac}.}
  \bibinfo{year}{2020}\natexlab{}.
\newblock \showarticletitle{Decolonial {AI}: {Decolonial} {Theory} as
  {Sociotechnical} {Foresight} in {Artificial} {Intelligence}}.
\newblock \bibinfo{journal}{\emph{Philosophy \& Technology}}
  (\bibinfo{date}{July} \bibinfo{year}{2020}).
\newblock
\showISSN{2210-5433, 2210-5441}
\urldef\tempurl%
\url{https://doi.org/10.1007/s13347-020-00405-8}
\showDOI{\tempurl}
\newblock
\shownote{arXiv: 2007.04068.}


\bibitem[\protect\citeauthoryear{Mohseni, Block, and Ragan}{Mohseni
  et~al\mbox{.}}{2020a}]%
        {mohseni_human-grounded_2020}
\bibfield{author}{\bibinfo{person}{Sina Mohseni}, \bibinfo{person}{Jeremy~E.
  Block}, {and} \bibinfo{person}{Eric~D. Ragan}.}
  \bibinfo{year}{2020}\natexlab{a}.
\newblock \showarticletitle{A {Human}-{Grounded} {Evaluation} {Benchmark} for
  {Local} {Explanations} of {Machine} {Learning}}.
\newblock \bibinfo{journal}{\emph{arXiv:1801.05075 [cs]}} (\bibinfo{date}{June}
  \bibinfo{year}{2020}).
\newblock
\urldef\tempurl%
\url{http://arxiv.org/abs/1801.05075}
\showURL{%
\tempurl}
\newblock
\shownote{arXiv: 1801.05075.}


\bibitem[\protect\citeauthoryear{Mohseni, Zarei, and Ragan}{Mohseni
  et~al\mbox{.}}{2020b}]%
        {mohseni_multidisciplinary_2020}
\bibfield{author}{\bibinfo{person}{Sina Mohseni}, \bibinfo{person}{Niloofar
  Zarei}, {and} \bibinfo{person}{Eric~D. Ragan}.}
  \bibinfo{year}{2020}\natexlab{b}.
\newblock \showarticletitle{A {Multidisciplinary} {Survey} and {Framework} for
  {Design} and {Evaluation} of {Explainable} {AI} {Systems}}.
\newblock \bibinfo{journal}{\emph{arXiv:1811.11839 [cs]}}
  (\bibinfo{date}{April} \bibinfo{year}{2020}).
\newblock
\urldef\tempurl%
\url{http://arxiv.org/abs/1811.11839}
\showURL{%
\tempurl}
\newblock
\shownote{arXiv: 1811.11839.}


\bibitem[\protect\citeauthoryear{Narayanan, Chen, He, Kim, Gershman, and
  Doshi-Velez}{Narayanan et~al\mbox{.}}{2018}]%
        {narayanan2018humans}
\bibfield{author}{\bibinfo{person}{Menaka Narayanan}, \bibinfo{person}{Emily
  Chen}, \bibinfo{person}{Jeffrey He}, \bibinfo{person}{Been Kim},
  \bibinfo{person}{Sam Gershman}, {and} \bibinfo{person}{Finale Doshi-Velez}.}
  \bibinfo{year}{2018}\natexlab{}.
\newblock \showarticletitle{How do humans understand explanations from machine
  learning systems? an evaluation of the human-interpretability of
  explanation}.
\newblock \bibinfo{journal}{\emph{arXiv preprint arXiv:1802.00682}}
  (\bibinfo{year}{2018}).
\newblock


\bibitem[\protect\citeauthoryear{Nunes and Jannach}{Nunes and Jannach}{2017}]%
        {nunes2017systematic}
\bibfield{author}{\bibinfo{person}{Ingrid Nunes} {and} \bibinfo{person}{Dietmar
  Jannach}.} \bibinfo{year}{2017}\natexlab{}.
\newblock \showarticletitle{A systematic review and taxonomy of explanations in
  decision support and recommender systems}.
\newblock \bibinfo{journal}{\emph{User Modeling and User-Adapted Interaction}}
  \bibinfo{volume}{27}, \bibinfo{number}{3-5} (\bibinfo{year}{2017}),
  \bibinfo{pages}{393--444}.
\newblock


\bibitem[\protect\citeauthoryear{Ogbonnaya-Ogburu, Smith, To, and
  Toyama}{Ogbonnaya-Ogburu et~al\mbox{.}}{2020}]%
        {ogbonnaya-ogburu_critical_2020}
\bibfield{author}{\bibinfo{person}{Ihudiya~Finda Ogbonnaya-Ogburu},
  \bibinfo{person}{Angela~D.R. Smith}, \bibinfo{person}{Alexandra To}, {and}
  \bibinfo{person}{Kentaro Toyama}.} \bibinfo{year}{2020}\natexlab{}.
\newblock \showarticletitle{Critical {Race} {Theory} for {HCI}}. In
  \bibinfo{booktitle}{\emph{Proceedings of the 2020 {CHI} {Conference} on
  {Human} {Factors} in {Computing} {Systems}}}. \bibinfo{publisher}{ACM},
  \bibinfo{address}{Honolulu HI USA}, \bibinfo{pages}{1--16}.
\newblock
\showISBNx{978-1-4503-6708-0}
\urldef\tempurl%
\url{https://doi.org/10.1145/3313831.3376392}
\showDOI{\tempurl}


\bibitem[\protect\citeauthoryear{Olah, Mordvintsev, and Schubert}{Olah
  et~al\mbox{.}}{2017}]%
        {olah2017feature}
\bibfield{author}{\bibinfo{person}{Chris Olah}, \bibinfo{person}{Alexander
  Mordvintsev}, {and} \bibinfo{person}{Ludwig Schubert}.}
  \bibinfo{year}{2017}\natexlab{}.
\newblock \showarticletitle{Feature visualization}.
\newblock \bibinfo{journal}{\emph{Distill}} \bibinfo{volume}{2},
  \bibinfo{number}{11} (\bibinfo{year}{2017}), \bibinfo{pages}{e7}.
\newblock


\bibitem[\protect\citeauthoryear{Olah, Satyanarayan, Johnson, Carter, Schubert,
  Ye, and Mordvintsev}{Olah et~al\mbox{.}}{2018}]%
        {olah2018building}
\bibfield{author}{\bibinfo{person}{Chris Olah}, \bibinfo{person}{Arvind
  Satyanarayan}, \bibinfo{person}{Ian Johnson}, \bibinfo{person}{Shan Carter},
  \bibinfo{person}{Ludwig Schubert}, \bibinfo{person}{Katherine Ye}, {and}
  \bibinfo{person}{Alexander Mordvintsev}.} \bibinfo{year}{2018}\natexlab{}.
\newblock \showarticletitle{The building blocks of interpretability}.
\newblock \bibinfo{journal}{\emph{Distill}} \bibinfo{volume}{3},
  \bibinfo{number}{3} (\bibinfo{year}{2018}), \bibinfo{pages}{e10}.
\newblock


\bibitem[\protect\citeauthoryear{Paudyal, Lee, Kamzin, Soudki, Banerjee, and
  Gupta}{Paudyal et~al\mbox{.}}{2019}]%
        {paudyal2019learn2sign}
\bibfield{author}{\bibinfo{person}{Prajwal Paudyal}, \bibinfo{person}{Junghyo
  Lee}, \bibinfo{person}{Azamat Kamzin}, \bibinfo{person}{Mohamad Soudki},
  \bibinfo{person}{Ayan Banerjee}, {and} \bibinfo{person}{Sandeep~KS Gupta}.}
  \bibinfo{year}{2019}\natexlab{}.
\newblock \showarticletitle{Learn2Sign: Explainable AI for Sign Language
  Learning.}. In \bibinfo{booktitle}{\emph{IUI Workshops}}.
\newblock


\bibitem[\protect\citeauthoryear{Peck, Ayuso, and El-Etr}{Peck
  et~al\mbox{.}}{2019}]%
        {peck2019data}
\bibfield{author}{\bibinfo{person}{Evan~M Peck}, \bibinfo{person}{Sofia~E
  Ayuso}, {and} \bibinfo{person}{Omar El-Etr}.}
  \bibinfo{year}{2019}\natexlab{}.
\newblock \showarticletitle{Data is personal: Attitudes and perceptions of data
  visualization in rural pennsylvania}. In
  \bibinfo{booktitle}{\emph{Proceedings of the 2019 CHI Conference on Human
  Factors in Computing Systems}}. \bibinfo{pages}{1--12}.
\newblock


\bibitem[\protect\citeauthoryear{Poursabzi-Sangdeh, Goldstein, Hofman, Vaughan,
  and Wallach}{Poursabzi-Sangdeh et~al\mbox{.}}{2019}]%
        {poursabzi-sangdeh_manipulating_2019}
\bibfield{author}{\bibinfo{person}{Forough Poursabzi-Sangdeh},
  \bibinfo{person}{Daniel~G. Goldstein}, \bibinfo{person}{Jake~M. Hofman},
  \bibinfo{person}{Jennifer~Wortman Vaughan}, {and} \bibinfo{person}{Hanna
  Wallach}.} \bibinfo{year}{2019}\natexlab{}.
\newblock \showarticletitle{Manipulating and {Measuring} {Model}
  {Interpretability}}.
\newblock \bibinfo{journal}{\emph{arXiv:1802.07810 [cs]}} (\bibinfo{date}{Nov.}
  \bibinfo{year}{2019}).
\newblock
\urldef\tempurl%
\url{http://arxiv.org/abs/1802.07810}
\showURL{%
\tempurl}
\newblock
\shownote{arXiv: 1802.07810.}


\bibitem[\protect\citeauthoryear{Preece, Harborne, Braines, Tomsett, and
  Chakraborty}{Preece et~al\mbox{.}}{2018}]%
        {preece_stakeholders_2018}
\bibfield{author}{\bibinfo{person}{Alun Preece}, \bibinfo{person}{Dan
  Harborne}, \bibinfo{person}{Dave Braines}, \bibinfo{person}{Richard Tomsett},
  {and} \bibinfo{person}{Supriyo Chakraborty}.}
  \bibinfo{year}{2018}\natexlab{}.
\newblock \showarticletitle{Stakeholders in {Explainable} {AI}}.
\newblock \bibinfo{journal}{\emph{arXiv:1810.00184 [cs]}}
  (\bibinfo{date}{Sept.} \bibinfo{year}{2018}).
\newblock
\urldef\tempurl%
\url{http://arxiv.org/abs/1810.00184}
\showURL{%
\tempurl}
\newblock
\shownote{arXiv: 1810.00184.}


\bibitem[\protect\citeauthoryear{Raji, Smart, White, Mitchell, Gebru,
  Hutchinson, Smith-Loud, Theron, and Barnes}{Raji et~al\mbox{.}}{2020}]%
        {raji2020closing}
\bibfield{author}{\bibinfo{person}{Inioluwa~Deborah Raji},
  \bibinfo{person}{Andrew Smart}, \bibinfo{person}{Rebecca~N White},
  \bibinfo{person}{Margaret Mitchell}, \bibinfo{person}{Timnit Gebru},
  \bibinfo{person}{Ben Hutchinson}, \bibinfo{person}{Jamila Smith-Loud},
  \bibinfo{person}{Daniel Theron}, {and} \bibinfo{person}{Parker Barnes}.}
  \bibinfo{year}{2020}\natexlab{}.
\newblock \showarticletitle{Closing the AI accountability gap: defining an
  end-to-end framework for internal algorithmic auditing}. In
  \bibinfo{booktitle}{\emph{Proceedings of the 2020 Conference on Fairness,
  Accountability, and Transparency}}. \bibinfo{pages}{33--44}.
\newblock


\bibitem[\protect\citeauthoryear{Ras, van Gerven, and Haselager}{Ras
  et~al\mbox{.}}{2018}]%
        {ras_explanation_2018}
\bibfield{author}{\bibinfo{person}{Gabriëlle Ras}, \bibinfo{person}{Marcel van
  Gerven}, {and} \bibinfo{person}{Pim Haselager}.}
  \bibinfo{year}{2018}\natexlab{}.
\newblock \showarticletitle{Explanation {Methods} in {Deep} {Learning}:
  {Users}, {Values}, {Concerns} and {Challenges}}.
\newblock In \bibinfo{booktitle}{\emph{Explainable and {Interpretable} {Models}
  in {Computer} {Vision} and {Machine} {Learning}}},
  \bibfield{editor}{\bibinfo{person}{Hugo~Jair Escalante},
  \bibinfo{person}{Sergio Escalera}, \bibinfo{person}{Isabelle Guyon},
  \bibinfo{person}{Xavier Baró}, \bibinfo{person}{Yağmur Güçlütürk},
  \bibinfo{person}{Umut Güçlü}, {and} \bibinfo{person}{Marcel van Gerven}}
  (Eds.). \bibinfo{publisher}{Springer International Publishing},
  \bibinfo{address}{Cham}, \bibinfo{pages}{19--36}.
\newblock
\showISBNx{978-3-319-98131-4}
\urldef\tempurl%
\url{https://doi.org/10.1007/978-3-319-98131-4_2}
\showDOI{\tempurl}


\bibitem[\protect\citeauthoryear{Renkl}{Renkl}{2014}]%
        {renkl2014toward}
\bibfield{author}{\bibinfo{person}{Alexander Renkl}.}
  \bibinfo{year}{2014}\natexlab{}.
\newblock \showarticletitle{Toward an instructionally oriented theory of
  example-based learning}.
\newblock \bibinfo{journal}{\emph{Cognitive science}} \bibinfo{volume}{38},
  \bibinfo{number}{1} (\bibinfo{year}{2014}), \bibinfo{pages}{1--37}.
\newblock


\bibitem[\protect\citeauthoryear{Ribeiro, Singh, and Guestrin}{Ribeiro
  et~al\mbox{.}}{2016}]%
        {ribeiro2016model}
\bibfield{author}{\bibinfo{person}{Marco~Tulio Ribeiro},
  \bibinfo{person}{Sameer Singh}, {and} \bibinfo{person}{Carlos Guestrin}.}
  \bibinfo{year}{2016}\natexlab{}.
\newblock \showarticletitle{Model-agnostic interpretability of machine
  learning}.
\newblock \bibinfo{journal}{\emph{arXiv preprint arXiv:1606.05386}}
  (\bibinfo{year}{2016}).
\newblock


\bibitem[\protect\citeauthoryear{Ribera and Lapedriza}{Ribera and
  Lapedriza}{2019}]%
        {ribera2019can}
\bibfield{author}{\bibinfo{person}{Mireia Ribera} {and} \bibinfo{person}{Agata
  Lapedriza}.} \bibinfo{year}{2019}\natexlab{}.
\newblock \showarticletitle{Can we do better explanations? A proposal of
  user-centered explainable AI.}. In \bibinfo{booktitle}{\emph{IUI Workshops}}.
\newblock


\bibitem[\protect\citeauthoryear{Roscher, Bohn, Duarte, and Garcke}{Roscher
  et~al\mbox{.}}{2020}]%
        {roscher_explainable_2020}
\bibfield{author}{\bibinfo{person}{Ribana Roscher}, \bibinfo{person}{Bastian
  Bohn}, \bibinfo{person}{Marco~F. Duarte}, {and} \bibinfo{person}{Jochen
  Garcke}.} \bibinfo{year}{2020}\natexlab{}.
\newblock \showarticletitle{Explainable {Machine} {Learning} for {Scientific}
  {Insights} and {Discoveries}}.
\newblock \bibinfo{journal}{\emph{IEEE Access}}  \bibinfo{volume}{8}
  (\bibinfo{year}{2020}), \bibinfo{pages}{42200--42216}.
\newblock
\showISSN{2169-3536}
\urldef\tempurl%
\url{https://doi.org/10.1109/ACCESS.2020.2976199}
\showDOI{\tempurl}


\bibitem[\protect\citeauthoryear{Samek, Wiegand, and M{\"u}ller}{Samek
  et~al\mbox{.}}{2017}]%
        {samek2017explainable}
\bibfield{author}{\bibinfo{person}{Wojciech Samek}, \bibinfo{person}{Thomas
  Wiegand}, {and} \bibinfo{person}{Klaus-Robert M{\"u}ller}.}
  \bibinfo{year}{2017}\natexlab{}.
\newblock \showarticletitle{Explainable artificial intelligence: Understanding,
  visualizing and interpreting deep learning models}.
\newblock \bibinfo{journal}{\emph{arXiv preprint arXiv:1708.08296}}
  (\bibinfo{year}{2017}).
\newblock


\bibitem[\protect\citeauthoryear{Schlegel, Arnout, El-Assady, Oelke, and
  Keim}{Schlegel et~al\mbox{.}}{2019}]%
        {schlegel_towards_2019}
\bibfield{author}{\bibinfo{person}{Udo Schlegel}, \bibinfo{person}{Hiba
  Arnout}, \bibinfo{person}{Mennatallah El-Assady}, \bibinfo{person}{Daniela
  Oelke}, {and} \bibinfo{person}{Daniel~A. Keim}.}
  \bibinfo{year}{2019}\natexlab{}.
\newblock \showarticletitle{Towards a {Rigorous} {Evaluation} of {XAI}
  {Methods} on {Time} {Series}}.
\newblock \bibinfo{journal}{\emph{arXiv:1909.07082 [cs]}}
  (\bibinfo{date}{Sept.} \bibinfo{year}{2019}).
\newblock
\urldef\tempurl%
\url{http://arxiv.org/abs/1909.07082}
\showURL{%
\tempurl}
\newblock
\shownote{arXiv: 1909.07082.}


\bibitem[\protect\citeauthoryear{Schneider and Handali}{Schneider and
  Handali}{2019}]%
        {schneider_personalized_2019}
\bibfield{author}{\bibinfo{person}{Johannes Schneider} {and}
  \bibinfo{person}{Joshua Handali}.} \bibinfo{year}{2019}\natexlab{}.
\newblock \showarticletitle{{PERSONALIZED} {EXPLANATION} {FOR} {MACHINE}
  {LEARNING}: {A} {CONCEPTUALIZATION}}. In
  \bibinfo{booktitle}{\emph{Proceedings of the 27th {European} {Conference} on
  {Information} {Systems} ({ECIS})}}. \bibinfo{address}{Stockholm \& Uppsala,
  Sweden}.
\newblock
\urldef\tempurl%
\url{https://aisel.aisnet.org/ecis2019_rp/171}
\showURL{%
\tempurl}


\bibitem[\protect\citeauthoryear{Shneiderman}{Shneiderman}{2020}]%
        {shneiderman2020human}
\bibfield{author}{\bibinfo{person}{Ben Shneiderman}.}
  \bibinfo{year}{2020}\natexlab{}.
\newblock \showarticletitle{Human-centered artificial intelligence: Reliable,
  safe \& trustworthy}.
\newblock \bibinfo{journal}{\emph{International Journal of Human--Computer
  Interaction}} \bibinfo{volume}{36}, \bibinfo{number}{6}
  (\bibinfo{year}{2020}), \bibinfo{pages}{495--504}.
\newblock


\bibitem[\protect\citeauthoryear{Spinner, Schlegel, Schäfer, and
  El-Assady}{Spinner et~al\mbox{.}}{2020}]%
        {spinner_explainer_2020}
\bibfield{author}{\bibinfo{person}{Thilo Spinner}, \bibinfo{person}{Udo
  Schlegel}, \bibinfo{person}{Hanna Schäfer}, {and}
  \bibinfo{person}{Mennatallah El-Assady}.} \bibinfo{year}{2020}\natexlab{}.
\newblock \showarticletitle{{explAIner}: {A} {Visual} {Analytics} {Framework}
  for {Interactive} and {Explainable} {Machine} {Learning}}.
\newblock \bibinfo{journal}{\emph{IEEE Transactions on Visualization and
  Computer Graphics}} \bibinfo{volume}{26}, \bibinfo{number}{1}
  (\bibinfo{date}{Jan.} \bibinfo{year}{2020}), \bibinfo{pages}{1064--1074}.
\newblock
\showISSN{1941-0506}
\urldef\tempurl%
\url{https://doi.org/10.1109/TVCG.2019.2934629}
\showDOI{\tempurl}
\newblock
\shownote{Conference Name: IEEE Transactions on Visualization and Computer
  Graphics.}


\bibitem[\protect\citeauthoryear{Spinuzzi}{Spinuzzi}{2005}]%
        {spinuzzi2005methodology}
\bibfield{author}{\bibinfo{person}{Clay Spinuzzi}.}
  \bibinfo{year}{2005}\natexlab{}.
\newblock \showarticletitle{The methodology of participatory design}.
\newblock \bibinfo{journal}{\emph{Technical communication}}
  \bibinfo{volume}{52}, \bibinfo{number}{2} (\bibinfo{year}{2005}),
  \bibinfo{pages}{163--174}.
\newblock


\bibitem[\protect\citeauthoryear{Sundararajan, Xu, Taly, Sayres, and
  Najmi}{Sundararajan et~al\mbox{.}}{2019}]%
        {sundararajan2019exploring}
\bibfield{author}{\bibinfo{person}{Mukund Sundararajan},
  \bibinfo{person}{Jinhua Xu}, \bibinfo{person}{Ankur Taly},
  \bibinfo{person}{Rory Sayres}, {and} \bibinfo{person}{Amir Najmi}.}
  \bibinfo{year}{2019}\natexlab{}.
\newblock \showarticletitle{Exploring Principled Visualizations for Deep
  Network Attributions.}. In \bibinfo{booktitle}{\emph{IUI Workshops}},
  Vol.~\bibinfo{volume}{4}.
\newblock


\bibitem[\protect\citeauthoryear{Suresh, Lao, and Liccardi}{Suresh
  et~al\mbox{.}}{2020}]%
        {sureshMisplaced}
\bibfield{author}{\bibinfo{person}{Harini Suresh}, \bibinfo{person}{Natalie
  Lao}, {and} \bibinfo{person}{Ilaria Liccardi}.}
  \bibinfo{year}{2020}\natexlab{}.
\newblock \showarticletitle{Misplaced Trust: Measuring the Interference of
  Machine Learning in Human Decision-Making}. In
  \bibinfo{booktitle}{\emph{WebSci '20: 12th {ACM} Conference on Web Science,
  Southampton, UK, July 6-10, 2020}}, \bibfield{editor}{\bibinfo{person}{Emilio
  Ferrara}, \bibinfo{person}{Pauline Leonard}, {and} \bibinfo{person}{Wendy
  Hall}} (Eds.). \bibinfo{publisher}{{ACM}}, \bibinfo{pages}{315--324}.
\newblock
\urldef\tempurl%
\url{https://doi.org/10.1145/3394231.3397922}
\showDOI{\tempurl}


\bibitem[\protect\citeauthoryear{Thatcher, O’Sullivan, and Mahmoudi}{Thatcher
  et~al\mbox{.}}{2016}]%
        {thatcher_data_2016}
\bibfield{author}{\bibinfo{person}{Jim Thatcher}, \bibinfo{person}{David
  O’Sullivan}, {and} \bibinfo{person}{Dillon Mahmoudi}.}
  \bibinfo{year}{2016}\natexlab{}.
\newblock \showarticletitle{Data colonialism through accumulation by
  dispossession: {New} metaphors for daily data}.
\newblock \bibinfo{journal}{\emph{Environment and Planning D: Society and
  Space}} \bibinfo{volume}{34}, \bibinfo{number}{6} (\bibinfo{date}{Dec.}
  \bibinfo{year}{2016}), \bibinfo{pages}{990--1006}.
\newblock
\showISSN{0263-7758, 1472-3433}
\urldef\tempurl%
\url{https://doi.org/10.1177/0263775816633195}
\showDOI{\tempurl}


\bibitem[\protect\citeauthoryear{Theodorou, Wortham, and Bryson}{Theodorou
  et~al\mbox{.}}{2017}]%
        {theodorou2017designing}
\bibfield{author}{\bibinfo{person}{Andreas Theodorou},
  \bibinfo{person}{Robert~H Wortham}, {and} \bibinfo{person}{Joanna~J Bryson}.}
  \bibinfo{year}{2017}\natexlab{}.
\newblock \showarticletitle{Designing and implementing transparency for real
  time inspection of autonomous robots}.
\newblock \bibinfo{journal}{\emph{Connection Science}} \bibinfo{volume}{29},
  \bibinfo{number}{3} (\bibinfo{year}{2017}), \bibinfo{pages}{230--241}.
\newblock


\bibitem[\protect\citeauthoryear{Tomsett, Braines, Harborne, Preece, and
  Chakraborty}{Tomsett et~al\mbox{.}}{2018}]%
        {tomsett_interpretable_2018}
\bibfield{author}{\bibinfo{person}{Richard Tomsett}, \bibinfo{person}{Dave
  Braines}, \bibinfo{person}{Dan Harborne}, \bibinfo{person}{Alun Preece},
  {and} \bibinfo{person}{Supriyo Chakraborty}.}
  \bibinfo{year}{2018}\natexlab{}.
\newblock \showarticletitle{Interpretable to {Whom}? {A} {Role}-based {Model}
  for {Analyzing} {Interpretable} {Machine} {Learning} {Systems}}. In
  \bibinfo{booktitle}{\emph{{arXiv}:1806.07552 [cs]}}.
\newblock
\urldef\tempurl%
\url{http://arxiv.org/abs/1806.07552}
\showURL{%
\tempurl}
\newblock
\shownote{arXiv: 1806.07552.}


\bibitem[\protect\citeauthoryear{Tonekaboni, Joshi, McCradden, and
  Goldenberg}{Tonekaboni et~al\mbox{.}}{2019}]%
        {tonekaboni_what_2019}
\bibfield{author}{\bibinfo{person}{Sana Tonekaboni}, \bibinfo{person}{Shalmali
  Joshi}, \bibinfo{person}{Melissa~D. McCradden}, {and} \bibinfo{person}{Anna
  Goldenberg}.} \bibinfo{year}{2019}\natexlab{}.
\newblock \showarticletitle{What {Clinicians} {Want}: {Contextualizing}
  {Explainable} {Machine} {Learning} for {Clinical} {End} {Use}}. In
  \bibinfo{booktitle}{\emph{Machine {Learning} for {Healthcare} {Conference}}}.
  \bibinfo{pages}{359--380}.
\newblock
\urldef\tempurl%
\url{http://proceedings.mlr.press/v106/tonekaboni19a.html}
\showURL{%
\tempurl}
\newblock
\shownote{ISSN: 1938-7228 Section: Machine Learning.}


\bibitem[\protect\citeauthoryear{Tullio, Dey, Chalecki, and Fogarty}{Tullio
  et~al\mbox{.}}{2007}]%
        {tullio_how_2007}
\bibfield{author}{\bibinfo{person}{Joe Tullio}, \bibinfo{person}{Anind~K. Dey},
  \bibinfo{person}{Jason Chalecki}, {and} \bibinfo{person}{James Fogarty}.}
  \bibinfo{year}{2007}\natexlab{}.
\newblock \showarticletitle{How it works: a field study of non-technical users
  interacting with an intelligent system}. In
  \bibinfo{booktitle}{\emph{Proceedings of the {SIGCHI} {Conference} on {Human}
  {Factors} in {Computing} {Systems}}} \emph{(\bibinfo{series}{{CHI} '07})}.
  \bibinfo{publisher}{Association for Computing Machinery},
  \bibinfo{address}{San Jose, California, USA}, \bibinfo{pages}{31--40}.
\newblock
\showISBNx{9781595935939}
\urldef\tempurl%
\url{https://doi.org/10.1145/1240624.1240630}
\showDOI{\tempurl}


\bibitem[\protect\citeauthoryear{Ustun, Spangher, and Liu}{Ustun
  et~al\mbox{.}}{2019}]%
        {ustun2019recourse}
\bibfield{author}{\bibinfo{person}{Berk Ustun}, \bibinfo{person}{Alexander
  Spangher}, {and} \bibinfo{person}{Yang Liu}.}
  \bibinfo{year}{2019}\natexlab{}.
\newblock \showarticletitle{Actionable Recourse in Linear Classification}. In
  \bibinfo{booktitle}{\emph{Proceedings of the Conference on Fairness,
  Accountability, and Transparency}} (Atlanta, GA, USA)
  \emph{(\bibinfo{series}{FAT* '19})}. \bibinfo{publisher}{Association for
  Computing Machinery}, \bibinfo{address}{New York, NY, USA},
  \bibinfo{pages}{10–19}.
\newblock
\showISBNx{9781450361255}
\urldef\tempurl%
\url{https://doi.org/10.1145/3287560.3287566}
\showDOI{\tempurl}


\bibitem[\protect\citeauthoryear{Vayena, Blasimme, and Cohen}{Vayena
  et~al\mbox{.}}{2018}]%
        {vayena_machine_2018}
\bibfield{author}{\bibinfo{person}{Effy Vayena}, \bibinfo{person}{Alessandro
  Blasimme}, {and} \bibinfo{person}{I.~Glenn Cohen}.}
  \bibinfo{year}{2018}\natexlab{}.
\newblock \showarticletitle{Machine learning in medicine: {Addressing} ethical
  challenges}.
\newblock \bibinfo{journal}{\emph{PLOS Medicine}} \bibinfo{volume}{15},
  \bibinfo{number}{11} (\bibinfo{date}{Nov.} \bibinfo{year}{2018}),
  \bibinfo{pages}{e1002689}.
\newblock
\showISSN{1549-1676}
\urldef\tempurl%
\url{https://doi.org/10.1371/journal.pmed.1002689}
\showDOI{\tempurl}


\bibitem[\protect\citeauthoryear{Wachter, Mittelstadt, and Russell}{Wachter
  et~al\mbox{.}}{2018}]%
        {wachter_counterfactual_2018}
\bibfield{author}{\bibinfo{person}{Sandra Wachter}, \bibinfo{person}{Brent
  Mittelstadt}, {and} \bibinfo{person}{Chris Russell}.}
  \bibinfo{year}{2018}\natexlab{}.
\newblock \showarticletitle{Counterfactual {Explanations} without {Opening} the
  {Black} {Box}: {Automated} {Decisions} and the {GDPR}}.
\newblock \bibinfo{journal}{\emph{Harvard Journal of Law \& Technology}}
  \bibinfo{volume}{31}, \bibinfo{number}{2} (\bibinfo{date}{March}
  \bibinfo{year}{2018}), \bibinfo{pages}{841--887}.
\newblock
\urldef\tempurl%
\url{http://arxiv.org/abs/1711.00399}
\showURL{%
\tempurl}
\newblock
\shownote{arXiv: 1711.00399.}


\bibitem[\protect\citeauthoryear{Wang, Yang, Abdul, and Lim}{Wang
  et~al\mbox{.}}{2019}]%
        {wang2019designing}
\bibfield{author}{\bibinfo{person}{Danding Wang}, \bibinfo{person}{Qian Yang},
  \bibinfo{person}{Ashraf Abdul}, {and} \bibinfo{person}{Brian~Y Lim}.}
  \bibinfo{year}{2019}\natexlab{}.
\newblock \showarticletitle{Designing theory-driven user-centric explainable
  AI}. In \bibinfo{booktitle}{\emph{Proceedings of the 2019 CHI conference on
  human factors in computing systems}}. \bibinfo{pages}{1--15}.
\newblock


\bibitem[\protect\citeauthoryear{Weber}{Weber}{2012}]%
        {hartelius_review_2012}
\bibfield{author}{\bibinfo{person}{Ryan Weber}.}
  \bibinfo{year}{2012}\natexlab{}.
\newblock \showarticletitle{Review of {The} {Rhetoric} of {Expertise}}.
\newblock \bibinfo{journal}{\emph{Rhetoric and Public Affairs}}
  \bibinfo{volume}{15}, \bibinfo{number}{1} (\bibinfo{year}{2012}),
  \bibinfo{pages}{193--196}.
\newblock
\showISSN{1094-8392}
\urldef\tempurl%
\url{https://www.jstor.org/stable/41955617}
\showURL{%
\tempurl}
\newblock
\shownote{Publisher: Michigan State University Press.}


\bibitem[\protect\citeauthoryear{Weller}{Weller}{2019}]%
        {weller_transparency_2019}
\bibfield{author}{\bibinfo{person}{Adrian Weller}.}
  \bibinfo{year}{2019}\natexlab{}.
\newblock \showarticletitle{Transparency: {Motivations} and {Challenges}}.
\newblock \bibinfo{journal}{\emph{arXiv:1708.01870 [cs]}} (\bibinfo{date}{Aug.}
  \bibinfo{year}{2019}).
\newblock
\urldef\tempurl%
\url{http://arxiv.org/abs/1708.01870}
\showURL{%
\tempurl}
\newblock
\shownote{arXiv: 1708.01870.}


\bibitem[\protect\citeauthoryear{{Wexler}, {Pushkarna}, {Bolukbasi},
  {Wattenberg}, {Viégas}, and {Wilson}}{{Wexler} et~al\mbox{.}}{2020}]%
        {wexler2020what}
\bibfield{author}{\bibinfo{person}{J. {Wexler}}, \bibinfo{person}{M.
  {Pushkarna}}, \bibinfo{person}{T. {Bolukbasi}}, \bibinfo{person}{M.
  {Wattenberg}}, \bibinfo{person}{F. {Viégas}}, {and} \bibinfo{person}{J.
  {Wilson}}.} \bibinfo{year}{2020}\natexlab{}.
\newblock \showarticletitle{The What-If Tool: Interactive Probing of Machine
  Learning Models}.
\newblock \bibinfo{journal}{\emph{IEEE Transactions on Visualization and
  Computer Graphics}} \bibinfo{volume}{26}, \bibinfo{number}{1}
  (\bibinfo{year}{2020}), \bibinfo{pages}{56--65}.
\newblock


\bibitem[\protect\citeauthoryear{Williams, Faulkner, and Fleck}{Williams
  et~al\mbox{.}}{1998}]%
        {williams_exploring_1998}
\bibfield{editor}{\bibinfo{person}{Robin Williams}, \bibinfo{person}{Wendy
  Faulkner}, {and} \bibinfo{person}{James Fleck}} (Eds.).
  \bibinfo{year}{1998}\natexlab{}.
\newblock \bibinfo{booktitle}{\emph{Exploring {Expertise}}}.
\newblock \bibinfo{publisher}{Palgrave Macmillan UK},
  \bibinfo{address}{London}.
\newblock
\showISBNx{978-1-349-13695-7 978-1-349-13693-3}
\urldef\tempurl%
\url{https://doi.org/10.1007/978-1-349-13693-3}
\showDOI{\tempurl}


\bibitem[\protect\citeauthoryear{Wobbrock and Kientz}{Wobbrock and
  Kientz}{2016}]%
        {wobbrock2016research}
\bibfield{author}{\bibinfo{person}{Jacob~O Wobbrock} {and}
  \bibinfo{person}{Julie~A Kientz}.} \bibinfo{year}{2016}\natexlab{}.
\newblock \showarticletitle{Research contributions in human-computer
  interaction}.
\newblock \bibinfo{journal}{\emph{interactions}} \bibinfo{volume}{23},
  \bibinfo{number}{3} (\bibinfo{year}{2016}), \bibinfo{pages}{38--44}.
\newblock


\bibitem[\protect\citeauthoryear{Wolf}{Wolf}{2019}]%
        {wolf_explainability_2019}
\bibfield{author}{\bibinfo{person}{Christine~T. Wolf}.}
  \bibinfo{year}{2019}\natexlab{}.
\newblock \showarticletitle{Explainability scenarios: towards scenario-based
  {XAI} design}. In \bibinfo{booktitle}{\emph{Proceedings of the 24th
  {International} {Conference} on {Intelligent} {User} {Interfaces}}}.
  \bibinfo{publisher}{ACM}, \bibinfo{address}{Marina del Ray California},
  \bibinfo{pages}{252--257}.
\newblock
\showISBNx{978-1-4503-6272-6}
\urldef\tempurl%
\url{https://doi.org/10.1145/3301275.3302317}
\showDOI{\tempurl}


\bibitem[\protect\citeauthoryear{Xie, Gao, and Chen}{Xie et~al\mbox{.}}{2019}]%
        {xie2019outlining}
\bibfield{author}{\bibinfo{person}{Yao Xie}, \bibinfo{person}{Ge Gao}, {and}
  \bibinfo{person}{Xiang~'Anthony' Chen}.} \bibinfo{year}{2019}\natexlab{}.
\newblock \bibinfo{title}{Outlining the Design Space of Explainable Intelligent
  Systems for Medical Diagnosis}.
\newblock
\newblock
\showeprint[arxiv]{1902.06019}~[cs.HC]


\bibitem[\protect\citeauthoryear{Yang, Steinfeld, Rosé, and Zimmerman}{Yang
  et~al\mbox{.}}{2020}]%
        {yang_re-examining_2020}
\bibfield{author}{\bibinfo{person}{Qian Yang}, \bibinfo{person}{Aaron
  Steinfeld}, \bibinfo{person}{Carolyn Rosé}, {and} \bibinfo{person}{John
  Zimmerman}.} \bibinfo{year}{2020}\natexlab{}.
\newblock \showarticletitle{Re-examining {Whether}, {Why}, and {How}
  {Human}-{AI} {Interaction} {Is} {Uniquely} {Difficult} to {Design}}. In
  \bibinfo{booktitle}{\emph{Proceedings of the 2020 {CHI} {Conference} on
  {Human} {Factors} in {Computing} {Systems}}} \emph{(\bibinfo{series}{{CHI}
  '20})}. \bibinfo{publisher}{Association for Computing Machinery},
  \bibinfo{address}{Honolulu, HI, USA}, \bibinfo{pages}{1--13}.
\newblock
\showISBNx{978-1-4503-6708-0}
\urldef\tempurl%
\url{https://doi.org/10.1145/3313831.3376301}
\showDOI{\tempurl}


\bibitem[\protect\citeauthoryear{Yielder}{Yielder}{2001}]%
        {yielder_professional_2001}
\bibfield{author}{\bibinfo{person}{Jill Yielder}.}
  \bibinfo{year}{2001}\natexlab{}.
\newblock \emph{\bibinfo{title}{Professional {Expertise}: {A} {Model} for
  {Integration} and {Change}}}.
\newblock Thesis. \bibinfo{school}{ResearchSpace@Auckland}.
\newblock
\urldef\tempurl%
\url{https://researchspace.auckland.ac.nz/handle/2292/2340}
\showURL{%
\tempurl}
\newblock
\shownote{Accepted: 2008-01-30T02:04:57Z.}


\bibitem[\protect\citeauthoryear{Yielder}{Yielder}{2004}]%
        {yielder_integrated_2004}
\bibfield{author}{\bibinfo{person}{Jill Yielder}.}
  \bibinfo{year}{2004}\natexlab{}.
\newblock \showarticletitle{An integrated model of professional expertise and
  its implications for higher education}.
\newblock \bibinfo{journal}{\emph{International Journal of Lifelong Education}}
  \bibinfo{volume}{23}, \bibinfo{number}{1} (\bibinfo{date}{Jan.}
  \bibinfo{year}{2004}), \bibinfo{pages}{60--80}.
\newblock
\showISSN{0260-1370, 1464-519X}
\urldef\tempurl%
\url{https://doi.org/10.1080/0260137032000172060}
\showDOI{\tempurl}


\bibitem[\protect\citeauthoryear{Yin, Wortman~Vaughan, and Wallach}{Yin
  et~al\mbox{.}}{2019}]%
        {yin_understanding_2019}
\bibfield{author}{\bibinfo{person}{Ming Yin}, \bibinfo{person}{Jennifer
  Wortman~Vaughan}, {and} \bibinfo{person}{Hanna Wallach}.}
  \bibinfo{year}{2019}\natexlab{}.
\newblock \showarticletitle{Understanding the {Effect} of {Accuracy} on {Trust}
  in {Machine} {Learning} {Models}}. In \bibinfo{booktitle}{\emph{Proceedings
  of the 2019 {CHI} {Conference} on {Human} {Factors} in {Computing}
  {Systems}}} \emph{(\bibinfo{series}{{CHI} '19})}.
  \bibinfo{publisher}{Association for Computing Machinery},
  \bibinfo{address}{Glasgow, Scotland Uk}, \bibinfo{pages}{1--12}.
\newblock
\showISBNx{978-1-4503-5970-2}
\urldef\tempurl%
\url{https://doi.org/10.1145/3290605.3300509}
\showDOI{\tempurl}


\bibitem[\protect\citeauthoryear{Yu and Shi}{Yu and Shi}{2018}]%
        {yu_user-based_2018}
\bibfield{author}{\bibinfo{person}{Rulei Yu} {and} \bibinfo{person}{Lei Shi}.}
  \bibinfo{year}{2018}\natexlab{}.
\newblock \showarticletitle{A user-based taxonomy for deep learning
  visualization}.
\newblock \bibinfo{journal}{\emph{Visual Informatics}} \bibinfo{volume}{2},
  \bibinfo{number}{3} (\bibinfo{date}{Sept.} \bibinfo{year}{2018}),
  \bibinfo{pages}{147--154}.
\newblock
\showISSN{2468502X}
\urldef\tempurl%
\url{https://doi.org/10.1016/j.visinf.2018.09.001}
\showDOI{\tempurl}


\bibitem[\protect\citeauthoryear{Zarsky}{Zarsky}{2013}]%
        {zarsky2013transparent}
\bibfield{author}{\bibinfo{person}{Tal~Z Zarsky}.}
  \bibinfo{year}{2013}\natexlab{}.
\newblock \showarticletitle{Transparent predictions}.
\newblock \bibinfo{journal}{\emph{U. Ill. L. Rev.}} (\bibinfo{year}{2013}),
  \bibinfo{pages}{1503}.
\newblock


\bibitem[\protect\citeauthoryear{Zeiler and Fergus}{Zeiler and Fergus}{2014}]%
        {fleet_visualizing_2014}
\bibfield{author}{\bibinfo{person}{Matthew~D. Zeiler} {and}
  \bibinfo{person}{Rob Fergus}.} \bibinfo{year}{2014}\natexlab{}.
\newblock \showarticletitle{Visualizing and {Understanding} {Convolutional}
  {Networks}}.
\newblock In \bibinfo{booktitle}{\emph{Computer {Vision} – {ECCV} 2014}},
  \bibfield{editor}{\bibinfo{person}{David Fleet}, \bibinfo{person}{Tomas
  Pajdla}, \bibinfo{person}{Bernt Schiele}, {and} \bibinfo{person}{Tinne
  Tuytelaars}} (Eds.). Vol.~\bibinfo{volume}{8689}.
  \bibinfo{publisher}{Springer International Publishing},
  \bibinfo{address}{Cham}, \bibinfo{pages}{818--833}.
\newblock
\showISBNx{978-3-319-10589-5 978-3-319-10590-1}
\urldef\tempurl%
\url{https://doi.org/10.1007/978-3-319-10590-1_53}
\showDOI{\tempurl}
\newblock
\shownote{Series Title: Lecture Notes in Computer Science.}


\bibitem[\protect\citeauthoryear{Zhang, Wu, and Zhu}{Zhang
  et~al\mbox{.}}{2018}]%
        {Zhang_2018_CVPR}
\bibfield{author}{\bibinfo{person}{Quanshi Zhang}, \bibinfo{person}{Ying~Nian
  Wu}, {and} \bibinfo{person}{Song-Chun Zhu}.} \bibinfo{year}{2018}\natexlab{}.
\newblock \showarticletitle{Interpretable Convolutional Neural Networks}. In
  \bibinfo{booktitle}{\emph{Proceedings of the IEEE Conference on Computer
  Vision and Pattern Recognition (CVPR)}}.
\newblock


\bibitem[\protect\citeauthoryear{Zhang, Liao, and Bellamy}{Zhang
  et~al\mbox{.}}{2020}]%
        {zhang2020effect}
\bibfield{author}{\bibinfo{person}{Yunfeng Zhang}, \bibinfo{person}{Q~Vera
  Liao}, {and} \bibinfo{person}{Rachel~KE Bellamy}.}
  \bibinfo{year}{2020}\natexlab{}.
\newblock \showarticletitle{Effect of confidence and explanation on accuracy
  and trust calibration in AI-assisted decision making}. In
  \bibinfo{booktitle}{\emph{Proceedings of the 2020 Conference on Fairness,
  Accountability, and Transparency}}. \bibinfo{pages}{295--305}.
\newblock


\bibitem[\protect\citeauthoryear{Zhao, Benbasat, and Cavusoglu}{Zhao
  et~al\mbox{.}}{2019}]%
        {zhao2019transparency}
\bibfield{author}{\bibinfo{person}{Ruijing Zhao}, \bibinfo{person}{Izak
  Benbasat}, {and} \bibinfo{person}{Hasan Cavusoglu}.}
  \bibinfo{year}{2019}\natexlab{}.
\newblock \showarticletitle{Transparency in Advice-Giving Systems: A Framework
  and a Research Model for Transparency Provision.}. In
  \bibinfo{booktitle}{\emph{IUI Workshops}}.
\newblock


\bibitem[\protect\citeauthoryear{Zilke, Menc{\'\i}a, and Janssen}{Zilke
  et~al\mbox{.}}{2016}]%
        {zilke2016deepred}
\bibfield{author}{\bibinfo{person}{Jan~Ruben Zilke},
  \bibinfo{person}{Eneldo~Loza Menc{\'\i}a}, {and} \bibinfo{person}{Frederik
  Janssen}.} \bibinfo{year}{2016}\natexlab{}.
\newblock \showarticletitle{Deepred--rule extraction from deep neural
  networks}. In \bibinfo{booktitle}{\emph{International Conference on Discovery
  Science}}. Springer, \bibinfo{pages}{457--473}.
\newblock


\end{thebibliography}
\end{document}